\documentclass[11pt]{article}

\usepackage[numbers]{natbib}

\usepackage[left=2.7cm,bottom=2.7cm,right=2.7cm,top=2.7cm]{geometry}

\usepackage{authblk,float}

\usepackage[small,bf]{caption}

\usepackage{amsmath, amsthm, amssymb,multirow,hyperref}

\usepackage{epsfig,psfrag,url,verbatim}
\usepackage{graphics}
\usepackage{graphicx, algorithm}
\usepackage[parfill]{parskip}
\usepackage{subcaption}

\usepackage{titlesec}
\usepackage{lipsum}

\titlespacing{\paragraph}{%
  0pt}{
  0.0\baselineskip}{
  .3em}

\usepackage{xargs,setspace}
\usepackage{amsfonts}

\usepackage{color}

\newtheorem{cor}{Corollary}
\newtheorem{thm}{Theorem}

\newtheorem{prop}{Proposition}

\newtheorem{definition}{Definition}
\newenvironment{remark}{\begin{quote}{\bf Remark.\quad}\rm\small}{\end{quote}}

\newcommand{\pparagraph}{\noindent\textbf}

\usepackage{array,graphicx,footnote,url}
\usepackage{booktabs}
\usepackage{pifont}

\usepackage{array,amsmath,amssymb, float}
\usepackage{epsfig,psfrag,url,verbatim, multirow}
\usepackage{graphics}
\usepackage{graphicx, algorithm}
\usepackage{xargs}
\usepackage{amsfonts}
\usepackage{color}

\newcommand{\mk}{{\mathcal K}}

\newcommand{\B}{\boldsymbol}
\newcommand{\M}{\mathbf}
\newcommand{\sgn}{\operatorname{sgn}}

\newcommand{\sbt}{\mathrm{s.t.}}

\newcommand{\half}{\mbox{$\frac12$}}

\newcommand{\MU}{{\mathcal M}_{U}}
\newcommand{\plots}{Figures}
\newcommand{\plotsP}{Figures}

\newcommand{\LDSF}{\emph{Discrete Dantzig Selector}}

 \newcommand{\Exampleone}{Example-A}
 \newcommand{\Examplethree}{Example-B}
 \newcommand{\Examplefour}{Example-C}

 \newcommand{\Exampleeight}{Example-D}

  \newcommand{\snrone}{$3$}
  
  \newcommand{\snrthree}{$10$}

\newcommand{\bet}{$\| \widehat{\B\beta} - \B\beta^*\|_{2}^2$}
\newcommand{\xbet}{Prediction Error}
\newcommand{\nnz}{Number of Nonzeros}
\newcommand{\err}{Metric}
\newcommand{\errzone}{Variable Selection Error}
\renewcommand{\c}{\psi}

\DeclareMathOperator*{\argmin}{arg\,min}
\DeclareMathOperator*{\argmax}{arg\,max}

\newenvironment{myarray}[2][1]
  {\def\arraystretch{#1}\array{#2}}
  {\endarray}
\newcommand{\mmk}{k}

\newcommand{\DSF}{Dantzig Selector}

\newcommand{\MIO}{{\texttt{MILO}}~}

\renewcommand{\v}{\text{v}}

\newcommand{\btheta}{\boldsymbol \theta}
\newcommand{\bepsilon}{\boldsymbol \epsilon}

\newcommand{\bbeta}{\boldsymbol \beta}

\newcommand{\bY}{{\bf y}}

\newcommand{\bX}{{\bf X}}

\newcommand{\bx}{{\bf x}}

\newcommand{\one}{{\bf 1}}

\def\RR{\mathbb{R}}

\begin{document}

\title{The Discrete Dantzig Selector: Estimating Sparse Linear Models via Mixed Integer Linear Optimization}

\date{June, 2016}


\author[1]{\sc Rahul Mazumder \thanks{Rahul Mazumder's research was partially supported by the Office of Naval Research: ONR-024511-00001 and a grant from the Moore Sloan Foundation. email: {\texttt{rahulmaz@mit.edu}} }}
\author[2]{\sc Peter Radchenko \thanks{Peter Radchenko's research was partially supported by NSF Grant DMS-1209057. email: {\texttt{radchenk@usc.edu}} }}
\affil[1]{Massachusetts Institute of Technology}
\affil[2]{University of Southern California}


\maketitle

\begin{abstract}
We propose a novel high-dimensional linear regression estimator: the \LDSF, which minimizes the number of nonzero regression coefficients subject to a budget on the maximal absolute correlation between the features and residuals.
Motivated by the significant advances in integer optimization over the past $10$-$15$ years, we present a Mixed Integer Linear Optimization ({\texttt{MILO}}) approach to obtain \emph{certifiably optimal} global solutions to this nonconvex optimization problem.
The current state of algorithmics in integer optimization makes our proposal substantially more computationally attractive than the
least squares subset selection framework based on integer \emph{quadratic} optimization, recently proposed in~\cite{bertsimas2015best} and the
continuous nonconvex quadratic optimization framework of~\cite{mipgo-li-2016}.
We propose new discrete first-order methods, which when paired with state-of-the-art {\texttt{MILO}} solvers,
lead to good solutions for the \LDSF~problem for a given computational budget.
We illustrate that our integrated approach provides globally optimal solutions in significantly shorter computation times, when compared to off-the-shelf {\texttt{MILO}} solvers.
We demonstrate both theoretically and empirically that in a wide range of regimes the statistical properties of the \LDSF~are superior to those of popular $\ell_1$-based approaches.
We illustrate that our approach can handle problem instances with $p =10,\!000$ features with certifiable optimality making it a highly scalable combinatorial variable selection approach in sparse linear modeling.
\end{abstract}

\section{Introduction}\label{sec:intro}
We consider the familiar linear regression framework, with
response vector $\M{y} \in \RR^{n \times 1}$,
model matrix $\M{X} \in  \mathbb{R}^{n \times p}$, regression coefficients $\B\beta \in  \mathbb{R}^{p \times 1}$
and errors $\B\epsilon \in  \mathbb{R}^{n \times 1}$: $\M{y} = \M{X} \B\beta + \B\epsilon.$
We assume, unless otherwise mentioned, that the columns of $\M{X}$, denoted by $\M{x}_{j}$ for $j=1,\ldots, p$, have been standardized to have zero means and unit $\ell_{2}$-norm.
In many modern statistical applications, the number of variables, $p$, is larger than the number of observations, $n$.  In such cases, to carry out statistically meaningful
estimation, it is often assumed that the number of nonzero elements in $\B\beta$ is quite small~\cite{hastie2015statistical}.
The task is to obtain a good estimate, $\widehat{\B\beta}$, which is sparse and
serves as a good approximation to the underlying \emph{true} regression coefficient.
Of course, the basic problem of obtaining a sparse model with good data-fidelity is also of interest when the number of observations is comparable to or larger than~$p$.
In the sparse high-dimensional setting described above, two estimation approaches that have been very popular among statisticians and researchers in related fields are the Lasso~\cite{Ti96} and the Dantzig Selector~\cite{candes2007dantzig}.
Both estimators can be expressed as solutions to convex optimization problems, which can be solved using computationally attractive procedures~\cite{BV2004,becker2011templates,james2009dasso,FHT2007}, and come with strong theoretical guarantees~\cite{candes2007dantzig,bickel1,buhlmann2011statistics}.
For reasons that are explained later in this section, the primary motivation for our investigation in this paper is the Dantzig Selector, which is defined as the solution to the following linear optimization problem:
\begin{equation}\label{L1-DZ-1}
\min\limits_{\B\beta} \;\;  \|\B\beta\|_1 \;\; ~~~\sbt~~~ \;\;  \|\M{X}^\top  (\M{y}-\M{X}\B\beta)\|_\infty \leq \delta.
\end{equation}
To distinguish this estimator from our proposed approach,  we refer to it as the $\ell_{1}$-\DSF.  This estimator seeks to minimize the $\ell_1$-complexity of the coefficient vector, subject to a constraint on the maximal absolute correlation between the corresponding residual vector and the predictors.  The tuning parameter $\delta$ controls the amount of data-fidelity: a small value of $\delta$ corresponds to a good fit, and a larger value of $\delta$ leads to heavy shrinkage of the estimated regression coefficients.
\cite{candes2007dantzig} point out several reasons as to why the  feasibility set in~\eqref{L1-DZ-1} might serve as a good measure for data-fidelity.  In particular, this set is invariant with respect to orthogonal transformations on the data $(\M{y}, \M{X})$.  It can also be shown that~$\delta$ controls\footnote{More formally,
we have: $$\|\M{X}^\top(\M{y} - \M{X} \B\beta)\|_{\infty} \geq   \frac{\lambda_{\text{pmin}}(\M{X})}{2np^{\frac12}} \left(\| \M{y} - \M{X} \B\beta\|_{2}^2 -  \| \M{y} - \M{X} \widehat{\B\beta}_{\text{LS}}\|_{2}^2\right)^{\half},$$
where, $\lambda_{\text{pmin}}(\M{X})$ is the minimum \emph{nonzero} singular value of $\M{X}$,  and $\widehat{\B\beta}_{\text{LS}}$ is any least-squares solution --- see Proposition A.1 in~\cite{freund2015new} for a proof of this result.}
the residual sum of squares: the latter can be made arbitrarily close to the minimal least-squares value by decreasing~$\delta$.
The $\ell_{1}$-\DSF, like the Lasso, is used extensively as a model fitting routine to obtain a path of sparse linear models, as the data-fidelity parameter is allowed to vary~\cite{james2009dasso}, and
allows a natural extension to more general response distributions \cite{PVR.gds}.
 Note that Problem~\eqref{L1-DZ-1} can be rewritten as a linear optimization problem and can be solved quite easily for problems with $p$ in the order of thousands.  Under some mild conditions, and even for~$p$ much larger than~$n$, the corresponding estimator achieves a loss within a  logarithmic factor of the ideal mean squared error achieved if the locations of the nonzero coordinates were known~\cite{candes2007dantzig, bickel1}.

The $\ell_{1}$-\DSF, however, has limitations. In the presence of highly correlated covariates, the estimator tends to choose a dense model, typically bringing in an important variable together with its correlated cousins, which does not significantly hurt the $\ell_{1}$-norm of the corresponding coefficient vector.  If one increases the data-fidelity threshold $\delta$, the selected model becomes sparser, however, in the process, important variables might get left out.  This is largely due to the nature of the bias imparted by the $\ell_{1}$-norm, which penalizes both large and small coefficients in a similar fashion. Similar issues also arise in the case of Lasso~\cite{mhf-09-jasa,FHT-09-new,ZH08,buhlmann2011statistics}.  If the $\ell_{0}$-pseudo-norm is used instead of the $\ell_{1}$-norm, the aforementioned problems can be ameliorated: given multiple representations of the model with similar data-fidelity, the $\ell_{0}$-pseudo-norm will always prefer the most parsimonious representation.  In addition, the $\ell_{0}$-pseudo-norm does not shrink the regression coefficients: once an important variable enters the model, it comes in unshrunk with its full effect, which, in turn, drains the effect of its correlated cousins and naturally leads to a sparser model.

\pparagraph{Our Proposal.}
The preceding discussion suggests a natural question: what if we replace $\|\B\beta\|_{1}$ in Problem~\eqref{L1-DZ-1} with $\|\B\beta\|_{0}:= \sum_{i=1}^{p} \one(\beta_{i} \neq 0)$ -- the number of nonzero entries in $\B\beta$?
This leads to the
following discrete optimization problem, which also happens to define the estimator that we propose:
 \begin{equation}\label{L0-DZ-1}
\min \limits_{\B\beta} \;\;  \|\B\beta\|_0 \;\;\; ~~\sbt ~~\;\;\; \|\M{X}^\top  (\M{y}-\M{X}\B\beta)\|_\infty \leq \delta.
\end{equation}
We refer to the above estimator as the \LDSF.
A couple of questions that may be asked at this point are:
\begin{itemize}
\item Is the estimator defined via Problem~\eqref{L0-DZ-1} computationally \emph{tractable}?
\item Does the \LDSF~lead to solutions with superior statistical properties, when compared to its $\ell_{1}$ counterpart?
\end{itemize}
Addressing these questions and answering them affirmatively is the main focus of this paper.

The objective function in Problem~\eqref{L1-DZ-1}, represented by $\|\B\beta\|_1$, may be thought of as a \emph{convexification} of the discrete quantity $\|\B\beta\|_{0}$, which counts the number of nonzeros in the regression coefficient vector $\B\beta$.
The corresponding estimator seeks solutions with small $\ell_{1}$-complexity.   While this often leads to sparse solutions, i.e. those with few nonzero coefficients, the sparsity is an indirect consequence of minimizing $\|\B\beta\|_1$.  The \LDSF~on the other hand, targets sparsity \emph{directly}, in its very formulation.
Problem~\eqref{L0-DZ-1} can be reformulated as a Mixed Integer Linear Optimization ({\texttt{MILO}}) problem ---
due to the major advances in algorithmic research in {\texttt{MILO}} over the past 10-15 years, these methods
are widely considered as a mature technology in a subfield of mathematical programming~\cite{vielma2015mixed,hemmecke2010nonlinear}.
Algorithmic advances coupled with hardware and software improvements have made
{\texttt{MILO}} problems solvable to \emph{certifiable} optimality for various problem sizes of practical interest.
In this sense, it is perhaps appropriate to perceive {\texttt{MILO}} as a computationally \emph{tractable} tool.
The view of computational tractability we
adopt here is \emph{not} polynomial time tractability, but the ability of a method to provide high quality solutions with \emph{provable} optimality certificates for problem types that are encountered
in practice, in times that are appropriate for the applications being addressed.  Our approach is aligned with an intriguing recent line of work in computational statistics: the use of Mixed Integer Optimization and, more broadly, modern optimization techniques to solve certain classes of discrete problems arising in statistical estimation tasks --- see, for example, the recent works of~\cite{bertsimas2015best,bertsimas2014least}. Further background on {\texttt{MILO}} appears in Section~\ref{sec:background-mio}.

In this paper, we bring together recent advances from \emph{diverse} areas of modern mathematical optimization methods: first-order techniques in convex optimization and {\texttt{MILO}} techniques.  We provide a novel
unified algorithmic approach that
\begin{itemize}
\item[(a)] performs favorably over standalone of-the-shelf {\texttt{MILO}} solvers applicable for Problem~\eqref{L0-DZ-1}, in terms of obtaining good quality solutions with provable certificates of optimality, and
\item[(b)] scales gracefully to problem sizes up to $p=10,\!000$ or even larger.
\end{itemize}
In an extensive series of experiments with synthetic and real data we demonstrate that our unified approach solves, to global optimality, instances of Problem~\eqref{L0-DZ-1} with $n \approx 500, p \approx 100$ in seconds, and underdetermined problems with $n \approx 900, p \approx 3,\!000$ in minutes.  While it takes marginally longer to provide \emph{certificates} or guarantees of global optimality, the corresponding times are quite reasonable: in all the aforementioned instances the certificates of optimality are available within an hour.  Our approach scales to several instances of problems with $n \leq p$ and $p$ in the range $5,\!000$ to $10,\!000$, delivering optimal solutions in approximately an
hour and proving optimality within at most two days, in all instances.
We also find that the statistical properties of our estimates are substantially better than those  of computationally friendlier alternatives, like the $\ell_{1}$-\DSF, in terms of both the estimation error and the variable selection properties. Detailed results appear in Section~\ref{sec:stats-prop-expt}.

\pparagraph{Examples.}
To provide the reader with some intuition, we present a set of three examples, which illustrate the differences between the solutions to Problems~\eqref{L1-DZ-1} and~\eqref{L0-DZ-1}.  The following simple example\footnote{this example was suggested to us by Emmanuel Candes} demonstrates how
the $\ell_{1}$-based method Dantzig Selector might experience difficulty in producing a sparse solution in cases where the signal predictors are highly correlated.

Example~1.  Let~$p=n+1$. Take the first feature as $\bx_1=(1,\tau,.....,\tau)^\top$, take the $(i,j)$th entry of the feature matrix, $\M{X}$, as
$x_{ij} = 1$ if $i = j-1$ and zero otherwise, for $i=1,...,n$ and $j \geq 2$, and set $\bY=\bx_1-\bx_2$.

%
The $\ell_{1}$-norm of the \textit{sparse} representation of the response, $\bY=\bx_1-\bx_2$, equals~$2$.   Note that, given the available predictors, the response admits only one other exact representation, $\bY=\tau\bx_3+...+\tau\bx_p$.  The $\ell_{1}$ cost for this \textit{dense} representation is $\tau(n-1)$, which is lower than the corresponding value for the sparse representation when~$\tau$ is small.  Consequently, as long as $\tau(n-1)<2$, both the Lasso and the $\ell_1$-Dantzig selector select the dense representation of the response.  Alternatively, $\ell_0$-based methods recover the sparse representation.  More specifically, consider the solution to Problem~\eqref{L0-DZ-1}:~if the tuning parameter $\delta$ is set below $\tau/(1+\tau)$, then the estimator exactly recovers the sparse representation of the response.

\begin{figure}[h!]
\centering
\resizebox{\textwidth}{0.3\textheight}
{\begin{tabular}{l c c c}
\multicolumn{4}{c}{ \sf \scriptsize {\LDSF~Coefficient Profiles}} \medskip \\
&\sf{\scriptsize{Example~1}} &  \sf {\scriptsize{Example~1$'$}} & \sf {\scriptsize{Diabetes data (n=442,p=10)}}\\
\rotatebox{90}{\sf {\scriptsize{~~~~~~~~~~~~~~~~~~~~~~~Regression Coefficients}}}&
\includegraphics[width=0.3\textwidth,height=0.25\textheight,  trim =1.0cm 2.5cm .2cm 1.5cm, clip = true ]{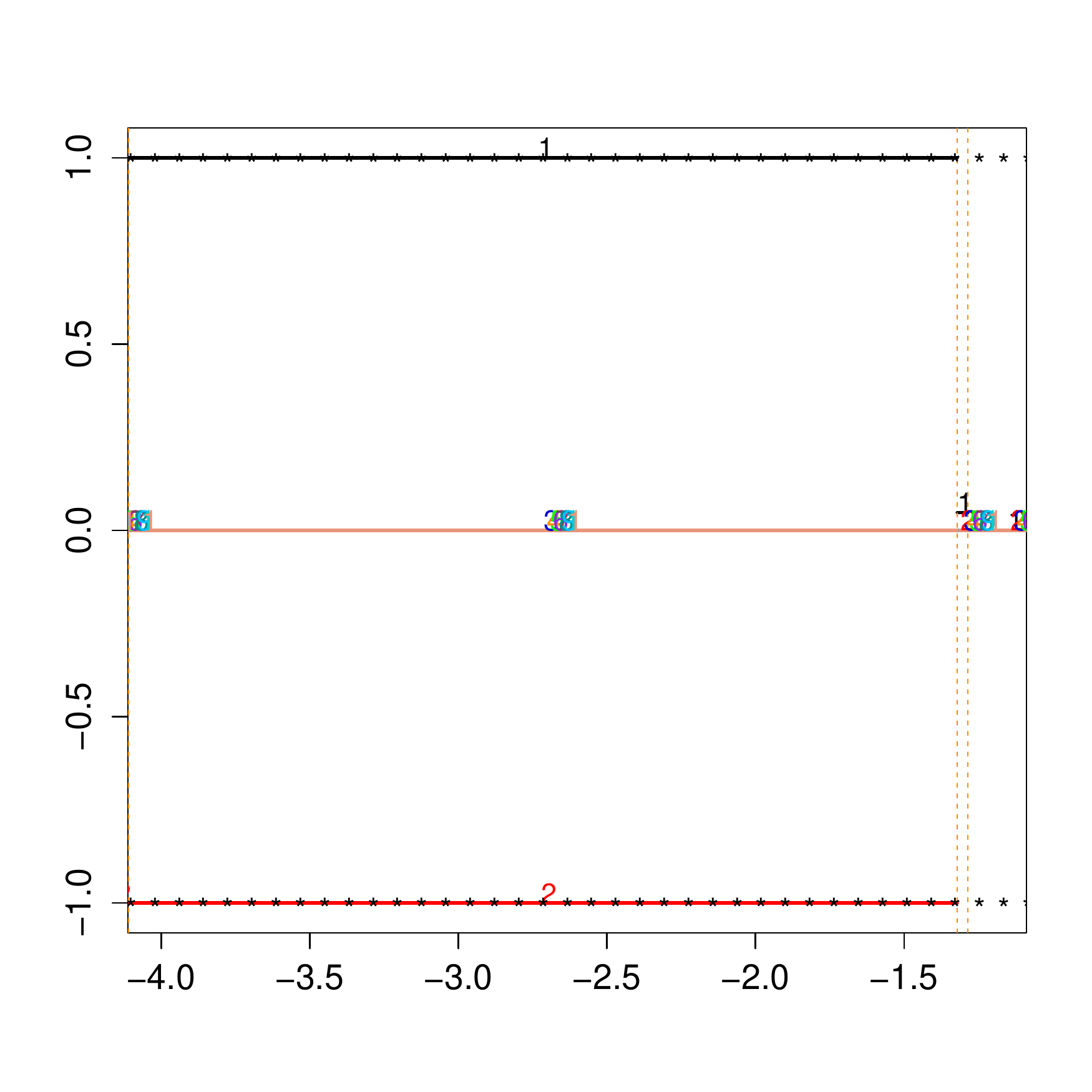}&
\includegraphics[width=0.3\textwidth,height=0.25\textheight,  trim =1.0cm 2.5cm .2cm 1.5cm, clip = true ]{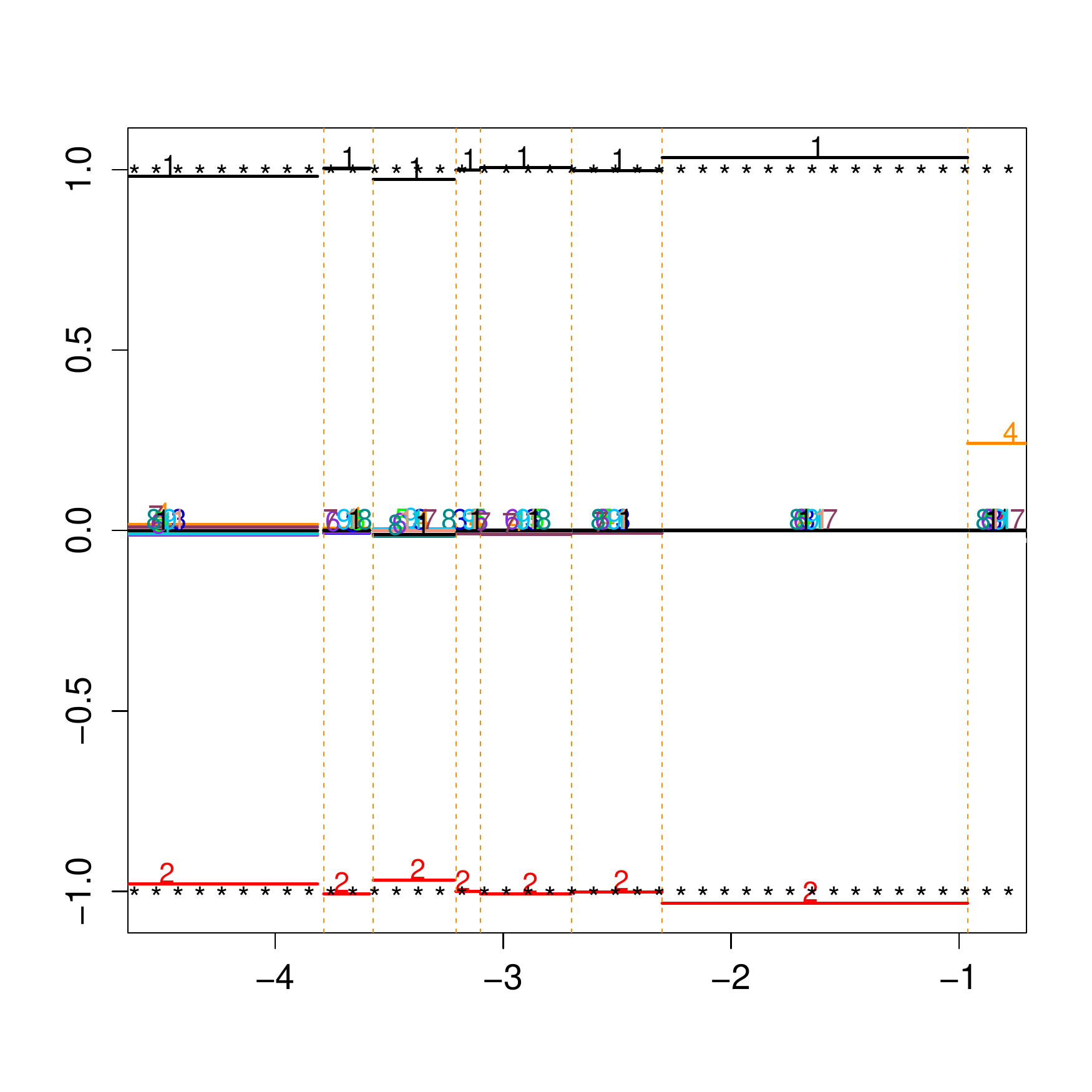}&
\includegraphics[width=0.3\textwidth,height=0.25\textheight,  trim =1.0cm 2.5cm .2cm 1.5cm, clip = true ]{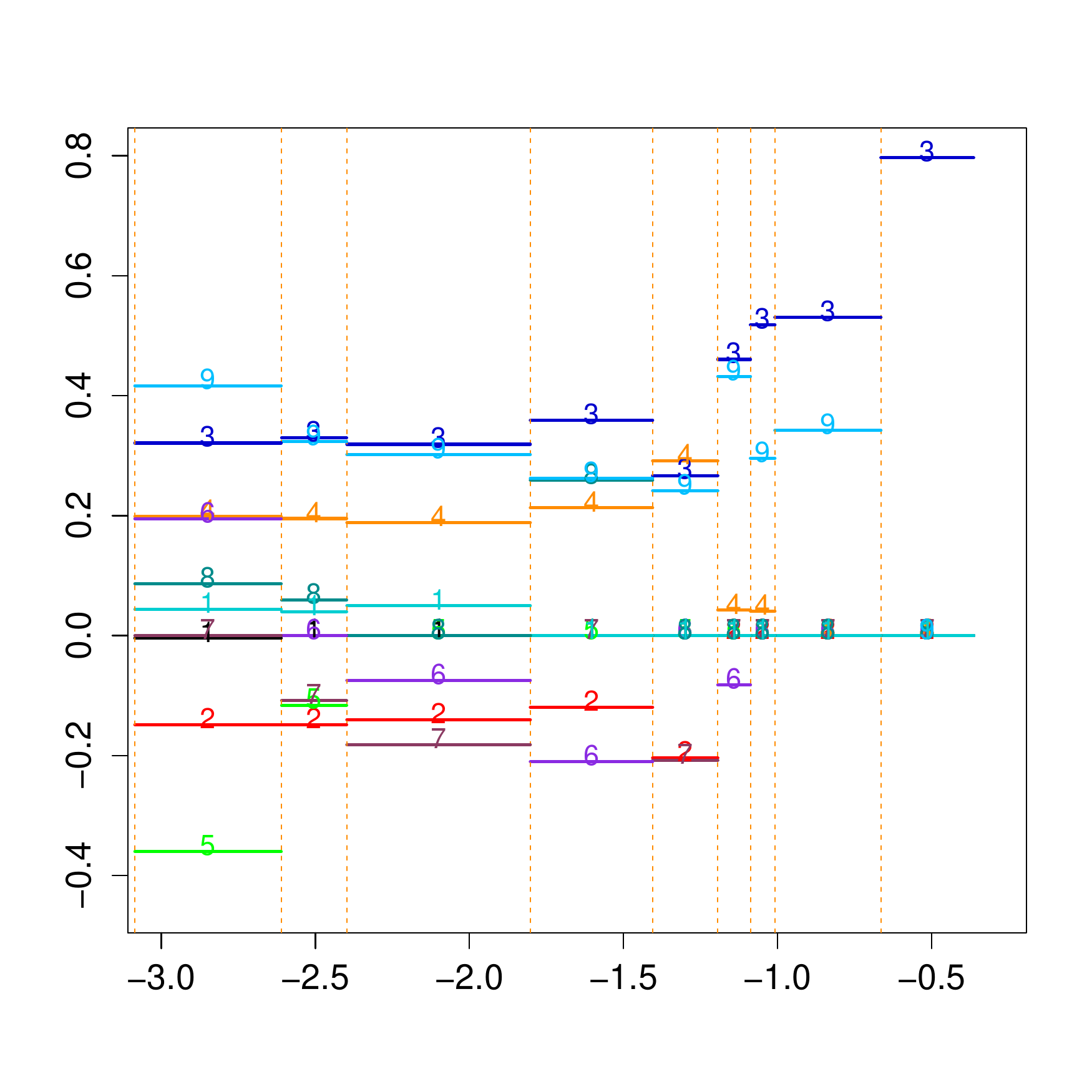} \medskip\medskip \medskip \medskip \\

\multicolumn{4}{c}{ \sf \scriptsize {$\ell_{1}$-\DSF~Coefficient Profiles}} \medskip \\
\rotatebox{90}{\sf {\scriptsize{~~~~~~~~~~~~~~~~~~~~~~Regression Coefficients}}}&
\includegraphics[width=0.3\textwidth,height=0.25\textheight,  trim =1.0cm 1.5cm .2cm 1.5cm, clip = true ]{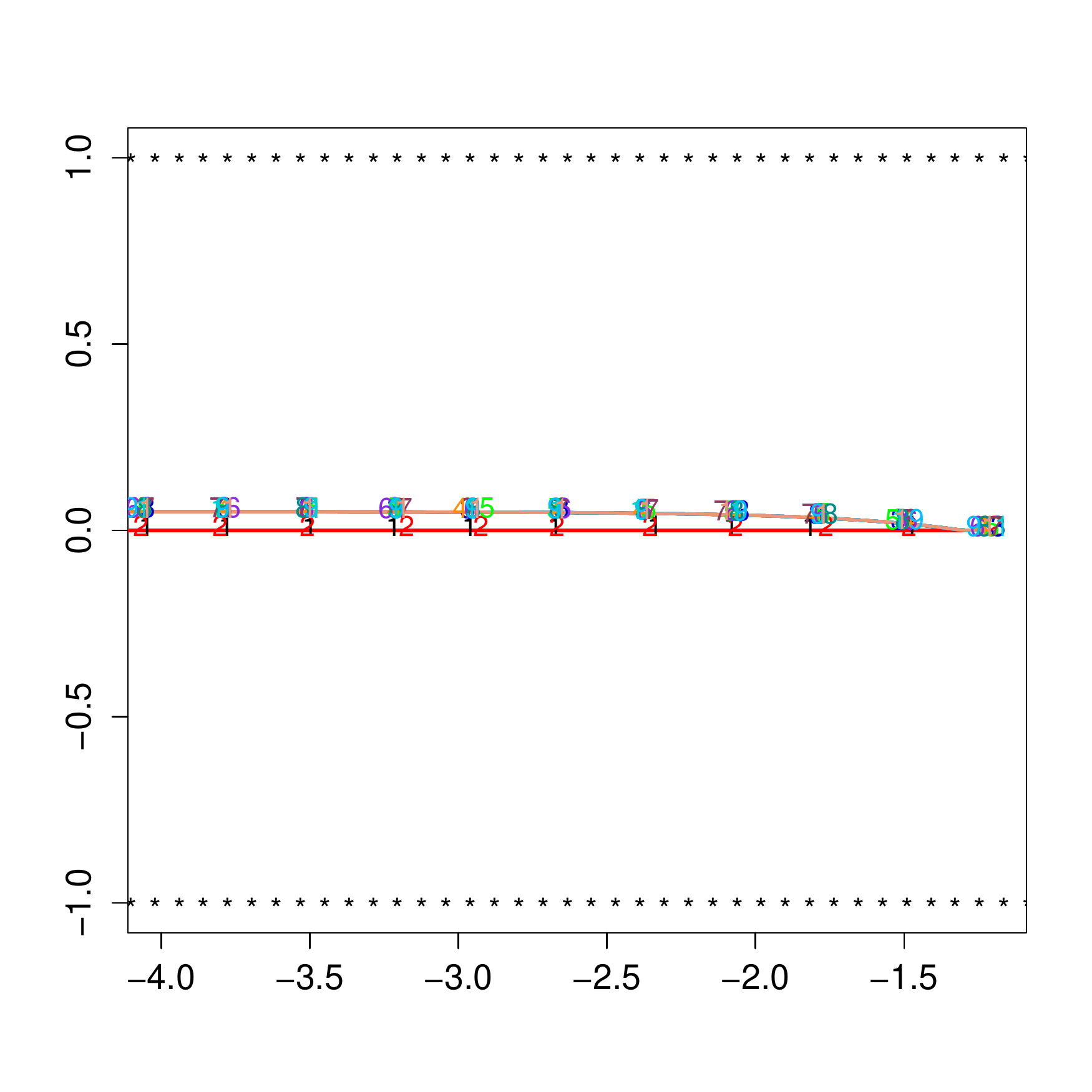}&
\includegraphics[width=0.3\textwidth,height=0.25\textheight,  trim =1.0cm 1.5cm .2cm 1.5cm, clip = true ]{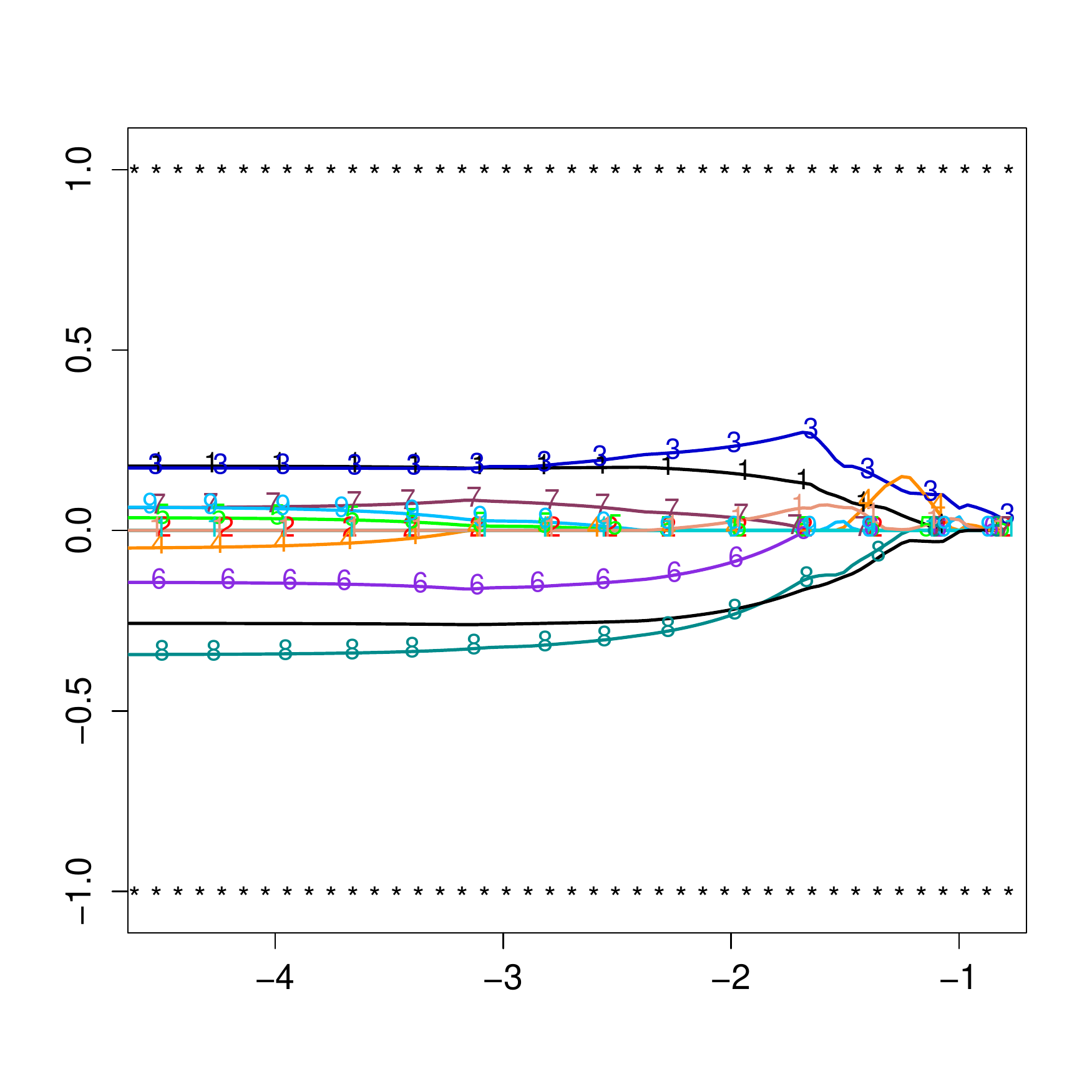}&
\includegraphics[width=0.3\textwidth,height=0.25\textheight,  trim =1.0cm 1.5cm .2cm 1.5cm, clip = true ]{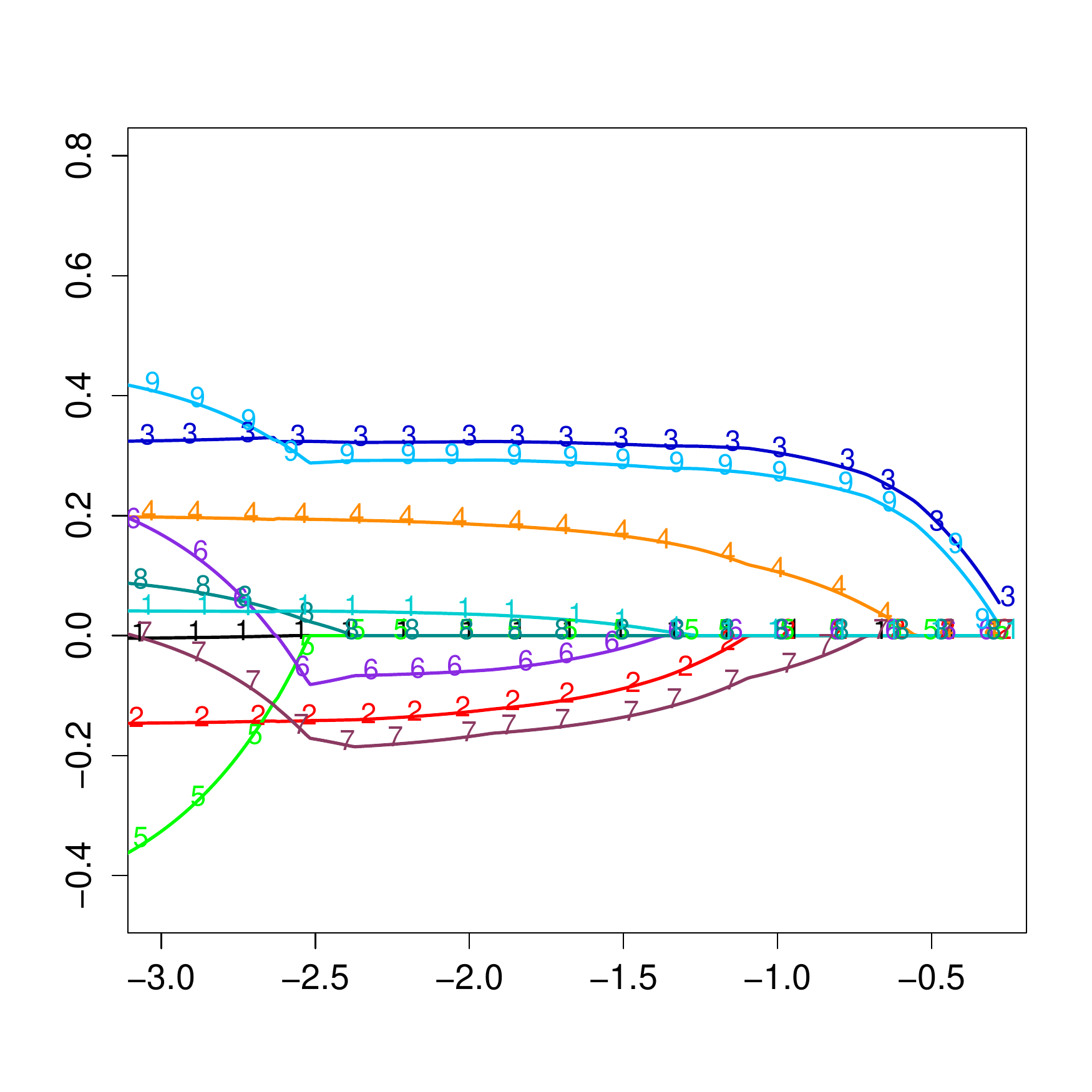}\\
 & \sf {\scriptsize{ $\log_{10}(\delta)$} } &  \sf {\scriptsize{ $\log_{10}(\delta)$ }}  & \sf {\scriptsize{ $\log_{10}(\delta)$ }} \\
\end{tabular}}
\caption{{\small{Coefficient profiles for the \LDSF~and the $\ell_{1}$-\DSF, as a function of the data-fidelity parameter, $\delta$.  The dashed vertical lines in the top row of plots indicate the locations where the number of active variables changes.
[Left Panel] corresponds to Example~1; [Middle Panel] corresponds to Example~1$'$; the ``true'' nonzero coefficients of $+1$ and $-1$ are shown as horizontal starred lines.  [Right panel] corresponds to the path for the Diabetes dataset. The numbers overlain on the profiles indicate
the different features.}}}\label{L1-L1-DS-path}
\end{figure}

Figure~\ref{L1-L1-DS-path} demonstrates the difference between the \LDSF~and the  $\ell_1$-Dantzig selector, by displaying the coefficient profiles for both methods.
Note that the profiles for the \LDSF~are constructed in a piece-wise constant fashion, where for each given model size, the displayed coefficients are taken from the solution corresponding to the lowest attainable value of~$\|\M{X}^\top  (\M{y}-\M{X}\widehat{\B\beta})\|_\infty$.
The left panel of Figure~\ref{L1-L1-DS-path} corresponds to Example~1, with $n=10,p=11$ and $\tau= 1/2p$, which we slightly modified by adding noise to the response: $\M{y} = \bx_1-\bx_2 + \B\epsilon$.  $\epsilon_{i}$'s are independently generated from a centered Gaussian distribution, corresponding to the Signal to Noise Ratio\footnote{For a model generated as $y_{i} = \mu_{i} + \epsilon_{i}, i = 1, \ldots, n$; we define SNR as follows: SNR= $\text{Var}(\mu)/\text{Var}(\epsilon)$.} (SNR) of
$1.3 \times 10^5$.

Now consider Example~1$'$, which is similar in spirit to Example~1.  Here, the first two features, $\M{x}_{1}$ and $\M{x}_{2}$, are drawn from a centered bivariate Gaussian distribution with correlation~$0.7$.
The remaining $p-2$ features are drawn from an independent standard Gaussian ensemble. All the features are standardized to have unit~$\ell_{2}$-norm, and the response is generated
with SNR=$1.4 \times 10^3$.
The middle panel in Figure~\ref{L1-L1-DS-path} displays the corresponding coefficient profiles with $n=10,p=12$. The \LDSF~exactly recovers the true model for a wide range of the tuning parameter, $\delta$.  As~$\delta$ is decreased, and noise variables come into the model, their coefficients remain highly shrunk, while the coefficients for the signal variables remain near their true values.  On the other hand, the $\ell_1$-Dantzig selector is unable to recover the true model, and produces a large estimation error for all values of the tuning parameter.

The right panel in Figure~\ref{L1-L1-DS-path} corresponds to the well known Diabetes dataset~\cite{LARS}, where $n=442$ and $p=10$; here all the variables including the response
were standardized to have unit $\ell_{2}$-norm and zero mean.  Note that, despite some similarities, the sequences of predictors entering the model are different for the two approaches.

We note that instead of formulation~\eqref{L0-DZ-1}, one may prefer to minimize the data-fidelity term subject to a constraint on the number of nonzeros in $\B\beta$:
 \begin{equation}\label{L0-DZ-const}
\min \limits_{\B\beta} \;\; \|\M{X}^\top  (\M{y}-\M{X}\B\beta)\|_\infty  ~~\;\;\sbt \;\;~~  \| \B\beta \|_{0}  \leq \mmk.
\end{equation}
The framework developed in this paper may be adapted to Problem~\eqref{L0-DZ-const}.
In this paper, however, we focus on Problem~\eqref{L0-DZ-1}.

\pparagraph{Context and Related Work.}
A primary motivation of our work is derived from the recent work on the least squares variable selection problem~\cite{bertsimas2015best}, where the authors study
 \begin{equation}\label{subset-l0-old}
\min_{\B\beta} \;\;\; \frac12 \| \M{y} - \M{X} \B\beta \|_{2}^2 \;\;\;  ~~\sbt~~ \;\;\; \| \B\beta \|_{0} \leq k,
 \end{equation}
using mixed integer convex \emph{quadratic} optimization ({\texttt{MIQO}}) methods. While
certain statistical properties of solutions from Problems~\eqref{L0-DZ-1} and~\eqref{subset-l0-old} are comparable;
we observed in our computational experiments (see Section~\ref{sec:com-mipgo}) that Problem~\eqref{L0-DZ-1}
 is orders of magnitude faster (often by a factor of hundreds) than Problem~\eqref{subset-l0-old} in obtaining solutions with \emph{certificates} of global optimality.
  In addition, a \MIO formulation for Problem~\eqref{L0-DZ-1} consumes much less memory than a comparable {\texttt{MIQO}} formulation for Problem~\eqref{subset-l0-old}.
 The aforementioned computational superiority of \MIO over {\texttt{MIQO}} should not come as a surprise.
Indeed, it is quite well known
 in the integer programming community (see, for example, the nice review papers~\cite{hemmecke2010nonlinear,burer2012milp}) that current algorithms for \MIO
problems are a much more \emph{mature} technology than {\texttt{MIQO}}.
Recently, \cite{mipgo-li-2016} proposed MIPGO for nonconvex penalized least squares regression based on
purely continuous nonconvex quadratic optimization. We demonstrate the substantial superiority of our proposal over MIPGO (in Section~\ref{sec:com-mipgo}) in obtaining high quality
statistical solutions within a given computational budget.

Thus, the superior computational scalability of the corresponding optimization methods forms a principal motivation to study Problem~\eqref{L0-DZ-1}, as an effective estimation procedure for sparse linear regression.  In addition, from a statistical viewpoint, in terms of estimating sparse regression models subject to good data-fidelity,
the \LDSF~may be perceived as a natural, interpretable and useful alternative to least squares with variable selection, Problem~\eqref{subset-l0-old} --- in the same way as the
$\ell_{1}$-\DSF~may be viewed as an appealing alternative to Lasso.



\pparagraph{Contributions.}
Our main contributions can be summarized as follows:
\begin{enumerate}
\item We propose a new high-dimensional linear regression estimator: the \LDSF, which minimizes the number of nonzero regression coefficients,
subject to a budget on the maximal absolute correlation between the features and the residuals.  We show that the estimator can be expressed as the
solution to a \MIO problem, a computationally \emph{tractable} framework that delivers certifiably optimal global solutions; and is computationally more scalable
than the recently proposed methods in~\cite{bertsimas2015best,bertsimas2015or,mipgo-li-2016}.
%
\item We develop new discrete first-order methods, motivated by recent algorithmic developments in first-order continuous \emph{convex} optimization, to
obtain high quality feasible solutions for the \LDSF~problem.  These solutions are passed onto \MIO solvers as warm-starts.  Our proposal leads to advantages over the off-the-shelf state-of-the-art integer programming algorithms in terms of
(a) obtaining superior upper bounds for a given computational budget and
(b) aiding \MIO solvers in obtaining tighter lower bounds and hence improved certificates of optimality.
%
Exploiting problem specific information, we also propose enhanced \MIO formulations, which further improve the algorithmic performance of \MIO solvers.

\item We characterize the statistical properties of the \LDSF~and demonstrate both theoretically and empirically its advantages over $\ell_1$-based approaches.
Our results also apply to \emph{approximate} solutions for the \LDSF~optimization problem.


\item Our approach obtains optimal solutions for $p \approx 500$ in a few minutes,
$p \approx 3,\!000$ within fifteen minutes and for problems with $p = 10,\!000$ in an hour. Certificates of optimality are obtained at the expense of higher
computation times---for instances with $p\approx 500$ they are obtained within half-hour, for $p\approx 3,\!000$ they are achieved around an hour
and for $p=10,\!000$ the certificates arrive in the range from three to forty hours. To the best of our knowledge, we present herein, the largest problem instances in subset selection for which
certifiably optimal solutions can be obtained.


\end{enumerate}


\pparagraph{Roadmap.}
The remainder of the paper is organized as follows. Section~\ref{sec:MIO} describes the optimization methodology behind the proposed approach, and discusses its connections with the $\ell_{1}$-Dantzig Selector optimization problem.  In Section~\ref{sec:theory}, the statistical properties of the \LDSF~are analyzed from a theoretical point of view; the results are compared to the $\ell_1$-Dantzig Selector and the Lasso.   
The framework of discrete first-order methods is described in Section~\ref{sec:FO-methods1}. Additional discussion of
 \MIO formulations together with problem specific enhancements, is presented in Section~\ref{sec:adv-formulations-1}.
Section~\ref{sec:numerics-algo} gathers numerical results on the computational performance of our algorithms in a variety of settings.
An empirical analysis of the statistical properties of the \LDSF~is conducted in Section~\ref{sec:stats-prop-expt}.
Some technical details  are provided in the Appendix.

\section{Overview of the Proposed Methodology}\label{sec:MIO}

Herein, we introduce and summarize the general aspects of the proposed methodology.  Further details and enhancements are provided in Sections~\ref{sec:FO-methods1} and~\ref{sec:adv-formulations-1}.

\subsection{Mixed Integer Linear Optimization ({\texttt{MILO}}) Preliminaries}\label{sec:background-mio}
The general form  of a  {\texttt{MILO}} problem is as follows:
$$\begin{myarray}[1.3]{c c c r}
\min\limits_{\B\alpha} &  \M{a}^\top \B{\alpha} & \\
\sbt & \;\; \M{A} \B\alpha  \B{\leq} \M{b}&\\
& \;\;\;\; \alpha_i \in \{ 0 , 1 \},& i =1, \ldots, m_{1}\\
& \;\;  \alpha_j  \geq 0,& j = m_{1}+1, \ldots, m,
\end{myarray}$$
where $\M{a}\in \mathbb{R}^{m \times 1}, \M{A} \in  \mathbb{R}^{d \times m}$ and $\M{b} \in  \mathbb{R}^{d \times 1}$  are the problem data, the symbol ``$\B{\leq}$'' denotes  element-wise inequalities, and we optimize over $(\alpha_{1}, \ldots, \alpha_{m}):=\B\alpha \in  \mathbb{R}^{m}$ containing both discrete ($\alpha_i, i =1, \ldots, m_{1} $)  and continuous ($\alpha_i,  i= m_{1}+1, \ldots, m$) variables.
For background on  {\texttt{MILO}}, we refer the reader to \cite{bertsimas2005optimization_new,junger200950}.
Some modern integer optimization solvers include
\textsc{Cplex}, \textsc{Glpk},   \textsc{Gurobi}, \textsc{Knitro}, \textsc{Mosek}, \textsc{Scip} --- see also~\cite{linderoth2010milp}.

\begin{figure}[]
\centering
\resizebox{\textwidth}{.14\textheight}{\begin{tabular}{l c l c}
\multicolumn{ 4}{c}{ { \sf {Diabetes Dataset $(n=442,p=64, \|\widehat{\B\beta}\|_{0} = 41)$} }} \medskip \\
\rotatebox{90}{\sf {\scriptsize{~~~~~~~~~~~~~~~~~~~~~~~~~~~~~~~~~~Bounds}}}&
\includegraphics[width=0.4\textwidth,height=0.25\textheight,  trim =1.0cm 1.5cm .2cm 1.5cm, clip = true ]{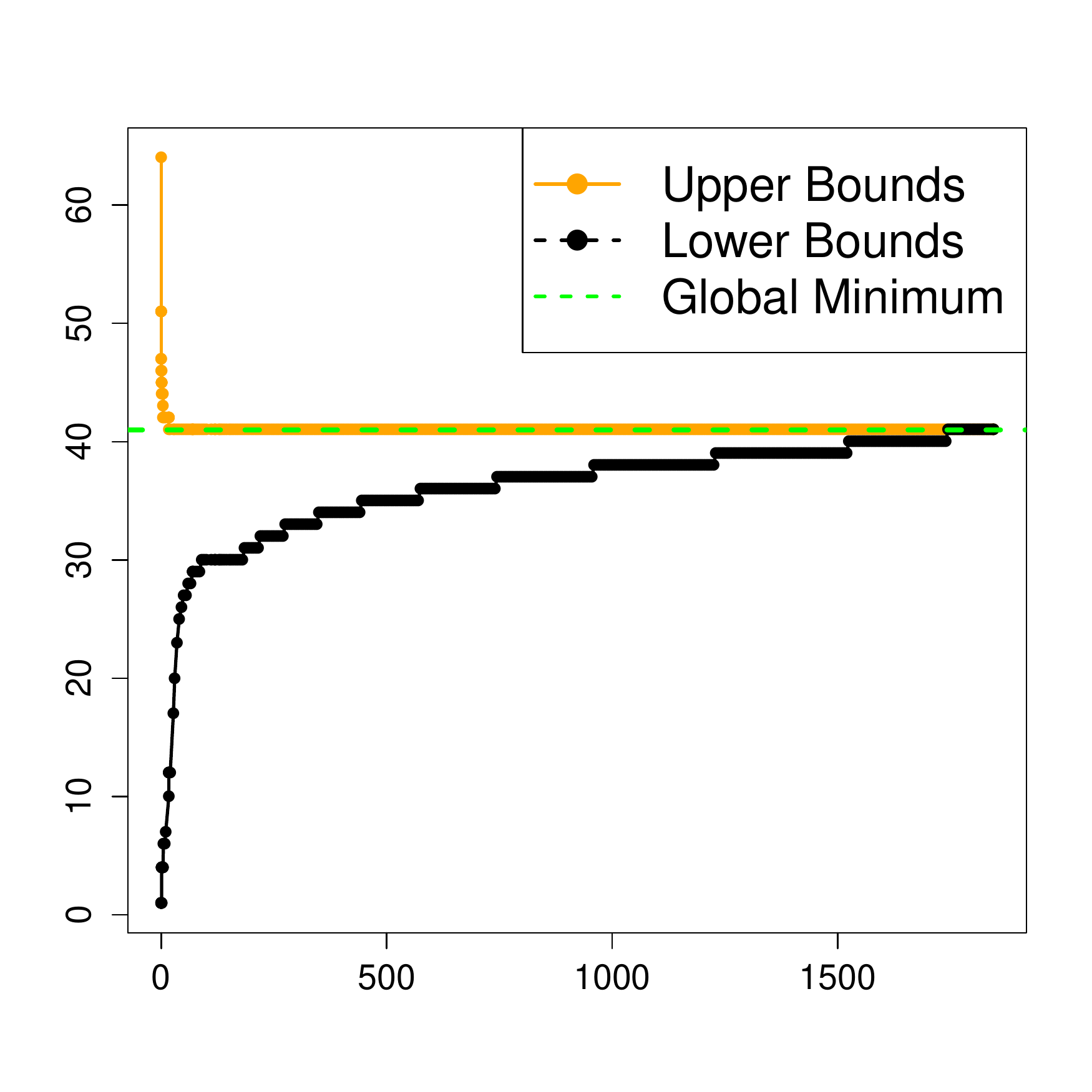}&
\hspace{4mm}
\rotatebox{90}{\sf {\scriptsize{~~~~~~~~~~~~~~~~~~~~\MIO~Optimality Gap (in \%)}}}&
\includegraphics[width=0.4\textwidth,height=0.25\textheight,  trim = 1.0cm 1.5cm .2cm 1.5cm, clip = true ]{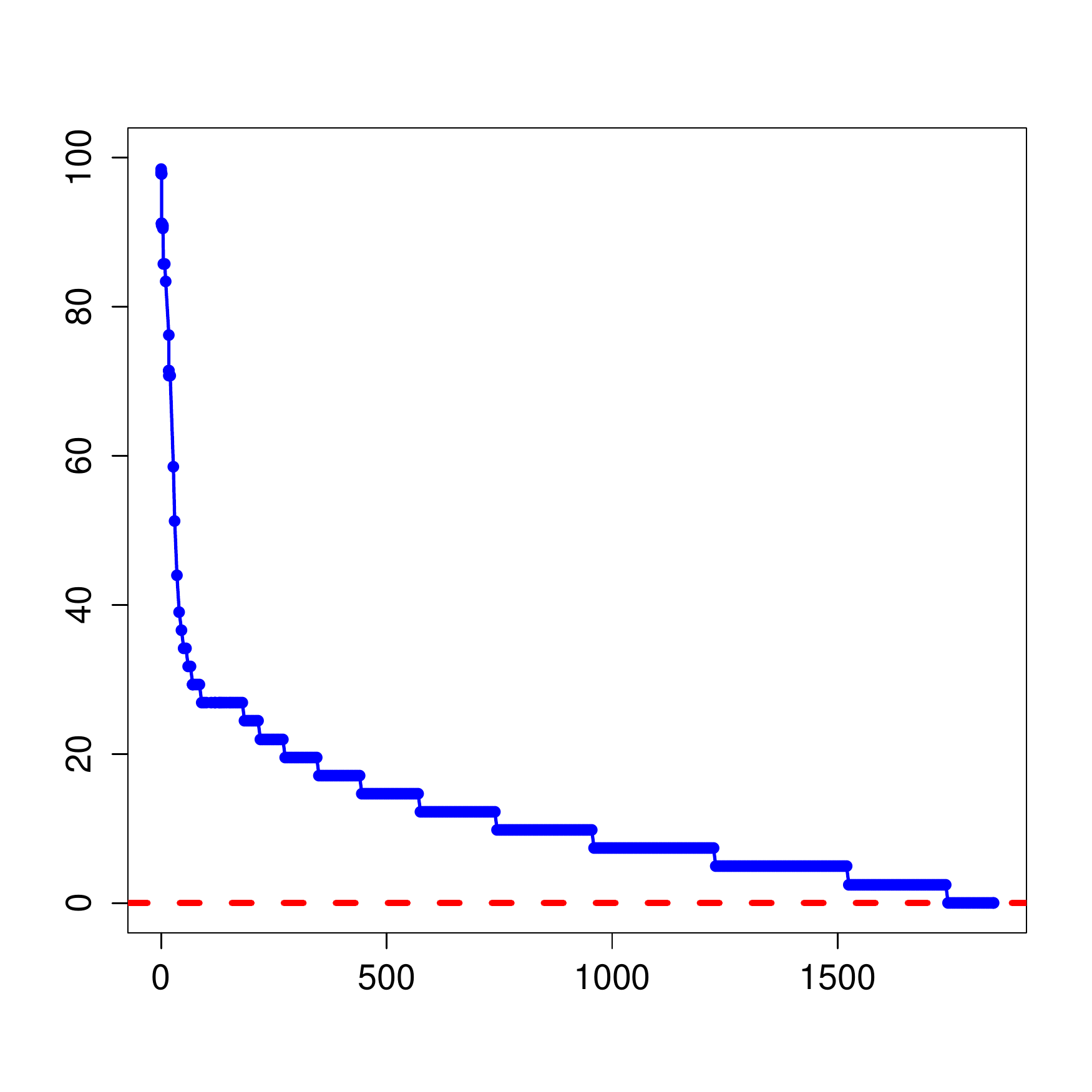}  \\
 & \sf { { Time (secs)}} &  & \sf { { Time (secs) }} \\
\end{tabular}}
\caption{{\small{Typical evolution of  a {\texttt {MILO}} algorithm for Problem~\eqref{L0-DZ-1}, as a function of time. [Left Panel] displays the progress of Upper Bounds (UB) and Lower Bounds (LB) for the optimal value of the objective function.
 The upper bounds, which correspond to feasible solutions for Problem~\eqref{L0-DZ-1} are seen to stabilize at the global minimum within a few seconds.
 The lower bounds provide a certificate of \emph{how far} the current solution might be from the global solution --- these bounds progressively improve as the \MIO algorithm explores more nodes in the branch and bound tree.
Observe that the certificate of global optimality arrives at a later stage, even though the algorithm finds the global solution very quickly.
 [Right Panel] displays the evolution of the corresponding \MIO~Optimality Gap (in \%), defined as (UB - LB)/UB, with time.}}}\label{mio-motivate-1}
\end{figure}

 As already alluded to in Section~\ref{sec:intro}, there has been significant progress in the theory and practice of \MIO over the past fifteen to twenty years.
 Specifically, the computational power of {\texttt {MILO} }solvers has undergone impressive advances over the past twenty-five years --- the
cumulative machine-independent speedup factor in {\texttt {MILO} }solvers between 1991 and 2015 is estimated to be $780,\!000$~\cite{bixby}.
This progress can be attributed to the inclusion of both theoretical and practical advances into {\texttt {MILO} }solvers.
Some of the main factors responsible for this speedup are advances in cutting plane theory, improved heuristic methods, disjunctive programming for branching rules,
techniques for preprocessing {\texttt {MILO}}s, using linear optimization  as a black box to be called by {\texttt {MILO} }solvers, and improved linear optimization  methods~\cite{bixby}.
In addition, there have been substantial improvements in hardware speed: the overall hardware speedup from 1993 to 2015 is approximately estimated to be
$10^{5.75}\sim 570,\!000$~\cite{supercomputer}. When both hardware and software advances are combined, the overall speedup for {\texttt{MILO}} problems is estimated to be around 450 billion!
One attractive feature of {\texttt {MILO} }solvers, which is a stark contrast to heuristic approaches, is that the former
provide (a) feasible solutions, which are also upper bounds to the minimum objective value and (b) lower bounds for the optimal value of the objective function.
As a {\texttt {MILO} }solver makes its way to the global optimum, the lower bounds become tighter, thereby
providing improved certificates of sub-optimality (see Figure~\ref{mio-motivate-1} for an illustration).  This aspect of {\texttt {MILO} }solvers is quite useful, especially if one decides to
stop the solver before reaching the global optimum. In the modern day world, {\texttt {MILO} }plays a key role in various impactful application areas of operations research: revenue management, air-traffic control, scheduling and matching tasks, production planning and
others~\cite{williams2013model,bertsimas2005optimization_new}.
In this paper, we show how the power of \MIO can be used in the context of a problem of fundamental importance in statistics, namely, sparse linear model estimation --- we build upon recent line of work in computational statistics, at the interface of modern discrete optimization and fundamental techniques in statistical modeling~\cite{bertsimas2014least,bertsimas2015best}.

\subsection{\MIO formulations for the \LDSF}\label{sec:mio-form-1}

Assuming without loss of generality that Problem~\eqref{L0-DZ-1} has a minimizer which is bounded, it can be obtained by solving
\begin{equation}\label{L0-DZ-1-M}
\begin{myarray}[1.3]{lc  c }
\Gamma_{1}:=& \min \limits_{\B\beta} &  \|\B\beta\|_0 \\
&\sbt&  \|\M{X}^\top  (\M{y}-\M{X}\B\beta)\|_\infty \leq \delta\\
&& \| \B\beta \|_{\infty} \leq \MU,
\end{myarray}
\end{equation}
 where, $\MU$ is a large but finite number~\cite{bertsimas2005optimization_new}.
We present a \MIO formulation for Problem~\eqref{L0-DZ-1-M} (and also, Problem~\eqref{L0-DZ-1})
\begin{equation}\label{L0-DZ-1-mio-M}
\begin{myarray}[1.3]{c  c  r}
\min \limits_{\B\beta, \M{z}} &  \sum\limits_{i=1}^{p} z_{i} & \\
\sbt&   - \delta  \leq d_{j}  - \langle \M{q}_{j}, \B\beta \rangle  \leq \delta, & j = 1, \ldots, p\\
& -\MU z_{j} \leq  \beta_{j} \leq \MU z_{j},& j = 1, \ldots, p\\
& z_{j} \in \{ 0 , 1 \}, & j = 1, \ldots, p,
\end{myarray}
\end{equation}
where the optimization variables are $\M{z}$ (binary) and $\B\beta$ (continuous); the problem data consists of $\M{d}:= \M{X}^\top  \M{y}=(d_1,\ldots,d_p)^\top $ and
$\M{Q}_{p \times p} := \M{X}^\top  \M{X} = [\M{q}_{1}, \ldots, \M{q}_{p}]$. Formulation~\eqref{L0-DZ-1-mio-M} is often referred to as a ``Big-M'' formulation due to the presence of the parameter $\MU$.
 The binary variable $z_{i}$ controls whether $\beta_{i}$ is zero or not: if $z_{i} =0$ then $\beta_{i} =0$ and if $z_{i}=1$ then $\beta_{i}$ is free to vary in the interval $[-\MU,\MU].$
The objective function $\sum_{i=1}^{p} z_{i}$ controls the number of nonzeros in the model.
Figure~\ref{mio-motivate-1} shows the performance of the above \MIO formulation on the
Diabetes dataset~\cite{LARS} with $n=442,p=64$ (here, we mean-centered and scaled $\M{y},\M{x}_{i}$'s to have unit $\ell_{2}$-norm).

Formulation~\eqref{L0-DZ-1-mio-M} has intriguing connections to the $\ell_{1}$-\DSF: the binary variables $z_{i} \in \{0, 1\}$ in Problem~\eqref{L0-DZ-1-mio-M} can be relaxed into continuous variables $z_{i} \in [0,1]$, leading to:
\begin{equation}\label{L1-DZ-1b}
\begin{myarray}[1.3]{cccc}
\Gamma_{2}:=&\min\limits_{\B\beta} &  \frac{1}{\MU} \sum\limits_{i=1}^{p} | \beta_{j}| \\
&\sbt&  \|\M{X}^\top  (\M{y}-\M{X}\B\beta)\|_\infty \leq \delta \\
&& -\MU \leq  \beta_{j} \leq \MU ,& j = 1, \ldots, p.\\
\end{myarray}
\end{equation}
Problem~\eqref{L1-DZ-1b} modifies the $\ell_{1}$-\DSF~problem:
$$ \Gamma_{3}:= \min\;\;  \frac{1}{\MU}\|\B\beta\|_{1}  \;\;\;  ~~\sbt~~ \;\;  \| \M{X}^\top  (\M{y} - \M{X}\B\beta)\|_{\infty} \leq \delta,$$
by including an $\ell_{\infty}$-constraint on $\B\beta$.
%
It follows from the above that: $ \Gamma_{3} \leq \Gamma_{2} \leq \Gamma_{1}$.  The last inequality  is typically strict, and, depending upon the data, the gap between the values, as well
as between the corresponding optimal solutions, can be substantial, as illustrated in Figure~\ref{L1-L1-DS-path}.  The above discussion provides another viewpoint for explaining the differences between the $\ell_{1}$-\DSF~and the \LDSF~estimators.
If $\MU$ is taken to be large enough
  then $\Gamma_{3} = \Gamma_{2}$.


We note that the constraint $\|\B\beta \|_{0} \leq k$ can also be expressed using
Specially Ordered Sets~\cite{bertsimas2005optimization_new} as an alternative to the ``Big-M'' formulation~\eqref{L0-DZ-1-mio-M}. This can be used when the user does not wish to specify
a-priori any
bound $\MU$ on the regression coefficients and/or the coefficients have widely varying amplitudes.
We discuss this further in Section~\ref{sec:adv-formulations-1}.

We emphasize that $\MU$ appearing in formulation~\eqref{L0-DZ-1-mio-M} should \emph{not} be interpreted as a statistical tuning parameter --- it appears from a purely algorithmic viewpoint and, as we saw,
has interesting connections to the $\ell_{1}$-\DSF~problem. $\MU$ might be taken to be arbitrarily large (but finite) to obtain a solution to Problem~\eqref{L0-DZ-1}.
In Section~\ref{sec:adv-formulations-1} we describe several data driven methods to estimate $\MU$, and we also
discuss other structured formulations of Problem~\eqref{L0-DZ-1-mio-M}, which lead to improved algorithmic performance: they deliver tighter computational lower bounds in smaller amounts of time.
We now proceed towards an analysis of the statistical properties of the \LDSF~and investigate its comparative advantages over its $\ell_{1}$-counterpart.

\section{Statistical Properties: Theory}\label{sec:theory}

In this section we study the statistical properties of the Discrete Dantzig estimator.  In particular, we characterize its connections with the Best Subset selection estimator in the case of the orthonormal design, we investigate its oracle properties in the classical asymptotic regime and, finally, we analyze its estimation, prediction and variable selection performance in the high dimensional setting.
To improve readability, all the technical proofs are presented in Section~\ref{sec:proofs-stat-theory} in the Appendix.

\subsection{Orthonormal Design}
Here we assume $\M{x}_j^\top\M{x}_k$ equals $1$ if $j=k$ and~$0$ otherwise.  Note that such an assumption requires $n\ge p$.  Our goal is to compare and connect the Discrete Dantzig estimator, $\widehat{\bbeta}$, which solves the optimization problem~\eqref{L0-DZ-1},
with the Best Subset selection estimator, $\widehat{\B\beta}^{BS}$, which solves \eqref{subset-l0-old}.  In particular, we want to understand the relationship between the tuning parameter $\delta$ in the Discrete Dantzig optimization problem and the tuning parameter $k$, which controls the $\ell_0$ norm of the Best Subset solution.
Note that Problem~\eqref{L0-DZ-1}  does not generally have a unique optimizer, so we use $\widehat{\B\beta}$ to refer to just one of the Discrete Dantzig solutions.

Define $c_j=\M{x}_j^\top\M{y}$ for $j=1,...,p$.  To simplify the presentation, and without loss of generality, suppose that the predictors are indexed in such a way that $|c_1|\ge|c_2|\ge...\ge |c_p|$.  Suppose also that $c_k\ne c_{k+1}$, to ensure uniqueness of the Best Subset solution.
\begin{thm}
\label{orthonorm.thm}
\begin{enumerate}
\item The Best Subset selection estimator is uniquely defined as
\begin{equation*}
\widehat{\B\beta}^{BS}=(c_1,\ldots,c_k,0,\ldots,0).
\end{equation*}
\item Suppose that $\delta\in[c_{k+1},c_k)$.  Then, the set of all the Discrete Dantzig solutions is
\begin{equation*}
\left\{\left(c_1+u_1,...,c_k+u_k,0,...,0\right),  \; |u_j|\le\delta, j=1,...,k\right\}.
\end{equation*}
\end{enumerate}
\end{thm}
It follows that each Best Subset estimator is a Discrete Dantzig solution for an appropriately chosen~$\delta$.  The coefficients of both estimators are obtained by the hard thresholding of the covariances $c_j$, however, the nonzero coefficients of the Discrete Dantzig estimator are allowed to deviate from the value of the covariance by an amount bounded above by $\delta$.

\subsection{Fixed $p$ asymptotics}
To avoid  confusion, we refer to the true coefficient vector as~$\bbeta^*$.  In this subsection we treat the number of predictors, $p$, as fixed and let the number of observations, $n$, tend to infinity.    The standard assumption for deriving asymptotic results in this setting is that $\B\beta^*$ does not depend on~$n$, and $n^{-1}\bX^\top\bX$ converges to a non-singular covariance matrix~$C$.  We rewrite the above assumption to be consistent with the scaling $\|\M{x}_j\|_2=1$ for $j=1,...,p$, which is used throughout this paper.  Thus, we require that
$\bX^\top\bX$ converges to $C$ as $n$ tends to infinity, and $\bbeta^*=n^{1/2}\tilde{\bbeta}^*$, for some fixed vector $\tilde{\bbeta}^*$.  We also impose a standard assumption that $\epsilon_i$ are i.i.d. with zero mean and finite variance.  Given an index set $J\subseteq \{1,\ldots,p\}$ we write  $\bX_J$ for the sub-matrix of~$\bX$ that consists of the columns identified by~$J$.  Let $J^*$ denote the support of the vector $\bbeta^*$.  We define the Oracle estimator, $\widehat{\bbeta}^O$, as the least-squares estimator computed using only the true predictors.  In other words, the support of $\widehat{\bbeta}^O$ equals~$J^*$, and $\widehat{\bbeta}^O_{J^*}=(\bX^\top_{J^*}\bX_{J^*})^{-1}\bX^\top_{J^*}\bY$.
\begin{thm}
\label{fixed.p.thm}
Let $\delta\rightarrow\infty$ and $\delta=o(n^{1/2})$.  Suppose that matrix $C$ is invertible.  Then, with probability tending to one,
\begin{enumerate}
\item The support of each solution to the Discrete Dantzig optimization problem equals $J^*$;
\item Both the true coefficient vector, $\bbeta^*$, and the Oracle estimator, $\widehat{\bbeta}^O$, belong to the set of solutions to the Discrete Dantzig optimization problem.
\end{enumerate}
\end{thm}
Consider the polished version of the Discrete Danzig estimator, which is defined as follows: given a Discrete Dantzig solution with support $\widehat{J}$, the support of the polished estimator, $\widehat{\bbeta}^P$, is set equal to $\widehat{J}$; on its support $\widehat{\bbeta}^P$ is defined as the least-squares estimator using the corresponding predictors, i.e. $\widehat{\bbeta}^P_{\widehat{J}}=(\M{X}^\top_{\widehat{J}}\M{X}_{\widehat{J}})^{-1}\M{X}^\top_{\widehat{J}}\M{y}$.  Note that the value of $\widehat{\bbeta}^P$ generally depends on the the choice of the Discrete Dantzig solution.  However, this choice becomes irrelevant under the setting of Theorem~\ref{fixed.p.thm}.  More specifically, with probability tending to one, every polished estimator coincides with the Oracle estimator.
\begin{cor}
Under the assumptions of Theorem~\ref{fixed.p.thm}, equality $\widehat{\bbeta}^P=\widehat{\bbeta}^O$ holds with probability tending to one.
\end{cor}
Consequently, estimator $\widehat{\bbeta}^P$ satisfies the oracle property in the sense of Fan and Li \cite{Fan01}.

\subsection{High Dimensional Setting}

Here we focus on the case where $p$ is large, possibly much larger than~$n$.  We  discuss the properties of the global
as well as approximate solutions to Problem~\eqref{L0-DZ-1}.
We also comment on the estimator obtained from the closely related Problem~\eqref{L0-DZ-const}, in which $\|\bbeta\|_0$ is constrained, rather than minimized.
We assume that the error terms in the underlying linear model are mean zero Gaussian with variance~$\sigma^2$ and, as before, use~$\bbeta^*$ to refer to the true regression coefficient vector.
We start with some notation.  For every vector $\btheta\in\RR^p$ and index set $J\subseteq \{1,...,p\}$ we write  $\btheta_J$ for the sub-vector of~$\btheta$ determined by~$J$.


\begin{definition}
Given positive integers~$k$ and~$m$, such that $m\in[k,p-k]$, and a positive~$c_0$, let
\begin{equation*}
\begin{aligned}
\gamma(k)=&\min_{\btheta\ne 0,\,\|\btheta\|_0\le k}\frac{\|\M{X}\btheta\|_2}{\|\btheta\|_2}\\
\kappa(k,c_0)=&\min_{|J_0|\le k}\quad \left\{\min_{\btheta\ne0,\,\|\btheta_{J_0^c}\|_1\le c_0\|\btheta_{J_0}\|_1}\frac{\|\M{X}\btheta\|_2}{\|\btheta_{J_0}\|_2} \right\}\\
\kappa(k,c_0,m)=& \min_{|J_0|\le k}\quad \left \{\min_{\btheta\ne0,\,\|\btheta_{J_0^c}\|_1\le c_0\|\btheta_{J_0}\|_1}\frac{\|\M{X}\btheta\|_2}{\|\btheta_{J_{01}}\|_2} \right\},
\end{aligned}
\end{equation*}
where $J_0\subseteq\{1,...,p\}$ and
 $J_{01}:=J_0\cup J_1$, with~$J_1$ identifying the~$m$ largest (in magnitude) coordinates of $\btheta$ outside of~$J_0$.
\end{definition}

We use~$s^*$ to denote $\|\bbeta^*\|_0$.  As we discuss in the next subsection,  quantities $\left[\kappa(s^*,c_0)\right]^{-1}$ and $\left[\kappa(s^*,c_0,m)\right]^{-1}$, for $m\ge s^*$ and $c_0\ge1$,  appear in the error bounds for the Lasso and the original Dantzig selector, while $\left[\gamma(2s^*)\right]^{-1}$ appears in the bounds for
\LDSF.  The following result establishes some useful relationships for these quantities.


\begin{prop}
\label{prop.ineqs}
For all positive integers~$k$ and~$m$, with $m\in[k,p-k]$, and all $c_0\ge1$ the following holds:
$\gamma(2k) \ge \kappa(k,c_0) / \sqrt{2}$ and $\gamma(2k)\ge \kappa(k,c_0,m).$
\end{prop}
Recall the setting of Example~1.  When $\tau(n-1)<2$, the $\ell_{1}$ methods, such as the original Dantzig selector, fail to recover the sparse representation of the response.  Note also that $\kappa(2,c_0)=\kappa(2,c_0,m)=0$, for $m\ge 2$ and $c_0\ge1$.   On the other hand, $\gamma(4)>0$ for $p>4$, and the
\LDSF~succeeds in recovering the correct sparse representation, for every sufficiently small value of the tuning parameter, $\delta$.


The following theorem establishes several useful bounds for the \LDSF.

\begin{thm}
\label{gen.thm}
Suppose that $\widehat \bbeta$ solves optimization problem~\eqref{L0-DZ-1} for~$\delta=\sigma\sqrt{2(1+a)\log p}$, where~$a\ge 0$. The following bounds hold with probability bounded below by $1-( p^a\sqrt{\pi\log p} )^{-1}$:
\begin{eqnarray*}
\|\widehat \bbeta\|_0 &\le& s^*\\
\|\widehat\bbeta-\bbeta^*\|_1 &\le& 4\left[\gamma(2s^*)\right]^{-2}s^*\delta\\
\|\widehat\bbeta-\bbeta^*\|^2_2 &\le& 8\left[\gamma(2s^*)\right]^{-4}s^*\delta^2\\
n^{-1}\|\M{X}(\widehat\bbeta-\bbeta^*)\|^2_2&\le& 8[\gamma(2s^*)]^{-2}s^*\delta^2.
\end{eqnarray*}
\end{thm}

\begin{remark}
It follows from the proof of Theorem~\ref{gen.thm} that the above result
\begin{enumerate}
\item[(i)] holds uniformly over the set $\{\bbeta^*:\,\|\bbeta^*\|_0\le s^*\}$;
\item[(ii)] also holds for the solution to Problem~\eqref{L0-DZ-const} with $\mmk = s^*$.
\end{enumerate}
\end{remark}
We now compare the above bounds to those established for the popular $\ell_{1}$-based approaches.   Under the assumed scaling of the predictors, and for every positive integer~$m$, such that $m\in[s^*,p-s^*]$, Theorem~7.1 in \cite{bickel1} gives the following error bounds for the $\ell_{1}$-Dantzig Selector estimator, $\widehat\bbeta_{\text{DS}}$:
 \begin{eqnarray}
\nonumber\|\widehat\bbeta_{\text{DS}}-\bbeta^*\|_1 &\le& 8\left[\kappa(s^*,1)\right]^{-2}s^*\delta\\
\label{DS.bnds}\|\widehat\bbeta_{\text{DS}}-\bbeta^*\|^2_2 &\le& 16\left(1+\sqrt{s/m}\right)^2 \left[\kappa(s^*,1,m)\right]^{-4}s^*\delta^2\\
\nonumber n^{-1}\|\M{X}(\widehat\bbeta_{\text{DS}}-\bbeta^*)\|^2_2&\le& 16[\kappa(s^*,1)]^{-2}s^*\delta^2.
\end{eqnarray}
By Proposition~\ref{prop.ineqs}, the right hand sides of the above inequalities are at least as large as the corresponding bounds in Theorem~\ref{gen.thm}.  Moreover, the differences in the two sets of bounds can potentially be quite significant.  Consider the setting of Example~1 for illustration.  The upper bounds in Theorem~\ref{gen.thm} are finite for $p>4$, while the three bounds in display~(\ref{DS.bnds}) are infinite.

Examining Theorem~7.2 in \cite{bickel1}, we conclude that the corresponding error bounds for the Lasso are at least as large as those given in  display~(\ref{DS.bnds}).   The same result also provides an upper bound on the $\ell_{0}$-pseudo-norm of the Lasso estimator,
$\widehat \bbeta_{\text{Lasso}}$:
\begin{equation*}
\|\widehat \bbeta_{\text{Lasso}}\|_0 \le \frac{64\phi_{\max}}{\left[\kappa(s^*,3)\right]^2} s^*,
\end{equation*}
where $\phi_{\max}$ is the maximum eigenvalue of the matrix $\M{X}^\top  \M{X}$.  Note that the right-hand side of the above bound is infinite in the setting of Example~1.  In general, this upper bound is at least~$64$ times as large as the one for the \LDSF~estimator.
We informally summarize the above findings as follows:  when compared to the $\ell_{1}$-based approaches, the \LDSF~satisfies as good or better estimation and prediction error bounds, while achieving significantly higher level of sparsity.

We can sharpen the bounds in Theorem~\ref{gen.thm} by making them dependent on the support of~$\bbeta^*$.  More specifically, given an index set~$J^*$ we define
\begin{equation*}
\tilde\gamma(J^*)=\min_{J\subset\{1,...,p\}, |J|= 2s^*, J\supseteq J^*}\quad\min_{\btheta\ne0,\,\btheta_{J^c}=0}\frac{\|\M{X}\btheta\|_2}{\|\btheta\|_2}.
\end{equation*}
Then, Theorem~\ref{gen.thm} holds with $\gamma(2s^*)$ replaced by $\tilde\gamma({\{k:\beta^*_k\ne0\}})$, and the corresponding result is uniform over~$\bbeta^*$.

The following corollary to Theorem~\ref{gen.thm} shows that the \LDSF~successfully recovers the support of the true coefficient vector, provided the nonzero coefficients are appropriately bounded away from zero.  Define $|\beta^*|_{\min}=\min\{|\beta^*_k|,\beta^*_k\ne0\}$.


\begin{cor}
\label{cor.gen.thm}
If
$|\beta^*|_{\min}>4\sigma\left[\gamma(2s^*)\right]^{-2}\sqrt{(1+a)s^*\log p}$,
 then the estimator from Theorem~\ref{gen.thm} exactly recovers the support of $\bbeta^*$, with probability bounded below by $1-( p^a\sqrt{\pi\log p} )^{-1}$.
\end{cor}

We now consider an estimator $\widehat\bbeta$ that is a feasible solution to the optimization problem~\eqref{L0-DZ-1}, but not necessarily the optimal solution.   Recall that our algorithms produce $\widehat\bbeta$ together with a lower bound on the minimum value of the objective function, $\|\bbeta\|_0$.  We denote this lower bound by $\widehat s_{LB}$.   The next result shows that if the algorithm is stopped when~$\|\widehat\bbeta\|_0$ is within a prespecified multiplicative factor of $\widehat s_{LB}$, the bounds from Theorem~\ref{gen.thm} continue to hold after an appropriate adjustment.  The corresponding proof follows the argument in the proof of Theorem~\ref{gen.thm}, making only minor modifications.


\begin{thm}
\label{lowb.thm}
Suppose that~$\widehat\bbeta$ is a feasible solution to the optimization problem~\eqref{L0-DZ-1}, corresponding to $\delta=\sigma\sqrt{2(1+a)\log p}$, where $a\ge0$, such that $\|\widehat\bbeta\|_0\le (1+\c)\widehat s_{LB}$.  Then, the following bounds hold with probability bounded below by $1-( p^a\sqrt{\pi\log p} )^{-1}$:
\begin{eqnarray*}
\|\widehat \bbeta\|_0 &\le& (1+\c)s^*\\
\|\widehat\bbeta-\bbeta^*\|_1 &\le& 2(2+\c)\left[\gamma\left([2+\c]s^*\right)\right]^{-2}s^*\delta\\
\|\widehat\bbeta-\bbeta^*\|^2_2 &\le& 4(2+\c)\left[\gamma([2+\c]s^*)\right]^{-4}s^*\delta^2\\
n^{-1}\|\M{X}(\widehat\bbeta-\bbeta^*)\|^2_2&\le& 4(2+\c)[\gamma([2+\c]s^*)]^{-2}s^*\delta^2.
\end{eqnarray*}
\end{thm}
Note that the constant~$\c$ is typically quite small in practice, for example, $0.1$ (see the right panel in Figure~\ref{mio-motivate-1} for an illustration of the evolution of~$\c$ over time for the Diabetes dataset.)  Thus, the corresponding effect on the error bounds is generally minor.

\section{Obtaining Good Solutions via Discrete First-Order Methods}\label{sec:FO-methods1}
In this section we propose new algorithms, referred to as \emph{discrete first-order methods}, which deliver good upper bounds for Problem~\eqref{L0-DZ-1}. It is important to note that
unlike the \MIO framework, these algorithms do not provide lower bounds.  Instead, the solutions obtained by our methods are passed to \MIO solvers as warm-starts. The proposed algorithms are inspired by recent advances of
first-order methods in convex optimization~\cite{nesterov2004introductorynew,nesterov2013gradient,parikh2013proximal}, and can be viewed as their nonconvex adaptations.
We summarize their key advantages:

\begin{itemize}
\item They provide excellent upper bounds to Problem~\eqref{L0-DZ-1} with low computational cost, time and memory requirements.


\item \MIO solvers accept these solutions as warm-starts and consequently improve upon them. This hybrid approach outperforms
the stand-alone capabilities of an off-the-shelf \MIO solver, producing high quality upper bounds in amounts of time that are orders of magnitude smaller.

\item The solutions obtained can be used to improve the overall run-time of \MIO solvers, including certificates of global optimality.
\end{itemize}

We validate our proposed methods on several synthetic and real-data datasets.
\subsection{Discrete First-Order Methods}\label{sec:FO-methods}
Problem~\eqref{L0-DZ-1} involves the minimization of a discontinuous objective function over a polyhedral set.  Thus, it is not directly amenable to simple proximal gradient type algorithms~\cite{nesterov2013gradient,nesterov2004introductorynew,bertsimas2015best}. We propose two algorithms: Algorithm~1 (see Section~\ref{sec:splitting-method1}) and Algorithm~2 (see Section~\ref{sec:weighted-l1}), both of which can be used as stand-alone solvers for obtaining good quality upper bounds to Problem~\eqref{L0-DZ-1}.
We also present a hybrid method,  Algorithm~3, which combines the strengths of both Algorithms~1 and~2, by using the solution obtained from Algorithm~1 as an initialization to Algorithm~2.
 In our experiments Algorithm~3 showed the best empirical performance. Section~\ref{sec:numerics-algo} presents numerical results illustrating the performance of our framework.
We emphasize that Algorithms~1---3  only provide good upper bounds, they do not certify the quality of solutions via lower bounds.
A  main purpose of these algorithms is to
provide good quality upper bounds to initialize the \MIO solvers --- the latter, in turn, are often found to improve upon the upper bounds obtained from these first-order algorithms, at the cost of more
(but still reasonable) computation times.

\subsubsection{The Variable Splitting Method}\label{sec:splitting-method1}
We present our first discrete first-order method based on a classical method in nonlinear optimization: the Alternating Direction Method of Multipliers (aka ADMM)~\cite{bertsekas-99-nourl} popularly used in the context of convex optimization--we refer the reader to~\cite{boyd-admm1nourl} for a nice exposition on this topic.
We choose this method because of its simplicity and good performance in practice, as seen in our experiments.
To apply this algorithm, we \emph{decouple} the feasible set $\{ \B\beta : \|\M{X}^\top (\M{y} - \M{X}\B\beta)\|_{\infty} \leq \delta_{} \}$ and the discontinuous function $\B\beta \mapsto \| \B\beta\|_{0}$.
Observe that Problem~\eqref{L0-DZ-1} can be equivalently rewritten as:
\begin{equation}
\begin{myarray}[1.3]{l c}
\min\limits_{\B\alpha, \B\beta}& ~ \| \B\beta\|_{0} \\
\sbt & \|\M{X}^\top (\M{y} - \M{X}\B\alpha)\|_{\infty} \leq \delta_{}\\
 &~ \B\alpha = \B\beta.
\end{myarray}
\end{equation}
We consider the Augmented Lagrangian given by:
\begin{equation}
{\mathcal L}_{\lambda}(\B\beta, \B\alpha; \B\nu):=  \| \B\beta\|_{0}  + \frac{\lambda}{2} \| \B\beta - \B\alpha\|_{2}^2 + \langle \B\nu, \B\alpha - \B\beta  \rangle
\end{equation}
for  some value of $\lambda>0$, where, $\B\nu$ may be thought of as a ``dual'' variable\footnote{Following the terminology
in~\cite{boyd-admm1nourl}, if instead of $\|\B\beta \|_{0}$
we had a convex function, then $\B\nu$ would be a dual variable, and its corresponding update step~\eqref{line-1-33} would be the dual update. We will with a slight abuse of terminology use the term ``dual'' here.} that along with $\lambda$ controls the proximity between $\B\alpha$ and $\B\beta$.
The ADMM procedure leads to the following update sequence:
\begin{align}
\B\beta_{k+1} \in& \argmin_{\B\beta} \; {\mathcal L}_{\lambda}(\B\beta, \B\alpha_{k}; \B\nu_{k})& \label{line-1-1}\\
\B\alpha_{k+1} \in& \argmin_{\B\alpha: \|\M{X}^\top (\M{y} - \M{X}\B\alpha)\|_{\infty} \leq \delta_{}} \; {\mathcal L}_{\lambda}(\B\beta_{k+1}, \B\alpha; \B\nu_{k}) & \label{line-1-2}\\
\B\nu_{k+1} =& \B\nu_{k} + \lambda\left( \B\alpha_{k+1} - \B\beta_{k+1}\right). \label{line-1-33}&
\end{align}
Step~\eqref{line-1-1} can be performed via a hard thresholding operation~\cite{dono:john:1994} and~\eqref{line-1-2} involves a simple projection onto the polyhedron
$\{ \B\alpha: \|\M{X}^\top (\M{y} - \M{X}\B\alpha)\|_{\infty} \leq \delta_{}\}$, which can be done efficiently, as detailed in Section~\ref{details-algo-1} (Appendix).

\smallskip

\begin{enumerate}
\item[] \noindent {\bf{ Algorithm~ 1}}
\item[(\bf 1.)] Input $( \B\beta_{1}, \B\alpha_{1},\B\nu_{1})$, choose $\lambda>0$, and repeat the following steps until convergence.
\item[(\bf 2.)] Update $(\B\beta_{k}, \B\alpha_{k}, \B\nu_{k})$ to  $(\B\beta_{k+1}, \B\alpha_{k+1}, \B\nu_{k+1})$ via~\eqref{line-1-1}--\eqref{line-1-33}.
\item[(\bf 3.)] Stop if $\|\B\beta_{k+1} - \B\beta_{k} \|_{2} \leq \tau_{1} \| \B\beta_{k} \|_{2}$ and\footnote{Here, $\tau_{1},\tau_{2}$ are tolerances for convergence, typically, taken to be equal and set to $10^{-4}$.}
 $\| \B\beta_{k} - \B\alpha_{k}\|_{2} \leq \tau_{2} \max \{ \| \B\beta_{k}\|_{2}, \| \B\alpha_{k} \|_{2} \}$, otherwise
go to Step~2.
\end{enumerate}

\smallskip

We found Algorithm~1 to work quite well in our experiments.
The algorithm may be sensitive to the choice of $\lambda$ --- affecting the solution and the time until convergence.
We recommend using multiple values of $\lambda$, and choosing the best solution among them.
In Section~\ref{sec:hybrid-algo-ubs} we address these shortcomings and describe modifications that lead to improvements in practice.

\subsubsection{Sequential Linear Optimization}\label{sec:weighted-l1}
We now describe another nonlinear optimization algorithm for obtaining upper bounds for Problem~\eqref{L0-DZ-1}, motivated by
ideas popularly used in nonconvex penalized regression (see, for example, \cite{mhf-09-jasa} and references therein).
Let us consider a family of nonconvex functions, $\rho_{\gamma} ( | \beta| )$, parametrized by $\gamma \in (0 , \bar{\gamma}]$, such that
$\gamma = \bar{\gamma}$ corresponds to $\rho_{\gamma}( | \beta|) = |\beta|$, and, as $\gamma$ decreases to $0$, $\rho_{\gamma} ( | \beta| )$ becomes a progressively better approximation to $\one(\beta \neq 0)$.  In other words,
\begin{equation}\label{approx-1-1}
\| \B\beta\|_{0} = \lim_{\gamma \rightarrow 0+} \;\; \sum_{i=1}^{p} \;\; \rho_{\gamma} ( | \beta_{i} | ).
\end{equation}

We make the following assumption about $\rho_{\gamma}(\cdot)$:

\medskip

{\noindent {\bf Assumption (A):}}
$\rho_{\gamma}(\beta)$ is  symmetric in $\beta$ around zero and continuous. Let $\rho_{\gamma}(\beta) \geq 0$ and $\rho_{\gamma}(0) =0$. For every $\gamma$, the map $|\beta| \mapsto  \rho_{\gamma} ( | \beta| )$ is concave and differentiable on $(0, \infty)$.

\medskip

Some popular choices of $\rho_{\gamma}(\cdot)$ are
$\rho_{\gamma}(t) = \log( \frac{t}{\gamma} + 1)/\log(\frac{1}{\gamma} + 1)$ and
$\rho_{\gamma}(t) = t^\gamma$ for $ t\geq 0$. We refer the reader to~\cite{mhf-09-jasa} (and references therein) for more context and examples of nonconvex penalty functions used in sparse linear regression.

We propose to compute good upper bounds for the following continuous nonconvex optimization problem:
\begin{equation}\label{obj-lla-1}
\begin{aligned}
\min\limits_{\B\beta}~~&~~ h(\B\beta):= \sum\limits_{i=1}^{p} \rho_{\gamma}( |\beta_{i}|)  \\
 \sbt ~~ &~~ \| \M{X}^\top  (\M{y} - \M{X} \B\beta)  \|_{\infty} \leq \delta,
\end{aligned}
\end{equation}
especially for $\gamma \approx 0+$.  In light of~\eqref{approx-1-1}, this leads to good upper bounds for Problem~\eqref{L0-DZ-1}.
Note that the concavity of $|\beta| \mapsto \rho_{\gamma}(|\beta|)$ leads to the following upper bound (for all $\B\beta$ and $\widetilde{\B\beta}$):
\begin{equation}\label{upper-bd-1-h}
\begin{aligned}
h(\B\beta) =& \sum_{i=1}^{p} \rho_{\gamma}( |\beta_{i}|)  \\
\leq& \underbrace{\sum_{i=1}^{p} \rho_{\gamma}( |\widetilde{\beta}_{i}|)+ \sum_{i=1}^{p}    \rho'_{\gamma}( |\widetilde{\beta}_{i}|)\left( | \beta_{i}| - | \widetilde{\beta}_{i}|  \right )}_{:=\bar{h}(\B\beta; \widetilde{\B\beta})},
\end{aligned}
\end{equation}
where $ \rho'_{\gamma}(\cdot)$ denotes the derivative of $|\beta| \mapsto \rho_{\gamma}(|\beta|)$, with the convention that
$ \rho'_{\gamma}(0)= \infty$ if the derivative is unbounded as $|\beta|\rightarrow 0+$.
Inequality~\eqref{upper-bd-1-h} suggests that we sequentially minimize an upper bound to the objective function in~\eqref{obj-lla-1}.  This leads to the
following iterative scheme:
\begin{equation}\label{obj-lla-2}
\begin{myarray}[1.3]{ccc}
\B\beta^{k+1} \in& \argmin\limits_{\B\beta}&  \sum\limits_{i=1}^{p}    \rho'_{\gamma}(|\beta^{k}_{i}|) | \beta_{i}| \\
&\sbt & \| \M{X}^\top  (\M{y} - \M{X} \B\beta)  \|_{\infty} \leq \delta,
\end{myarray}
\end{equation}
where we assume, without loss of generality, that $\B\beta^1$ is feasible for Problem~\eqref{obj-lla-2}.
The above sequential approximation of the function $h(\B\beta)$ is similar to the popular reweighted $\ell_{1}$-minimization method, used in
signal processing~\cite{boyd08-new} for sparse linear model estimation.

We now present a simple finite time convergence rate of the iterative process~\eqref{obj-lla-2} in terms of reaching an approximate first order stationary point of
Problem~\eqref{obj-lla-1}.
Towards this end, we introduce the following quantity:
\begin{equation}\label{obj-lla-2-defn}
\begin{myarray}[1.3]{ccc}
\Delta (\B\theta):= & \min\limits_{\B\beta}&  \sum\limits_{i=1}^{p}   \rho'_{\gamma}(|\theta_{i}|) \left(| \beta_{i}|  - |\theta_{i}| \right) \\
&\sbt & \| \M{X}^\top  (\M{y} - \M{X} \B\beta)  \|_{\infty} \leq \delta,
\end{myarray}
\end{equation}
which we use to define a first-order stationary point for Problem~\eqref{obj-lla-1}.


\begin{definition}\label{def-1}
Suppose $\widehat{\B\theta}$ is feasible for Problem~\eqref{obj-lla-1}.
We say that $\widehat{\B\theta}$ is  a first-order stationary point for Problem~\eqref{obj-lla-1}, if $\Delta(\widehat{\B\theta}) = 0$.
 For $\phi>0$, $\widehat{\B\theta}$ is said to be an $\phi$ accurate first-order stationary point if $ \Delta(\widehat{\B\theta}) \geq - \phi$.
\end{definition}

%
%

We note that $\Delta(\B\theta )$ (assuming that $\B\theta$ is
feasible for Problem~\eqref{obj-lla-1}) is a measure of how \emph{far} $\widehat{\B\theta}$ is from a first order stationary point of Problem~\eqref{obj-lla-1}--if $\Delta(\B\theta)<0$,
then the current estimate $\B\theta$ can be improved, if  $\Delta(\B\theta)=0$ then the solution cannot be improved via update~\eqref{obj-lla-2}. We refer the reader to Section~\ref{sec:thm:conv-rate-1} for a more detailed explanation.
The following theorem (for a proof see Section~\ref{sec:thm:conv-rate-1}) presents a finite time convergence rate analysis of the sequence~\eqref{obj-lla-2} to a first-order stationary point for Problem~\eqref{obj-lla-1}.
\begin{thm}\label{thm:conv-rate-1}
Consider Problem~\eqref{obj-lla-1} for a fixed $\gamma>0$, with Assumption (A) on $\rho_{\gamma}(\cdot)$ in place. The update sequence $\B\beta^{k}$,
defined via~\eqref{obj-lla-2}, leads to a decreasing sequence of objective values for Problem~\eqref{obj-lla-1}: $h(\B\beta^{k+1}) \leq h(\B\beta^k)$ for all $k \geq 1$.
In addition, for every $\mk >0$ we have the following finite-time convergence rate:
$$ \min_{1\leq k\leq \mk}\left \{ -\Delta(\B\beta^{k}) \right \} \leq \frac{1}{\mk} \left(h(\B\beta^1) - \widehat{h} \right), $$
where the sequence of objective function values satisfies $h(\B\beta^{k})\downarrow \widehat{h}$ as $k \rightarrow \infty$.
\end{thm}

\medskip

We emphasize that the result in Theorem~\ref{thm:conv-rate-1} pertains to the performance of the sequence~\eqref{obj-lla-2} as a numerical optimization scheme, and has no direct implication on the statistical properties of the sequence.
Theorem~\ref{thm:conv-rate-1} implies that for any $\phi>0$, it takes at most $\mk = O(\frac{1}{\phi})$ many iterations to reach a $\phi$-accurate first-order stationary point, i.e.,
there exists a $1\leq k^* \leq \mk$ such that $\Delta(\B\beta^{k^*}) > - \phi$.
 The sequence $\B\beta^k$ leads to an estimate $\widehat{\B\beta}_{\gamma}$, an upper bound for Problem~\eqref{obj-lla-1} for a fixed $\gamma$.
 Since our intent is to obtain a good solution to Problem~\eqref{L0-DZ-1}, we make use of property~\eqref{approx-1-1}. This suggests that
we obtain a good upper bound to Problem~\eqref{obj-lla-1} for a small value of $\gamma \approx 0+$.
Instead of applying iteration~\eqref{obj-lla-2} for a pre-specified (small) value of $\gamma$,
we recommend using a continuation strategy in practice. We take a sequence of decreasing values of $\gamma \in \{ \gamma_{1}, \ldots, \gamma_{N} \}$, where
$\gamma_{i} > \gamma_{i+1}$. We use $\widehat{\B\beta}_{\gamma_{i}}$ as a warm-start for obtaining a good solution (upper bound) to Problem~\eqref{obj-lla-1} for a smaller value of $\gamma  = \gamma_{i+1}$. In our numerical experiments, this
continuation strategy seems to work well.
The  method is summarized below.

\begin{enumerate}
\item[] \noindent {\bf {Algorithm~2}}
\item[(\bf 1.)] Take a decreasing sequence of $\gamma$ values $\{ \gamma_{1}, \ldots, \gamma_{N}\}$;
initialize with $\B\beta^0 = \M{0}$; and fix a value of $\text{Tol} = 10^{-5}$ (say).
Set $\kappa =1$ and $\gamma = \gamma_{\kappa}$.
\item[(\bf 2.)] Use the update sequence rule~\eqref{obj-lla-2} until some convergence criterion is met:
$\left(-\Delta(\B\beta^{k})\right)  < \text{Tol}.$ Let $\widehat{\B\beta}_{\gamma}$ denote the estimate of $\B\beta$, upon convergence.
\item[(\bf 3.)] Set $\kappa \leftarrow \kappa + 1$, $\gamma = \gamma_{\kappa}$ and $\B\beta^0 \leftarrow \widehat{\B\beta}_{\gamma}$. If $\kappa \leq N$, then
 goto Step~2. If $\kappa > N$, exit with $\widehat{\B\beta}_{\gamma}$ as an upper bound to Problem~\eqref{L0-DZ-1}.
\end{enumerate}


The linear optimization Problem~\eqref{obj-lla-2} can be solved quite efficiently using simplex methods.
For larger problems, i.e. $p$ larger than a few thousand, we recommend using modern first-order method as described in Section~\ref{sec:solvingproblem-22}.
Since Algorithm~2 requires solving several instances of related problems of the form~\eqref{obj-lla-2}, the warm-start capabilities of simplex methods and
first-order methods lead to computational benefits.

\subsubsection{Algorithm~3: Combining the Strengths of Algorithm~1 and Algorithm~2}\label{sec:hybrid-algo-ubs}
In our empirical studies we observed that Algorithm~1 is more effective in obtaining good upper bounds than Algorithm~2 for a given time limit.
Algorithm~2, on the other hand, has stronger convergence guarantees than Algorithm~1.
Algorithm~1 leads to an estimate of $\B\beta$ that is sparse but approximately satisfies\footnote{This is because Algorithm~1 delivers a pair, $\B\alpha, \B\beta$, which are approximately equal:
$\B\alpha \approx \B\beta$; $\B\alpha$ is feasible for Problem~\eqref{L0-DZ-1} but need not be exactly sparse; $\B\beta$, on the other hand, is sparse but approximately feasible.}
 the feasibility constraint of Problem~\eqref{L0-DZ-1}. Algorithm~2, in contrast, leads to solutions that are both sparse and feasible --- these advantages make Algorithm~2 an important tool in our framework. We propose to combine the best features of Algorithms~1 and 2 to develop a hybrid variant: Algorithm~3, which we recommend to use in practice. Algorithm~3 is simple but very effective: it \emph{uses} the solution obtained from Algorithm~1, say, $\widehat{\B\beta}^{(1)}$, to create a set ${\mathcal I} \subset \{1, \ldots, p\}$, which includes the nonzeros in $\widehat{\B\beta}^{(1)}$, and then
applies Algorithm~2 on this set ${\mathcal I}$ --- the details of this method are presented in Section~\ref{sec:details-algo-3} (in the Appendix).
\section{Structured \MIO Formulations and Certificates of Optimality}\label{sec:adv-formulations-1}
This section is dedicated to enhancements of the basic \LDSF~formulation~\eqref{L0-DZ-1-mio-M}, presented in Section~\ref{sec:mio-form-1}.  These are particularly useful in delivering tighter lower bounds, thereby providing certificates of global optimality in shorter times.

Note that formulation~\eqref{L0-DZ-1-mio-M} requires the specification of $\MU$ large enough to include the solution of \LDSF.
We mention another \MIO formulation for Problem~\eqref{L0-DZ-1}, based on
Specially Ordered Sets~\cite{bertsimas2005optimization_new}.
We introduce binary variables $z_{i} \in \{0,1\}$, which satisfy the condition $(1 - z_{i}) \beta_{i} = 0$ for all $i = 1, \ldots, p$ --- in other words,
if $z_{i}=0$, then $\beta_{i} = 0$, and if $z_{i} =1$, then $\beta_{i}$ is unconstrained.
This condition can be modeled via integer optimization using Specially Ordered Sets of Type 1 (SOS-1).  More specifically,
$$ (1 - z_{i}) \beta_{i} = 0 \;\;  \iff  \;\; (\beta_{i}, 1 - z_{i}) : \text{SOS-1},$$
for every $i = 1, \ldots, p.$  This leads to the following \MIO formulation for Problem~\eqref{L0-DZ-1}:
\begin{equation}\label{L0-DZ-sos}
\begin{myarray}[1.3]{l  l  r}
\min\limits_{\B\beta, \M{z}} &  \sum\limits_{i=1}^{p} z_{i} & \\
\sbt&   - \delta  \leq d_{j}  - \langle \M{q}_{j}, \B\beta \rangle  \leq \delta & j = 1, \ldots, p\\
& (\beta_j, 1 - z_{j}): \text{SOS-1}& j = 1, \ldots, p\\
& z_{j} \in \{ 0 , 1 \} & j = 1, \ldots, p,
\end{myarray}
\end{equation}
where, we use the notation as used in Problem~\eqref{L0-DZ-1-mio-M}.
Observe that unlike~\eqref{L0-DZ-1-mio-M}, Problem~\eqref{L0-DZ-sos} does not contain
any parameter $\MU$ in its formulation.
Problem~\eqref{L0-DZ-sos} may be
preferred over Problem~\eqref{L0-DZ-1-mio-M} when the different nonzero values of $|\widehat{\beta}_{i}|$'s have widely different amplitudes.
In general, however, we found empirically that the algorithmic performances of formulations~\eqref{L0-DZ-sos} and~\eqref{L0-DZ-1-mio-M} are comparable.
The \MIO formulations~\eqref{L0-DZ-1-mio-M} and~\eqref{L0-DZ-sos} are found to work quite well in obtaining good \emph{upper bounds} for
up to $p=10,\!000$, once they are warm-started via the discrete first-order methods described in Section~\ref{sec:FO-methods}.
If additional problem-specific
information which we refer to as ``intelligence'' is supplied to the \MIO formulations~\eqref{L0-DZ-1-mio-M} and~\eqref{L0-DZ-sos},
the results are found to improve substantially --- 
as shown in
Section~\ref{sec:cert-glob-opt-1}.
More specifically,  we use the term ``intelligence'' to broadly refer to two components:
\begin{itemize}
\item[(a)] providing an advanced warm-start to the \MIO solver, obtained via our
discrete first-order methods
\item[(b)] arming the \MIO solver with information in the form of
interval bounds on the regression coefficients $\beta_j$, predictions $\M{x}^\top_{i}\B\beta$, and also
bounds on $\|\B\beta\|_{1}$ and $\|\M{X}\B\beta\|_{1}$.
\end{itemize}
We note that the resulting formulation with the additional bounds as suggested in (b), above, should lead to a solution for Problem~\eqref{L0-DZ-1}.   We, thus, present the following
\emph{structured} version of formulation~\eqref{L0-DZ-1}:
\begin{subequations}\label{L0-DZ-sos-bounds1}
\begin{align}
\min\limits_{\B\beta, \M{z}}~~~&  \sum_{i=1}^{p} z_{i} &  \nonumber\\
\sbt~~~&   - \delta  \leq d_{j}  - \langle \M{x}_{j} , \B\xi \rangle  \leq \delta & j = 1, \ldots, p& \label{line-11-1}\\
& - \MU^j z_{j} \leq  \beta_j \leq \MU^j  z_{j} & j = 1, \ldots, p& \label{line-11-2} \\
& z_{j} \in \{ 0 , 1 \} & j = 1, \ldots, p& \nonumber \\
& \B\xi = \M{X} \B\beta &&  \label{line-11-3}\\
& - \MU^j \leq \beta_{j} \leq \MU^{j}& j = 1, \ldots, p&  \label{line-11-4} \\
& - \MU^{\xi, i} \leq \xi_{i} \leq \MU^{\xi, i}& i = 1, \ldots, n& \label{line-11-5} \\
&  \|\B\beta\|_{1} \leq  {\mathcal M}_{\ell} && \nonumber \\
&  \sum_{i=1}^{n} |\xi_{i}| \leq  {\mathcal M}^{\xi}_{\ell}, \label{line-11-6} &&
\end{align}
\end{subequations}
where the optimization variables are $\B\beta\in \RR^{p}, \M{z} \in \{0, 1\}^p,   \B{\xi} \in \RR^n$, and the parameters
$\MU^{i},\MU^{\xi, i},{\mathcal M}_{\ell},{\mathcal M}^{\xi}_{\ell}$ control, respectively,
upper bounds on $|\beta_{i}|$, $| \langle \M{x}_{i}, \B\beta \rangle |$, $\| \B\beta \|_{1}$ and $\| \M{X} \B\beta \|_{1}$.
We note that the parameter $\MU$ in Problem~\eqref{L0-DZ-1-mio-M} is such that $\MU \geq \max_{j} \MU^{j}$ for $j=1, \ldots, p$.
Problem~\eqref{L0-DZ-sos-bounds1} is equivalent to the following constrained version of Problem~\eqref{L0-DZ-1}:
 \begin{equation}\label{L0-DZ-1-allbounds}
\begin{myarray}[1.3]{cc r }
\min \limits_{\B\beta} &  \|\B\beta\|_0 &\\
\sbt&  \|\M{X}^\top (\M{y}-\M{X}\B\beta)\|_\infty \leq \delta_{}&\\
& - \MU^j \leq \beta_{j} \leq \MU^{j}& j = 1, \ldots, p \\
& | \langle \M{x}_{i},  \B\beta \rangle | \leq  {\mathcal M}^{\xi,i}_{U}& i= 1, \ldots, n  \\
&  \|\B\beta\|_{1} \leq  {\mathcal M}_{\ell}&\\
 & \| \M{X} \B\beta\|_{1} \leq  {\mathcal M}^{\xi}_{\ell}.&
\end{myarray}
\end{equation}
Section~\ref{sec:bounds-compute} presents several strategies to compute these parameters such that
a solution to Problem~\eqref{L0-DZ-1-allbounds} is also a solution to Problem~\eqref{L0-DZ-1}.

We present a few variations of  formulation~\eqref{L0-DZ-sos-bounds1} that might be preferred from a computational viewpoint, depending upon the problem instance under consideration.
For large values of $p$ and $n$ (approximately a few thousand), the constraints appearing in~\eqref{line-11-1} and~\eqref{line-11-3} may be replaced by:
$$ - \delta  \leq d_{j}  - \langle \tilde{\M{x}}_{j} , \B\xi \rangle  \leq \delta,\;\; \B\xi  = \tilde{\M{X}} \B\beta,$$
where, $\tilde{\M{X}}^\top\tilde{\M{X}} = \M{X}^\top \M{X}$ and $\tilde{\M{X}}$ is triangular --- this leads to a sparse representation of the constraints appearing in~\eqref{L0-DZ-sos-bounds1}.
When $n$ is large and $p$ is smaller, it may be useful to perform a variable reduction by removing the variable $\B\xi$ from~\eqref{L0-DZ-sos-bounds1}.
This will replace constraint~\eqref{line-11-1}
by $ - \delta  \leq d_{j}  - \langle \M{q}_{j} , \B\beta \rangle  \leq \delta$, and constraints~\eqref{line-11-3},~\eqref{line-11-5}
and~\eqref{line-11-6} will be dropped.
Constraints~\eqref{line-11-2} imply the bounds indicated in the constraints~\eqref{line-11-4}, hence the constraints~\eqref{line-11-4} may be dropped in favor of a formulation with fewer constraints.

We note that formulation~\eqref{L0-DZ-sos-bounds1} is an optimization problem with many more continuous variables than
formulation~\eqref{L0-DZ-1-mio-M}. This implies that the \MIO solver needs to do more work at every node, by solving larger convex linear
programs. However, the advantage is that the resulting formulation is more structured, and, thus, tighter
lower bounds may be obtained by exploring fewer nodes.
Section~\ref{sec:cert-glob-opt-1} presents some computational results illustrating the performance of the above framework.

\subsection{Specification of Parameters}\label{sec:bounds-compute}
We present herein, several data-driven ways to compute the parameters in formulation~\eqref{L0-DZ-sos-bounds1}.
The methods presented here are quite different from those proposed in~\cite{bertsimas2015best}, where, the authors
rely crucially on being able to compute analytic expressions for least squares solutions for a given subset size---such expressions are not available for Problem~\eqref{L0-DZ-1}.

\subsubsection{Specification of Parameters via Linear Optimization}\label{subsub-opt-cvx-1}
We present several methods based on linear optimization that can be used to estimate the parameters appearing in Problem~\eqref{L0-DZ-sos-bounds1}, in such a way that these estimates lead to $\widehat{\B\beta}$, a  solution to Problem~\eqref{L0-DZ-1}.


\pparagraph{Bounds on $\widehat{\beta}_{i}$'s.}
Consider the following pair of  linear optimization problems:
\begin{equation}\label{ubs-data-1}
\begin{aligned}
\begin{myarray}[1.3]{c c  c }
\mu^+_{i} :=&  \max\limits_{\B\beta}  \;\; \beta_{i} & \\
\sbt\;\;\; & \|\M{X}^\top ( \M{y} - \M{X} \B\beta) \|_{\infty} \leq \delta_{},&
\end{myarray}   \\
& & \\
\begin{myarray}[1.3]{ccc }
\mu^-_{i} :=&  \min \limits_{\B\beta}  \;\; \beta_{i} & \\
\sbt\;\;\; & \|\M{X}^\top ( \M{y} - \M{X} \B\beta) \|_{\infty} \leq \delta_{},&
\end{myarray}
\end{aligned}
\end{equation}
for $i = 1, \ldots, p$. Note that $\mu^+_{i}$ and $\mu_{i}^{-}$ provide upper and lower bounds on $\widehat{\beta}_{i}$ for every $i$.
$\mu_{i}^+$ is typically a strict upper bound to $\widehat{\beta}_{i}$, because~\eqref{ubs-data-1} does not account for the fact that
solutions to Problem~\eqref{L0-DZ-1} are sparse.
Similarly, $\mu^{-}_{i}$ is a lower bound to $\widehat{\beta}_{i}$, and it is easy to see that ${\mathcal M}^i_{U} = \max \{ |\mu^+_{i}| , |  \mu^{-}_{i} |  \}$ is an upper bound to $| \widehat{\beta}_{i} |$.
Note that solutions to Problem~\eqref{ubs-data-1} are finite only if the feasible set is bounded.
If $n >p$ and if the entries of $\M{X}$ are drawn from a continuous probability measure, then the bounds are finite with probability one.
The above bounds can be made tighter by using information about upper bounds on Problem~\eqref{L0-DZ-1} as obtained via the discrete first-order methods. We describe such methods in
Section~\ref{sec:tight-bounds-bet-1} (Appendix).
Once upper bounds on $|\widehat{\beta}_{i}|$, i.e. $\MU^{i}$, are obtained, they can be used to compute
bounds on $\|\widehat{\B\beta}\|_{\infty}$
and $\|\widehat{\B\beta}\|_{1}$ as follows:
\begin{equation*}
\begin{aligned}
\|\widehat{\B\beta}\|_{\infty} \leq {\mathcal M}_{U} =  \max_{i=1, \ldots, p} \;\; {\mathcal M}^i_{U}  ~~\text{and}~~  \|\widehat{\B\beta}\|_{1}  \leq  \sum_{i=1}^{\alpha_{0}}  {\mathcal M}^{(i)}_{U},
\end{aligned}
\end{equation*}
where, $\alpha_{0}$ denotes an upper bound to Problem~\eqref{L0-DZ-1} and ${\mathcal M}^{(1)}_{U} \geq {\mathcal M}^{(2)}_{U} \geq \ldots  \geq {\mathcal M}^{(p)}_{U}$.

\pparagraph{Bounds on $\langle \M{x}_{i}, \widehat{\B\beta} \rangle$'s.}
Bounds on $\langle \M{x}_{i}, \widehat{\B\beta} \rangle$ can be obtained by solving the following pair of linear optimization problems:
\begin{equation}\label{ubs-data-2-new}
\begin{aligned}
\begin{myarray}[1.3]{ ccc }
v^+_{i}(\alpha_0) :=&  \max\limits_{\B\beta}  \;\; \langle \M{x}_{i}, \B\beta \rangle & \\
\sbt & \|\M{X}^\top ( \M{y} - \M{X} \B\beta) \|_{\infty} \leq \delta_{}&\\
& \| \B\beta\|_{\infty} \leq \MU& \\
& \| \B\beta\|_{1} \leq \MU\alpha_{0}&
\end{myarray}  \\
&& \\
\begin{myarray}[1.3]{ ccc}
v^-_{i} (\alpha_0):=&  \min\limits_{\B\beta}  \;\;\langle \M{x}_{i}, \B\beta \rangle & \\
\sbt &\|\M{X}^\top ( \M{y} - \M{X} \B\beta) \|_{\infty} \leq \delta_{}&\\
& \| \B\beta\|_{\infty} \leq \MU& \\
& \| \B\beta\|_{1} \leq \MU\alpha_{0},&
\end{myarray}
\end{aligned}
\end{equation}
for every $i = 1, \ldots, n$.

Analogous to the bounds derived via Problem~\eqref{ubs-data-1}, it is also possible to compute more conservative bounds on $\langle \M{x}_{i}, \widehat{\B\beta} \rangle$, by dropping the constraints $\| \B\beta\|_{\infty} \leq \MU$ and $\| \B\beta\|_{1} \leq \MU\alpha_{0}$ in Problem~\eqref{ubs-data-2-new}.
This gives nontrivial bounds even for the under-determined $n<p$ case, as long as $\M{X}$ has rank $n$ (this is in contrast with the bounds
from Problem~\eqref{ubs-data-1} being vacuous when $n <p$).
It is also possible to estimate bounds on $\langle \M{x}_{i}, \widehat{\B\beta}  \rangle$ by including an additional constraint: $\| \M{X} \B\beta \|_{\infty} \leq \MU^{\xi}$, and using
an iterative method as described in the Appendix, Section~\ref{sec:tight-bounds-bet-1} (see Step-1--Step-4) while computing bounds on the regression coefficients.


The quantity ${ \v }_{i} = \max \{ v^+_{i}(\alpha_0),  -v^{-}_{i} (\alpha_0)  \}$ provides an upper bound
to $| \langle \M{x}_{i}, \widehat{\B\beta} \rangle |$.  In particular, this leads to the following upper bounds:
$$\|\M{X}\widehat{\B\beta}\|_{\infty} \leq  \max_{i=1, \ldots, n} { \v }_{i}~~~~\text{and}~~~~\|\M{X}\widehat{\B\beta}\|_{1}  \leq  \sum_{i=1}^{n} { \v }_{i},$$
leading to a data-driven method to estimate bounds appearing in~\eqref{L0-DZ-sos-bounds1}.


\pparagraph{Computational Cost.}
Computing the quantities appearing in~\eqref{ubs-data-1},~\eqref{bound-s-1-2} and~\eqref{ubs-data-2-new} requires solving  at least $2(p+n)$ linear optimization problems.
However, these individual problems are quite simple to parallelize and they need to be solved once, before proceeding to solve Problem~\eqref{L0-DZ-sos-bounds1}.
These linear optimization problems can be solved by simplex based solvers quite easily
for $p$ in the lower thousands (typically less than a minute with \textsc{Gurobi}'s simplex solver).

\subsubsection{Specification of Parameters from warm-starts}\label{subsub-opt-cvx-2}
We present herein, simple practical methods to compute the parameter values by using good upper bounds to Problem~\eqref{L0-DZ-1}.
Let $\widehat{\B\beta}^0$ denote a solution that corresponds to a good upper bound to Problem~\eqref{L0-DZ-1}.
$\widehat{\B\beta}^0$ can be obtained from Algorithm~3, for example. One can also use the solution obtained from Algorithm~3 as a warm-start to
Problem~\eqref{L0-DZ-sos}  and allow it to run for a few
minutes --- the resulting estimate may be used as $\widehat{\B\beta}^0$.
The parameters appearing in the bounds can be based on  $\widehat{\B\beta}^0$, as follows.
To be on the conservative side, we recommend setting the same bound for all the $\MU^{i}$'s: for example, they can all be assigned the value $\tau \| \widehat{\B\beta}^0 \|_{\infty}$.
Similarly, a conservative bound for all the ${\mathcal M}^{\xi,i}_{U}$'s is given by $\tau \|\M{X} \widehat{\B\beta}^0 \|_{\infty}$.
In addition,
 we can set ${\mathcal M}_{\ell} = \min \left\{ \tau \| \widehat{\B\beta}^0 \|_{0}\| \widehat{\B\beta}^0 \|_{\infty}, \tau \| \widehat{\B\beta}^0 \|_{1} \right\}$
and ${\mathcal M}^{\xi}_{\ell} =  \tau \| \M{X} \widehat{\B\beta}^0 \|_{\infty}$, for some value of $\tau \in \{ 1.5, 2 \}$.

The method described above leads to parameter specific bounds as a simple by-product
of our general algorithmic framework.  Unlike the methods in Section~\ref{subsub-opt-cvx-1}, it requires no additional computation.  On the other hand, the bounds in Section~\ref{subsub-opt-cvx-1} are conservative, because they are implied by the bounds from Problem~\eqref{L0-DZ-1}.


\begin{table}[]
\begin{center}
\resizebox{0.95\textwidth}{.3\textheight}{\begin{tabular}{c  c}
\scalebox{1}{\begin{tabular}{| cccc |}
\multicolumn{4}{c}{{\bf Type-1}  $(n=100,p=1000)$}  \medskip \\ \hline
Data Fidelity & Time &  \multicolumn{2}{c|}{Quality of Upper Bounds}\\
 Parameter & (in secs) & With Warm & Vanilla\\   \hline
\multirow{3}{*}{$1.5\bar{\delta}$} & 2 (*) & 0 & -- \\
  & 120 & 0 & 0 \\
  & 500 & 0 & 0 \\ \hline
\multirow{3}{*}{$\bar{\delta}$} & 90 (*) & 0& 42.85\\
 & 120 & 0 & 42.85 \\
 & 500 & 0 & 28.57 \\ \hline
\multirow{3}{*}{$0.5\bar{\delta}$} & 57 (*) & 0 & 200 \\
& 120 & 0 &  86.66 \\
 & 500 & 0 &26.66\\  \hline
\multirow{3}{*}{$0.2\bar{\delta}$} & 120 & 3.12 & 25 \\
& 210 (*) & 0 & 15.62 \\
 & 500 & 0 & 15.62\\
   \hline
\end{tabular}}&
\scalebox{1}{\begin{tabular}{|cccc|}
\multicolumn{4}{c}{{\bf Type-2}  $(n=300,p=1000)$}  \medskip  \\ \hline
Data Fidelity & Time &  \multicolumn{2}{c|}{Quality of Upper Bounds}\\
 Parameter & (in secs) & With Warm & Vanilla\\   \hline
\multirow{3}{*}{$1.5\bar{\delta}$} & 50 (*) & 0 & -- \\
 & 120 & 0 & 214.28 \\
& 500 & 0 & 14.28 \\ \hline

  \multirow{3}{*}{$\bar{\delta}$} & 120 & 6.66 & 146.66 \\
  & 132(*) & 0 & 73.33 \\
& 500 & 0 & 0 \\ \hline

 \multirow{3}{*}{$0.5\bar{\delta}$} & 35 (*) & 0 & -- \\
 & 120 & 0 & 29.62 \\
   & 500 & 0 & 25.92 \\ \hline
 \multirow{3}{*}{$0.2\bar{\delta}$}& 40 (*) & 0& -- \\
 & 120 & 0 &73.44 \\
 & 500 & 0 & 23.43 \\
   \hline
\end{tabular}}  \medskip \\
& \\
\scalebox{1}{\begin{tabular}{|cccc|}
\multicolumn{4}{c}{{\bf Type-3}  $(n=600,p=2000)$}  \medskip  \\ \hline
Data Fidelity & Time &  \multicolumn{2}{c|}{Quality of Upper Bounds}\\
 Parameter & (in secs) & With Warm & Vanilla\\   \hline

\multirow{3}{*}{$1.5\bar{\delta}$} & 500 & 7.14 & -- \\
& 530 (*) & 0 & -- \\
 & 950 & 0 & 28.57\\ \hline

 \multirow{3}{*}{$\bar{\delta}$}   & 500 & 11.11 & -- \\
 & 875 (*) & 0 & 137.03 \\
 & 950 & 0 & 33.33 \\ \hline

  \multirow{3}{*}{$0.5\bar{\delta}$} & 55 (*)& 0 & -- \\
  & 500 & 0 & -- \\
 & 950 & 0 & 120.37 \\ \hline
  \multirow{3}{*}{$0.2\bar{\delta}$} & 500 & 0.8 & -- \\
& 560 (*)& 0 & -- \\
& 950 & 0 & 77.6\\
   \hline
\end{tabular}}&

\scalebox{1}{\begin{tabular}{|cccc|}
\multicolumn{4}{c}{{\bf Type-4}  $(n=58,p=2000)$}   \medskip  \\ \hline
Data Fidelity & Time &  \multicolumn{2}{c|}{Quality of Upper Bounds}\\
 Parameter & (in secs) & With Warm & Vanilla\\   \hline

\multirow{3}{*}{$1.2\bar{\delta}$}  & 300 (*) & 0 & 220 \\
 & 370& 0 & 0\\
& 600 & 0& 0 \\ \hline

 \multirow{3}{*}{$\bar{\delta}$}   & 300 & 16.66 & -- \\
 & 367 (*)& 0 & 216.66\\
 & 600 & 0& 0 \\ \hline

\multirow{3}{*}{$0.5\bar{\delta}$}  & 300 & 5 & -- \\
   & 560 (*)& 0 & 95 \\
  & 600 & 0 & 95 \\ \hline
 \multirow{3}{*}{$0.2\bar{\delta}$} & 145 (*)& 0 & -- \\
   & 300 & 0 & 165 \\
 & 600 & 0 & 60 \\
   \hline
\end{tabular}}
\end{tabular}}\end{center}
\caption{\upshape{Tables showing ``Quality of Upper Bounds'', defined as $100 \times (h_{\text{alg}} - \widehat{h})/\widehat{h}$, where $h_{\text{alg}}$ refers to the objective value obtained by
algorithm ``alg'' (at the given time), and $\widehat{h}$ is the best objective value found in the entire run-time duration of all the algorithms. Two cases of ``alg'' $\in \{\text{``With Warm'', ``Vanilla''} \}$  have been considered: ``With Warm'' denotes \MIO
warm-started with a solution from Algorithm~3 (the structured formulations of Section~\ref{sec:adv-formulations-1} are not used here);
``Vanilla'' denotes a \MIO solver without any warm-start specification. ``With Warm'' is found to obtain the best upper bound for a given computation time-limit in all the instances.
In several instances, \MIO is found to improve the solution obtained via Algorithm~3, after accepting it as a warm-start.
  For Type-1,2 the total time limit was 500 secs; for Type-3 it was 950 secs, and for Type-4 the algorithms were considered for a total time limit of 600 secs. For method ``With Warm'', the times
reported show the overall time taken by Algorithm~3 and the \MIO algorithm. An asterisk ``(*)'' indicates that the best solution is obtained at that time. A ``--''  means that no feasible solution was obtained by the algorithm in that time.}}\label{tab:upper-bounds1}
\end{table}

\section{Numerical Experiments: Algorithmic Performance} \label{sec:numerics-algo}
In this section, we report extensive numerical experiments that demonstrate:
(a) the usefulness of the discrete first-order methods (Section~\ref{sec:FO-methods}) in obtaining good quality upper bounds, especially when they are
used to provide warm-starts to \MIO solvers --- this is shown in  Section~\ref{sec-upper-bounds-1}; and
(b) how advanced warm-starts, coupled with the enhanced  formulations presented in Section~\ref{sec:bounds-compute}, can be used to
improve the overall run-time for off-the-shelf \MIO solvers, when proving global optimality for the \LDSF~problem --- this is shown in Section~\ref{sec:cert-glob-opt-1}.

All computations were carried out on Columbia University's high performance computing (HPC) facility, \url{http://hpc.cc.columbia.edu/}, on the {\texttt{Yeti}} cluster computing environment.
The discrete first-order methods were implemented in \textsc{Matlab} 2014a, and we used \textsc{Gurobi}~\cite{gurobi} version 6.0.3.
For all experiments in Sections~\ref{sec-upper-bounds-1} and~\ref{sec:cert-glob-opt-1} (except the large scale examples) we used 16GB of memory.

\subsection{Obtaining Good Quality Upper Bounds} \label{sec-upper-bounds-1}
From a practical viewpoint, being able to obtain good quality upper bounds to Problem~\eqref{L0-DZ-1}
is, perhaps, of foremost importance. To demonstrate
the effectiveness of our computational framework in this regard,
we perform a series of experiments on the data-types described below.

\medskip

\noindent {\texttt{Type-Synth:}} We generate a Gaussian ensemble  $\M{X}_{n \times p} \sim \text{MVN}(\M{0}, \B\Sigma)$, where $\sigma_{ij} = \rho^{|i-j|}$ for some value of $\rho \in [0,1)$, with the convention that $0^0=1$.
 The underlying true regression coefficient vector, $\B\beta^*\in \RR^p$,
has $\beta^*_{j} = 1$ for $k^*$ equi-spaced values of $j\in\{1, \ldots, p\}$ and $\beta^*_{j}=0$ for the remaining values of~$j$.


\noindent {\bf{ Type-1:}} This is of {\texttt{Type-Synth}} with $n=100$,  $p = 1000$,  $\rho=0$, $k^*=10$. We studied Problem~\eqref{L0-DZ-1} for four different values
of the parameter $\delta$ set at $\bar{\delta} (1.5, 1, 0.5, 0.2)$ with $\bar{\delta}$ being defined below.

\noindent {\bf{ Type-2:}} This is of {\texttt{Type-Synth}} with $n=300$,  $p = 1000$, $\rho=0.8$, $k^*= 25$. Here $\delta$ values were set as $\bar{\delta} (1.5, 1, 0.5, 0.2)$.

\noindent {\bf{ Type-3:}}  This is of {\texttt{Type-Synth}} with $n=600$,  $p = 2000$, $\rho=0.8$, $k^*= 40$. Here $\delta$ values were set as $\bar{\delta} (1.5, 1, 0.5, 0.2)$.

\noindent {\bf{ Type-4:}} This is a semi-synthetic dataset: we considered the Radiation sensitivity gene expression dataset\footnote{We downloaded the dataset from the website~\url{http://statweb.stanford.edu/~tibs/ElemStatLearn/datasets/}}
from Ch. 16 of the book~\cite{FHT-09-new}.
 The features were randomly downsampled
to $p=2000$ and there were $n=58$ observations.  We generated response $\M{y}$ based on a linear model with
 $\|\B\beta^*\|_{0} = 10$, $\beta^*_{j}=1$ for $j \leq 10$, and $\beta_{j}^0=0$ for $j > 10$.
Here $\delta$ values were set as $\bar{\delta} (1.2, 1, 0.5, 0.2)$.

\noindent {\bf{ Type-5:}}  This is of {\texttt{Type-Synth}} with $n=1000$,  $p = 3000$,  $\rho=0$, $k^*=10$;
we considered one value of $\delta$, which was set to $\overline{\delta}$.

In each of the above examples, after $\M{X}$ was generated, we standardized its columns to have zero mean and unit $\ell_{2}$-norm.  Then, the response was generated as
$\M{y} = \M{X} \B\beta^* + \B{\epsilon}$, where $\epsilon_{i} \stackrel{\text{iid}}{\sim} N(0, \sigma^2)$, and $\sigma^2$ was adjusted to match the selected value of
SNR$=\text{Var}( \M{x}^\top \B\beta^*)/\sigma^2$ (taken as 3 in all the above cases); the reference value of the tuning parameter was set to $\bar{\delta} = \| \M{X}^\top (\M{y}- \M{X}\B\beta^*) \|_{\infty}$.

We studied different first-order algorithms described in Section~\ref{sec:FO-methods1}.
Algorithm~3 was empirically seen to have the best performance over its constituents, Algorithms~1 and~2, when used separately.
Hence, we used Algorithm~3 in all the experiments to obtain good upper bounds to Problem~\eqref{L0-DZ-1}.
The solution obtained from Algorithm~3 was passed as a warm-start to the \MIO formulation~\eqref{L0-DZ-1-mio-M} (for a large value of $\MU=10^3$) --- this hybrid \MIO approach is denoted by ``With Warm'' in Table~\ref{tab:upper-bounds1}. We compared this method with the vanilla
 \MIO formulation~\eqref{L0-DZ-1-mio-M} (for a large value of $\MU=10^3$), which was implemented without any warm-start information.
Table~\ref{tab:upper-bounds1} shows the objective values obtained by these two methods --- the \MIO algorithm aided with advanced warm-starts was found to perform the best across all the examples. For the hybrid approach (``With Warm''), in many of the instances, the solution obtained by Algorithm~3 was further improved by {\texttt{MILO}}. In some cases, the vanilla \MIO approach took a while before it was able to find a feasible solution.
For example, in the Type-5 setting (which does not appear in Table~\ref{tab:upper-bounds1}) the
best solution  was delivered by Algorithm~3 within one minute; in contrast, the vanilla \MIO algorithm failed to find a feasible solution within 1000 seconds.
\begin{figure}[h!]
\centering
\resizebox{\textwidth}{0.14\textheight}{\begin{tabular}{l c c c}
\multicolumn{ 4}{c}{{ \sf {Diabetes Dataset $(n=442,p=64)$ } }} \smallskip \\
 & \sf { { $\|\widehat{\B\beta}\|_{0} = 31$}} & \sf { { $\|\widehat{\B\beta}\|_{0} = 14$ }} & \sf { { $\|\widehat{\B\beta}\|_{0} = 7$ }} \\
\rotatebox{90}{\sf {\scriptsize{~~~~~~~~~~~~~~\MIO~Optimality Gap (in \%)}}}&
\includegraphics[width=0.33\textwidth,height=0.3\textheight,  trim =1.0cm 1.5cm .2cm 1.5cm, clip = true ]{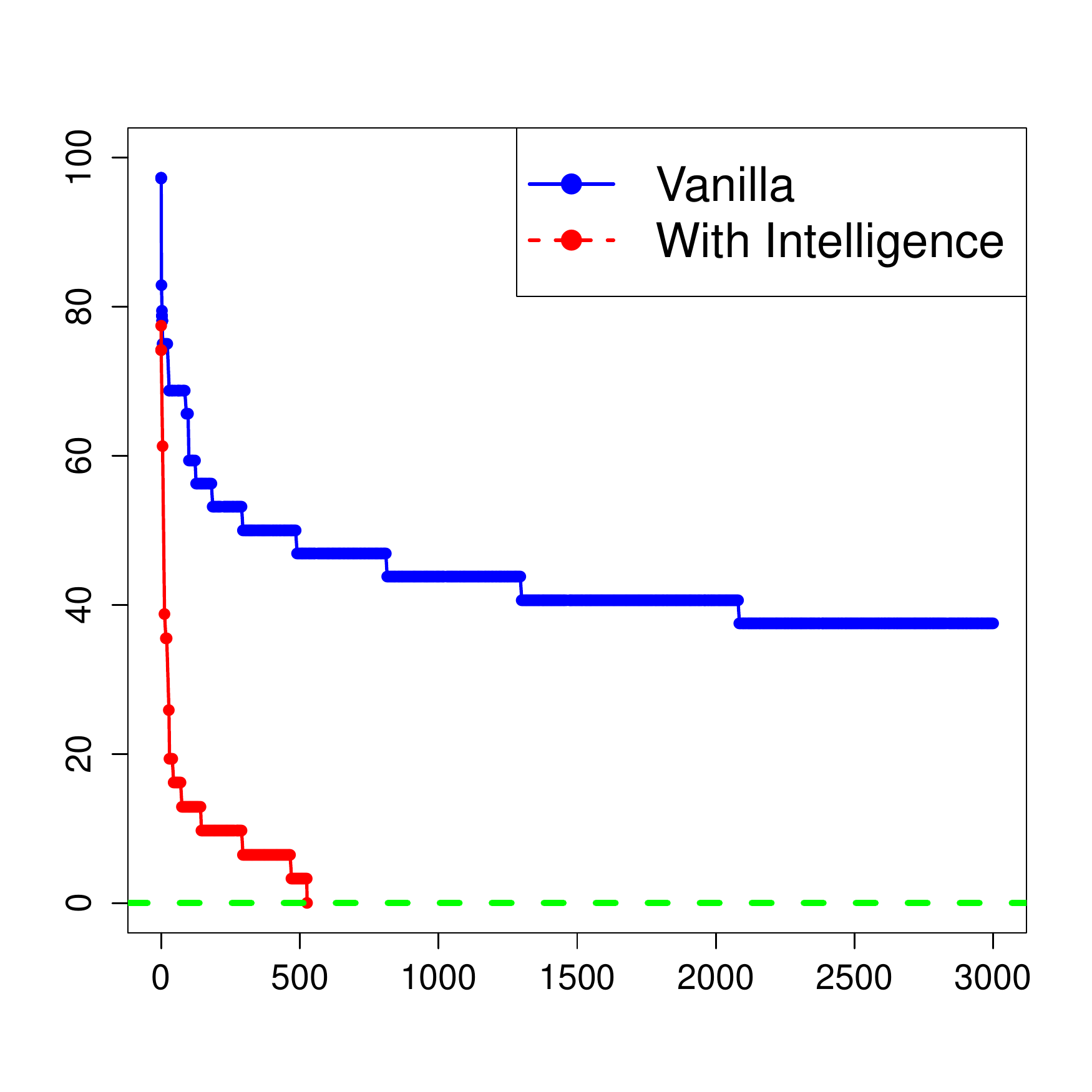}&
\includegraphics[width=0.33\textwidth,height=0.3\textheight,  trim = 1.6cm 1.5cm .2cm 1.5cm, clip = true ]{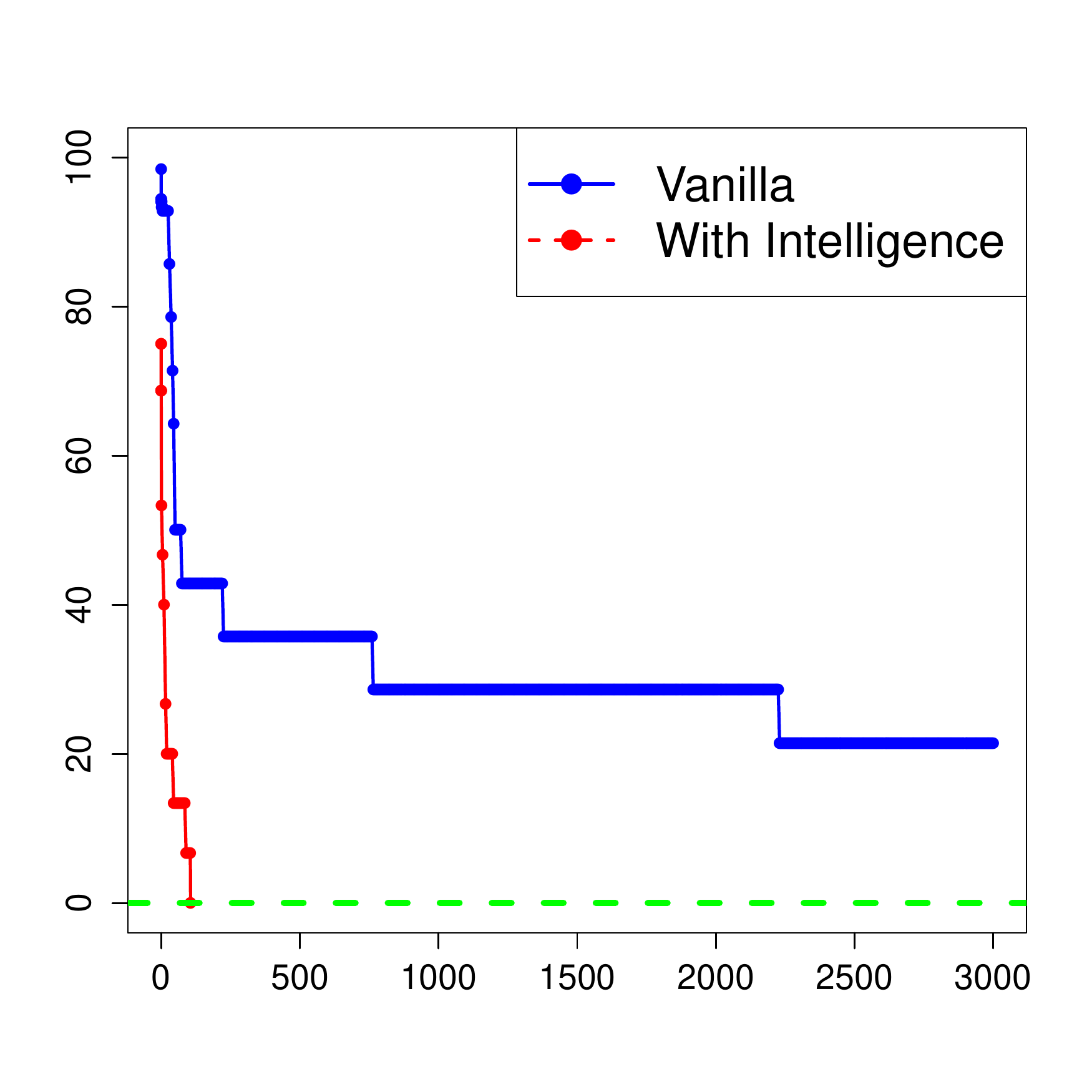}&
\includegraphics[width=0.33\textwidth,height=0.3\textheight,  trim = 1.6cm 1.5cm .2cm 1.5cm, clip = true ]{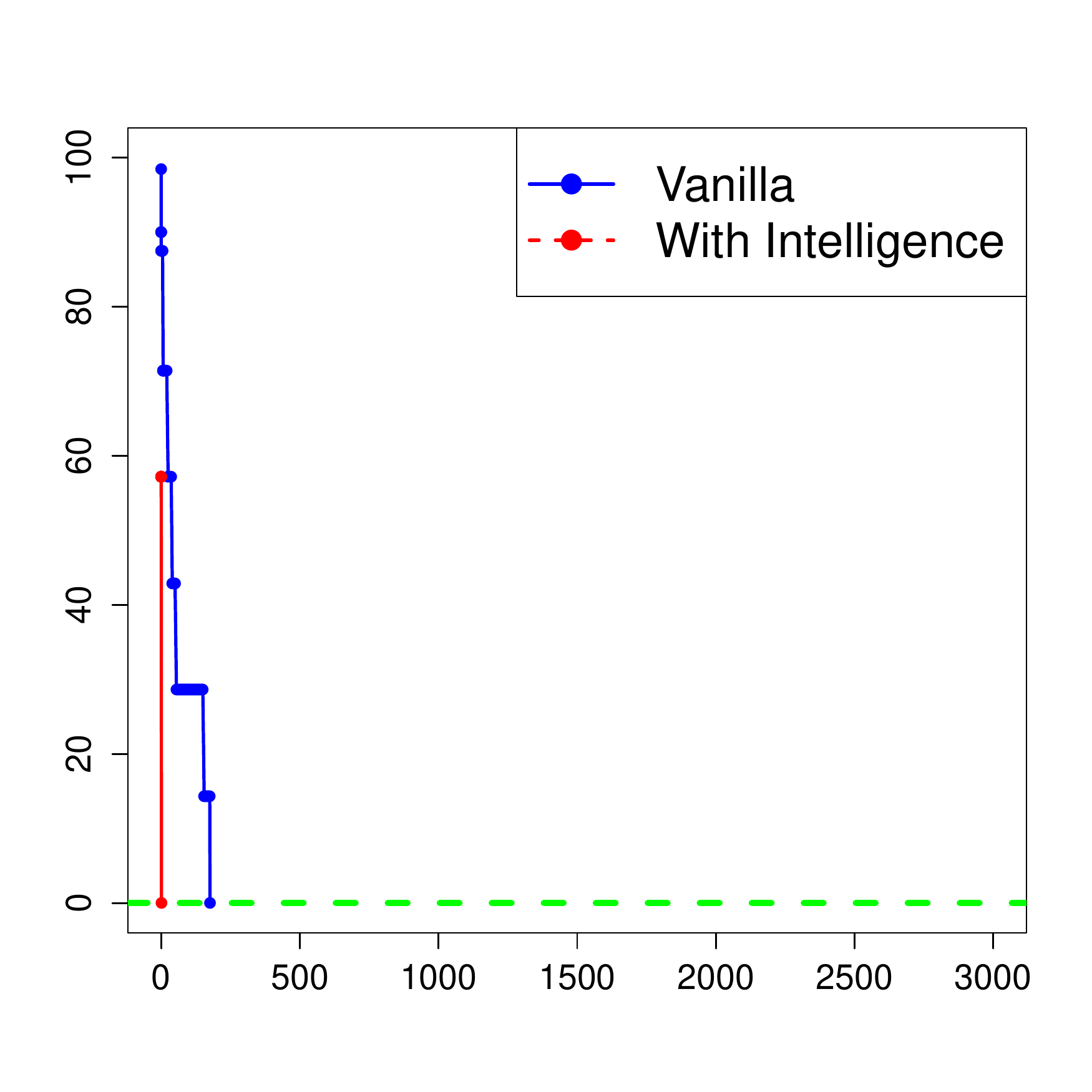}  \\
 & \sf {\scriptsize{ Time (secs)}} & \sf {\scriptsize{ Time (secs) }} & \sf {\scriptsize{ Time (secs) }} \\
 \end{tabular}}
\caption{ {  The evolution of \MIO Optimality gaps (defined in Figure~\ref{mio-motivate-1}) as functions of time (in secs)
for the \MIO
methods with and without problem-specific information.
We consider three different values of the data fidelity parameter~$\delta$, leading to solutions with different number of nonzeros as mentioned in the figure panels.
Here, ``With Intelligence'' refers to a \MIO algorithm aided with an advanced warm-start, in addition to the bounds specified in Section~\ref{sec:adv-formulations-1},
and ``Vanilla'' refers to a \MIO algorithm without any such additional information.  \MIO is found to benefit significantly from  additional problem-specific information.}}\label{diabetes-cold-warm}
\end{figure}

\begin{figure*}[h!]
\centering
\resizebox{\textwidth}{0.16\textheight}{\begin{tabular}{l c c c c}
 & \sf { \scriptsize{ $(n,p, \| \widehat{\B\beta}\|_{0}) = (400,1000,20)$ } } &\sf {\scriptsize{ $(n,p, \| \widehat{\B\beta}\|_{0}) = (600,1500,15)$ } } & \sf { \scriptsize{ $(n,p, \| \widehat{\B\beta}\|_{0}) = (750,2000,20)$ } }&
 \sf { \scriptsize{ $(n,p, \| \widehat{\B\beta}\|_{0}) = (900,2500,20)$ } }\\
\rotatebox{90}{\sf {\scriptsize{~~~~~~~~~~~~~~~~~~~~~~~~\MIO~Optimality Gap (in \%)}}}&
\includegraphics[width=0.24\textwidth,height=0.27\textheight,  trim =1.0cm 1.5cm 1cm 1.5cm, clip = true ]{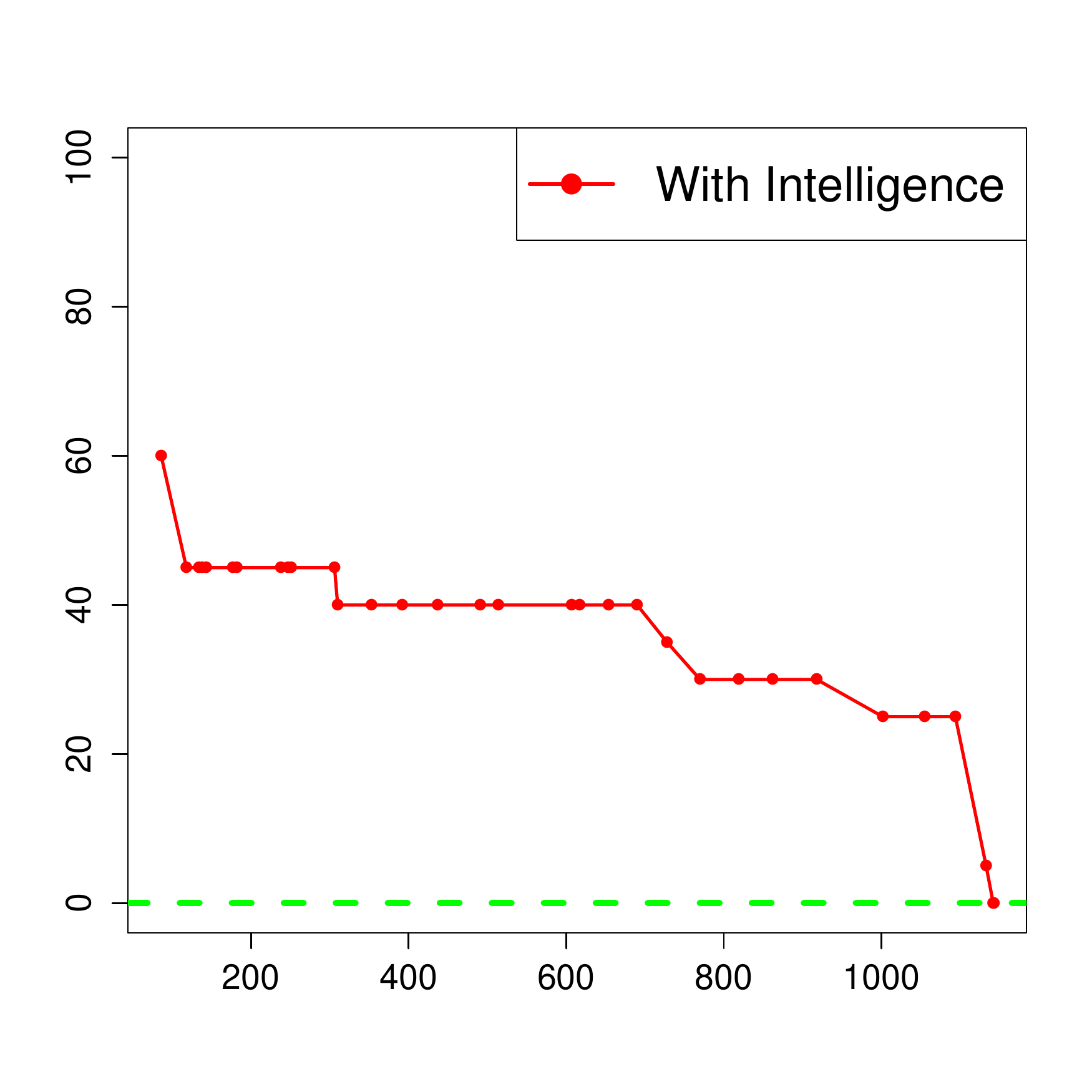}&
\includegraphics[width=0.24\textwidth,height=0.27\textheight,  trim = 1.6cm 1.5cm 1cm 1.5cm, clip = true ]{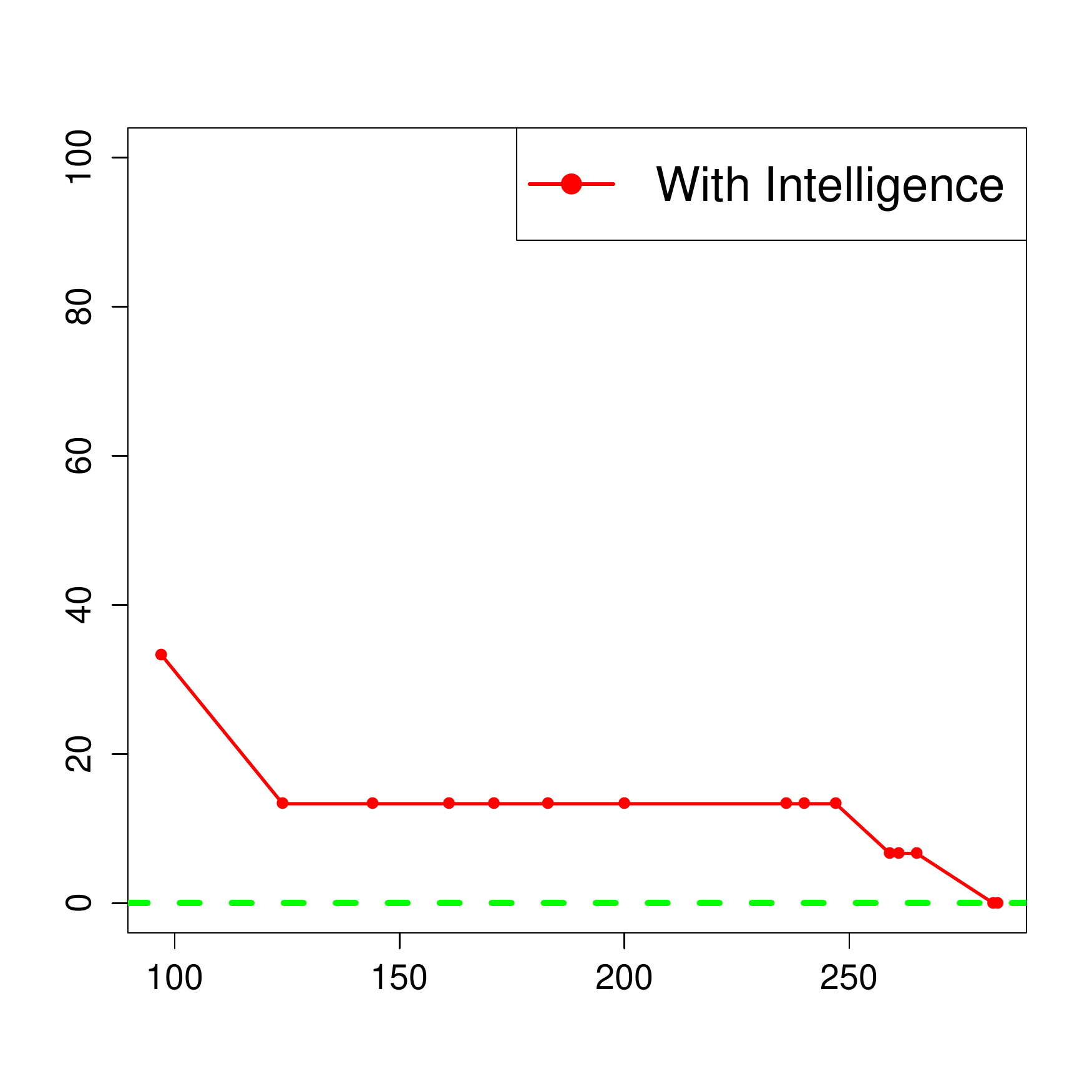}&
\includegraphics[width=0.24\textwidth,height=0.27\textheight,  trim = 1.6cm 1.5cm 1cm 1.5cm, clip = true ]{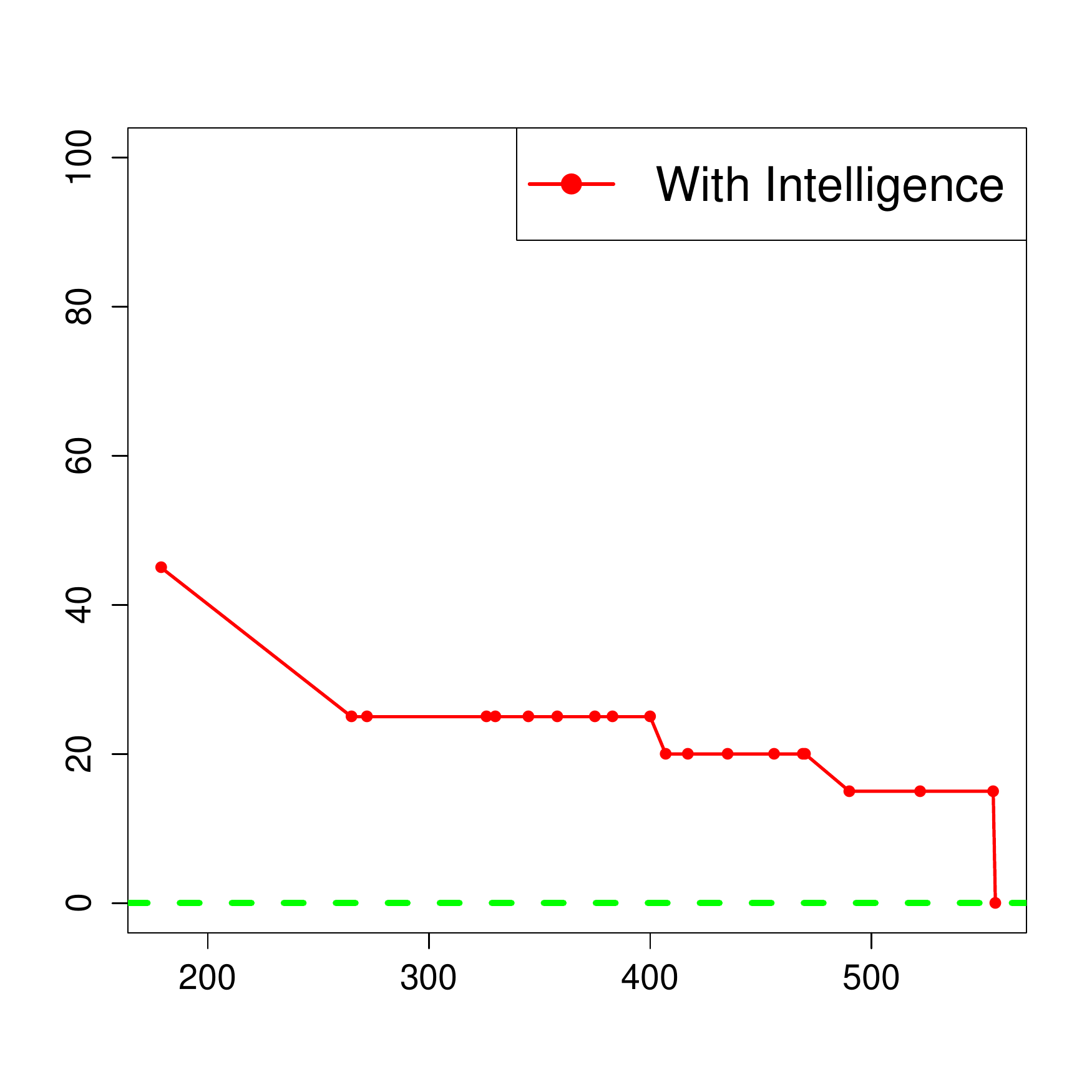} &
\includegraphics[width=0.24\textwidth,height=0.27\textheight,  trim = 1.6cm 1.5cm 1cm 1.5cm, clip = true ]{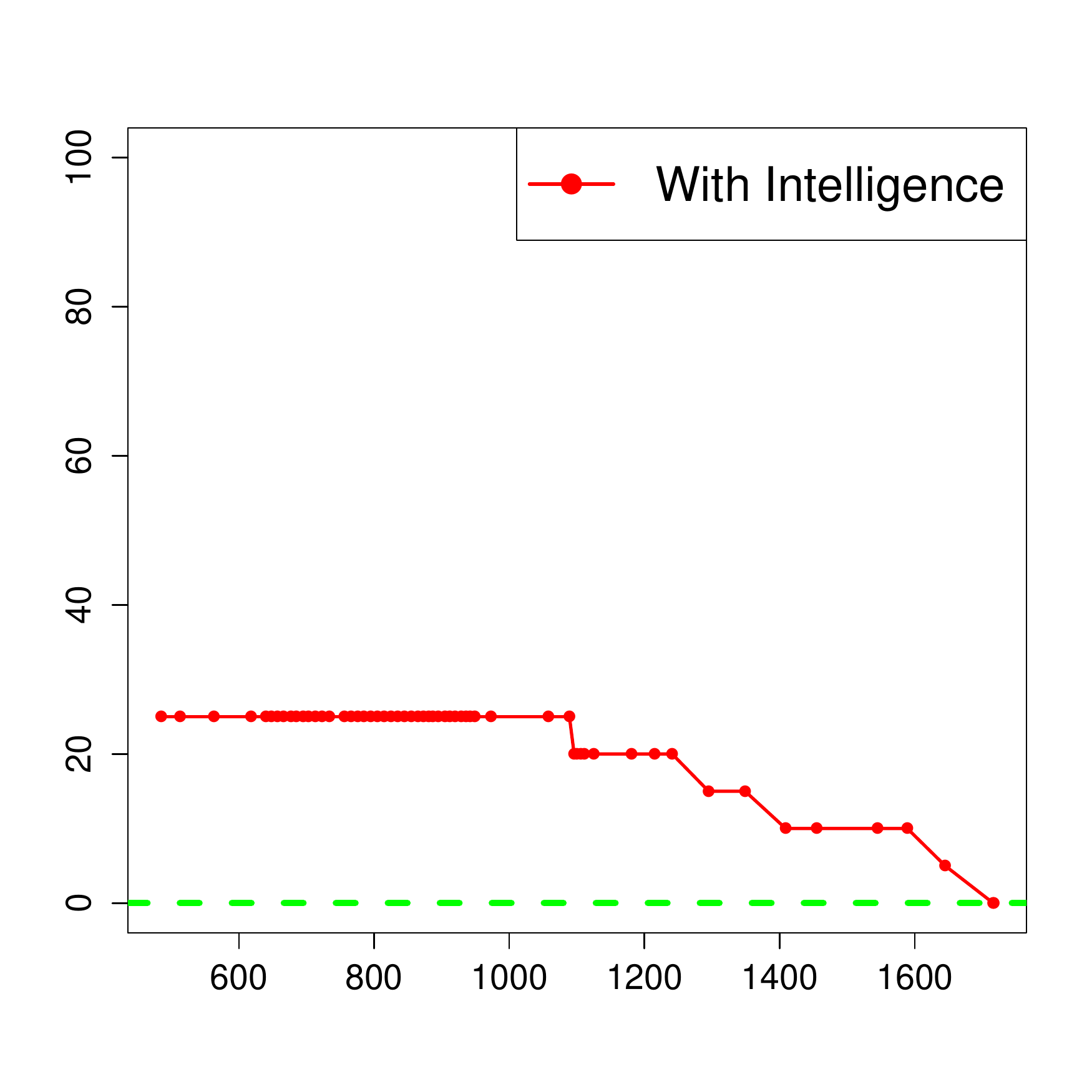} \\
 & \sf {\scriptsize{ Time (secs)}} & \sf {\scriptsize{ Time (secs) }} & \sf {\scriptsize{ Time (secs) }} &\sf {\scriptsize{ Time (secs) }} \\
\end{tabular}}
\caption{\upshape{The evolution of \MIO Optimality gaps (defined in Figure~\ref{mio-motivate-1}) as a function of time (after a warm-start was supplied to it) for different synthetic examples with varying problem-sizes.
``With Intelligence'' refers to \MIO aided with an advanced warm-start information, in addition to the bounds specified in Section~\ref{sec:adv-formulations-1}, as described in the text.
The ``Vanilla'' version of the \MIO was found to perform worse, and is, thus, not shown in the figures. In all these instances, the $\ell_{1}$-\DSF~resulted in denser solutions.}}
\label{synthetic-warm-1}
\end{figure*}

\subsection{Lower Bounds and Certificates of Optimality}\label{sec:cert-glob-opt-1}
Here we demonstrate how our framework delivers certifiably optimal solutions to Problem~\eqref{L0-DZ-1}.
In our first series of experiments we considered the popular diabetes dataset~\cite{LARS}, which we examined  with interaction terms included, giving us $p=64$ and $n=442$.  All the features
and the response were mean-centered and standardized to have unit $\ell_{2}$-norm. Figure~\ref{diabetes-cold-warm}  shows the performance of two versions of \MIO -- ``With Intelligence'' and ``Vanilla''.  ``With Intelligence'' refers to \MIO formulation~\eqref{L0-DZ-sos-bounds1}, where a \MIO solver is provided with an advanced warm-start, say, $\widehat{\B\beta}^0$.
The parameter specifications are obtained based on the method in Section~\ref{subsub-opt-cvx-2}; here,
we used the box constraints and $\ell_{1}$-constraint on $\B\beta$.
The ``Vanilla'' version of \MIO was not provided with any such problem-specific information --- we used formulation~\eqref{L0-DZ-1-mio-M}, as in Section~\ref{sec-upper-bounds-1}. Our experimental results (Figure~\ref{diabetes-cold-warm}) show that ``Intelligence'' significantly enhances the performance of the \MIO solver, in terms of proving global optimality.
Usually, we observe that for a fixed $n,p$ with $n >p$ the time to certify optimality is smaller when $k$ is small or close to $p$ -- intuitively, this is due to the ``search-space'' being small.
The computation time increases as $k$ becomes closer to $p/2$. This is reflected in Figures~\ref{mio-motivate-1} and~\ref{diabetes-cold-warm}.

\subsubsection{Moderate Scale Examples}\label{sec:moderate-scale}
We considered some examples of {\texttt{Type-Synth}} for $\rho=0$ and different values of $n,p,k^*$; in all the examples, we set $\delta = \| \M{X}^\top(\M{Y} - \M{X} \B\beta^*)\|_{\infty}.$
The results for \MIO with intelligence are displayed in Figure~\ref{synthetic-warm-1}.
 We obtained an advanced warm-start ($\widehat{\B\beta}^0$) from a
combination of Algorithm~3 and \MIO formulation~\eqref{L0-DZ-sos}, where the latter was allowed to run for an overall time limit of 500 seconds.  The warm start $\widehat{\B\beta}^0$ was used to
initialize formulation~\eqref{L0-DZ-sos-bounds1} --- the parameter specifications in the formulation were obtained based on the method in Section~\ref{subsub-opt-cvx-2}.
We also experimented with the version of formulation~\eqref{L0-DZ-sos-bounds1} that considers only
box constraints on $\B\beta$; the results were often found to be roughly similar --- both methods certified optimality, though
there were some differences in the total run-time (roughly around a few minutes).
For all the synthetic examples presented in Figure~\ref{synthetic-warm-1}, the vanilla version of \MIO took much longer to prove optimality and hence they are not shown in Figure~\ref{synthetic-warm-1}.

 In all these instances, the $\ell_{1}$-\DSF,~not surprisingly, resulted in a solution that was more dense
than the corresponding \LDSF.

\begin{table}[ht!]
\begin{center}
\resizebox{.75\textwidth}{0.15\textheight}{\begin{tabular}{ |ccc ccc c|} \hline
\multicolumn{7}{|c|}{(Synthetic Examples)} \\
$n$ & $p$ & $k^*$ & Upper Bound & Lower Bound & \MIO Gap & Time (hrs) \\
2,000  & 5,000 &  30     &  30&      30     &0&  8.3\\
4,000 & 5,000  &60       & 60      & 60  & 0 & 20.0  \\
7,000  &7,000  &20  &     20    &  20 & 0&   21.7 \\
6,000 & 7,000 & 60    &     60&   60    &0&  20.0 \\
4,000 & 8,000&  20&        20&     20&  0&   41.9  \\
3,000&  8,000 & 20&         20   &20& 0&   18.3 \\
3,000  & 9,000 & 20&    20  & 20&    0& 47.2 \\
4,000 & 9,000 & 20 &      20 &     20&  0&  44.4 \\
1,000&  10,000&    10& 10 &      10    &  0   & 14.2 \\
5,000&  10,000&    10& 10 &      10    &  0   & 2.5 \\
5,000&  10,000 &    20& 20 &      20    &  0   & 47.5 \\
10,000 &10,000  &  30      & 30     & 27  &$10\%$ & 42.5  \\
&&& &&& \\
\multicolumn{7}{|c|}{(Real Data Examples)} \\
$n$ & $p$ & $k^*$ & Upper Bound & Lower Bound & \MIO Gap & Time (hrs)\\
6,000 & 4,500&  20&     20&  20& 0 & 5.0 \\
6,000 & 4,500&  40 &  40&  37& $10\%$ &  12.5 \\  \hline
\end{tabular}}\end{center}\caption{{\upshape{Table showing times taken to reach global optimal solutions for several problem instances (both synthetic and real data), with $p$ up to $10,\!000$.
For each instance, we list the upper bound, the lower bound, the corresponding \MIO (Optimality) Gap and the time taken to reach the listed lower bound by
the \MIO solver equipped with ``Intelligence'' (as described in Figure~\ref{synthetic-warm-1}). The times (in hours) refer to those taken by the \MIO solver after being provided with a warm-start.
 In all the instances, the best upper bound was obtained around approximately one hour, however, it took much longer to
obtain a certificate of global optimality via the (almost) matching lower bounds.
The \MIO Gap was found to be zero in all the instances apart from the two where the algorithm was terminated upon obtaining a lower bound within 10\% of the upper bound.
The results demonstrate that certifiably optimal \LDSF~solutions can be obtained for large scale instances.}}}\label{tab:large-scale}
\end{table}

\subsubsection{Large Scale Examples}\label{sec-large-scale}
We consider several large scale examples with $p$ ranging in $4,\!500$ to $10,\!000$:  these problem-sizes are orders of magnitude greater
than those considered
in~\cite{bertsimas2015best}. These computations were performed with 100 GB memory.

We studied a host of synthetic examples, all generated as in Section~\ref{sec:moderate-scale}.
We also considered a semi-synthetic dataset derived from the well-known Gisette data~\url{http://archive.ics.uci.edu/ml/datasets/Gisette}.
Here,  we generated a response $\M{y}$, based on the Gisette data covariates (each feature was standardized to have zero mean and unit $\ell_{2}$ norm) --- here, $n=6000$ and $p=4500$, we set
$\beta^*_{i}=1,i=1, \ldots, k^*$ and the remaining $\beta_{i}^* =0$; SNR=3 and considered two instances with $k^* = 20, 30$.

The algorithmic set-up was similar to that used in Section~\ref{sec:moderate-scale}. The results are presented in Table~\ref{tab:large-scale}.
 In all these instances, the $\ell_{1}$-\DSF~resulted in a solution that was more dense
than those obtained via the \LDSF. For all the synthetic examples, the \MIO solver delivered solutions that matched the optimal solution of the data generating mechanism.
Typically, the time taken to prove optimality marginally increases with larger values of $k^*$ (for a fixed $n,p$); for a fixed $k^*,p$ the times taken to certify optimality increases with decreasing
values of $n$. The examples demonstrated in this paper, show the largest instances of discrete optimization problems for exact variable selection,
that can be solved to provable optimality.

\section{Numerical Experiments: Statistical Properties}\label{sec:stats-prop-expt}


We conducted a series of synthetic experiments to understand the statistical properties of the \LDSF~and compare them to those of the
$\ell_{1}$-\DSF~and variants.

We used the following datasets in our analysis.

\noindent {\bf { \Exampleone }:}
This is of {\texttt{Type-Synth}} with $n=200$, $p=500$, $\rho=0$ and $k^*=20$.

\noindent  {\bf { \Examplethree}:}
This dataset was similar to the one taken in \Exampleone, but the amplitudes and signs of the true regression coefficients were allowed to vary:
the twenty nonzero $\beta^*_{j}$'s were equally spaced in the interval $[-10,10]$.

\noindent  {\bf { \Examplefour}:}
This is of {\texttt{Type-Synth}} with $n=100,p=500,\rho=0.85, k^*=10$.

\noindent  {\bf { \Exampleeight}:}
We set $n=100,p=300$ and let  $\M{X} \sim \text{MVN}(\M{0}, \B\Sigma)$, where $\sigma_{12} = \sigma_{21} =  0.7$, and all the remaining $\sigma_{jk}$ are equal to zero.
We also took $\beta^*_{1} = 1$ and $\beta^*_{2} = -1$, with the remaining coefficients $\beta^*_{j}$ set to zero, resulting in $k^* = 2$.
(This example is a larger version of Example~$1'$ described in Section~\ref{sec:intro} and illustrated in Figure~\ref{L1-L1-DS-path}.)




In each of the above cases, after $\M{X}$ was generated, we standardized its columns to have unit $\ell_{2}$-norm.  Then, the response was generated as
$\M{y} = \M{X} \B\beta^* + \B{\epsilon}$, where $\epsilon_{i} \stackrel{\text{iid}}{\sim} N(0, \sigma^2)$, and $\sigma^2$ was adjusted to match the selected value of SNR, which was
varied across $\{3, 10 \}$ in the examples.

%
%
%
%
%
%

We considered the following estimators in our analysis:
\begin{itemize}


\item ``Warm'' --- this method applies a heuristic strategy to obtain upper bounds to Problem~\eqref{L0-DZ-1}. We used\footnote{This is similar to a re-weighted $\ell_{1}$-minimization~\cite{boyd08-new} method
applied to Problem~\eqref{L0-DZ-1}. We took the penalty $\rho_{\gamma}(|\beta|) \propto \log({|\beta|}/{\gamma} +1)$  on a geometrically decreasing grid of ten $\gamma$ values: $\gamma_{i}= 10^{-2}\times 0.8^{i-1}$ for $i = 1, \ldots, 10$.}
Algorithm~2, described in Section~\ref{sec:weighted-l1}.


\item  ``L0-DS'' --- the solution obtained from ``Warm'' is taken as a warm-start to a \MIO solver and subsequently allowed to run with a time limit of 4000 seconds.

\item  ``L0-DS-Pol'' --- this is a ``polished'' version of the \LDSF~estimator ``L0-DS'' and is obtained by performing a simple least squares fit
on the support of the ``L0-DS'' estimate.

\item  ``L1-DS'' --- this is the original $\ell_{1}$-\DSF.

\item ``L1-DS-Pol'' --- this is a polished version of the ``L1-DS''.
\end{itemize}

Each of the above estimators were computed on a range of approximately thirty different $\delta$ values around
$\bar{\delta} = \| \M{X}^\top(\M{y} - \M{X}^\top \B\beta^*)\|_{\infty}$.
We considered ten different replications (based on different $\B\epsilon$ realizations) and took the median of the results.
The optimal tuning parameter ($\delta^\text{opt}$) for every model was selected based on the value of
$\delta$ that minimized the estimation error with respect to the true regression coefficients. For this chosen value of $\delta^\text{opt}$ we considered
different metrics to assess the performance of the different estimators.
We computed the squared $\ell_{2}$-error in estimating the regression coefficients: $\| \widehat{\B\beta} - \B\beta^*\|_{2}^2$.
We also considered the ``Variable Selection error'', which is defined as $\sum_{j=1}^p 1(\widehat{S}_{j} \neq S^*_j)$, where $\widehat{S}_{j}$ is the $j$th coordinate of $\widehat{\M{S}}:= \text{Supp}(\widehat{\B\beta})$, and $S^*_j$ is the $j$th element of $\M{S}^* =  \text{Supp}({\B\beta}^*)$.
Finally, we computed the ``number of nonzeros'', which refers to the number of nonzero coefficients in $\widehat{\B\beta}$.

\begin{figure}[]
\centering
\resizebox{0.95\textwidth}{!}{\begin{tabular}{l c c c }
 & \sf {\scriptsize{ {\bf \Exampleone}}} &  \sf {\scriptsize {\bf \Examplefour }} & \sf {\scriptsize { \bf \Exampleeight }}  \vspace{0mm} \\
 & \sf {\scriptsize $(n=200,p=500)$} &  \sf {\scriptsize {  $(n=100,p=500)$ }} & \sf {\scriptsize { $(n=100,p=300)$}}   \\
\rotatebox{90}{\sf {\scriptsize{~~~~~~~~~~~~~~~~~~~~~~$\|\widehat{\B\beta} - \B\beta^*\|_{2}^2$ }}}&
\includegraphics[width=0.23\textwidth,height=0.2\textheight,  trim =0.2cm 1.5cm .2cm 1.5cm, clip = true ]{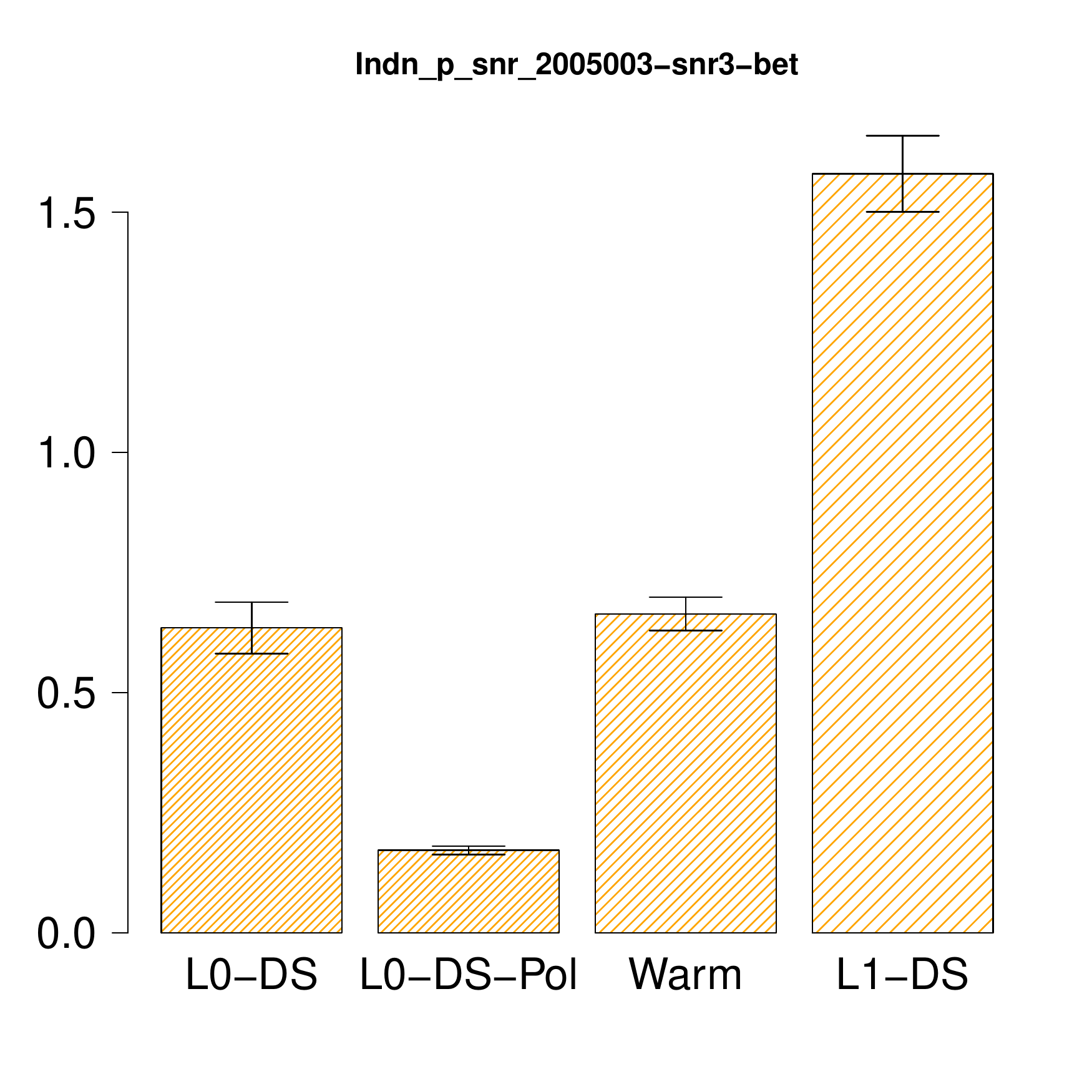}&
\includegraphics[width=0.23\textwidth,height=0.2\textheight,  trim = 0.2cm 1.5cm .2cm 1.5cm, clip = true ]{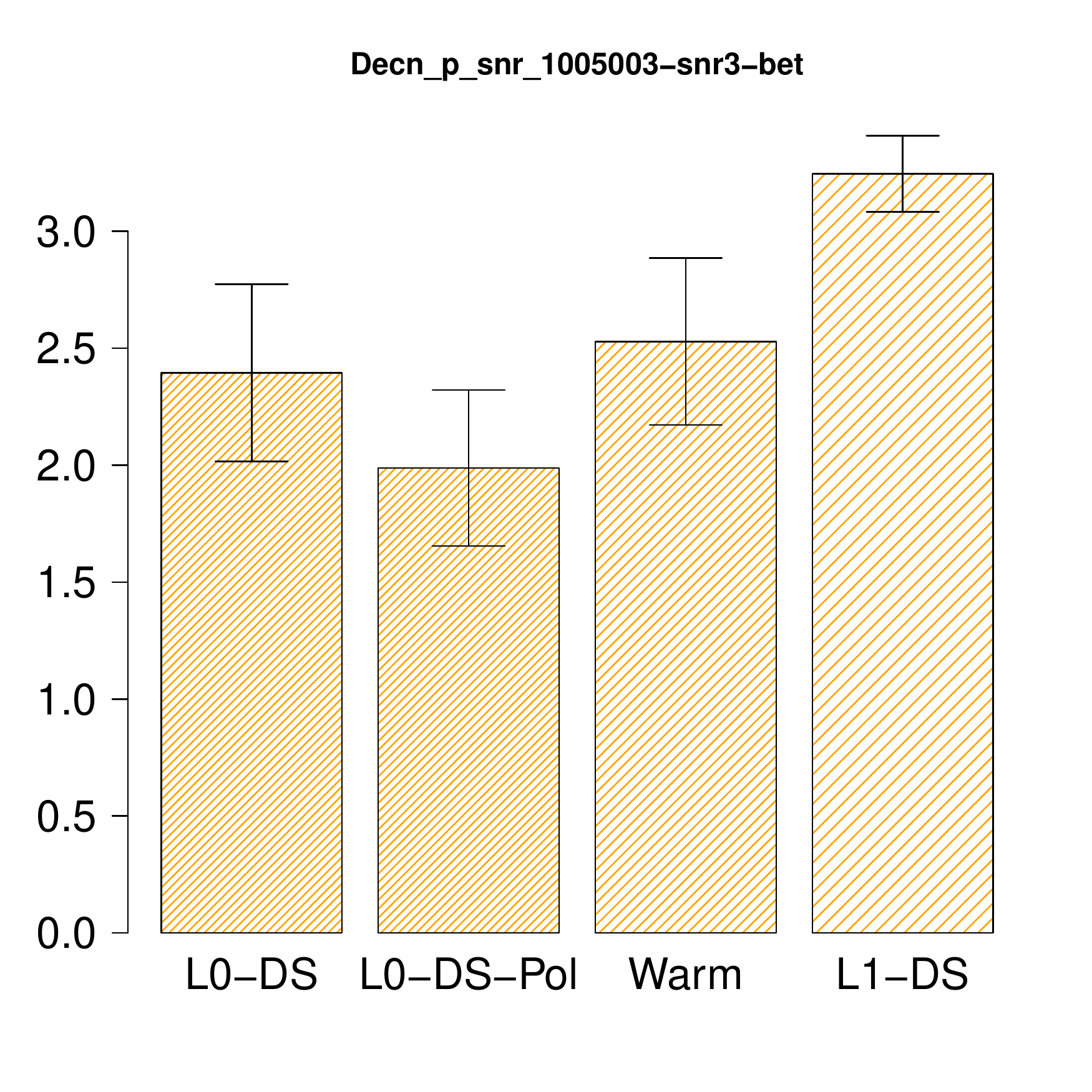}&
\includegraphics[width=0.23\textwidth,height=0.2\textheight,  trim = 0.2cm 1.5cm .2cm 1.5cm, clip = true ]{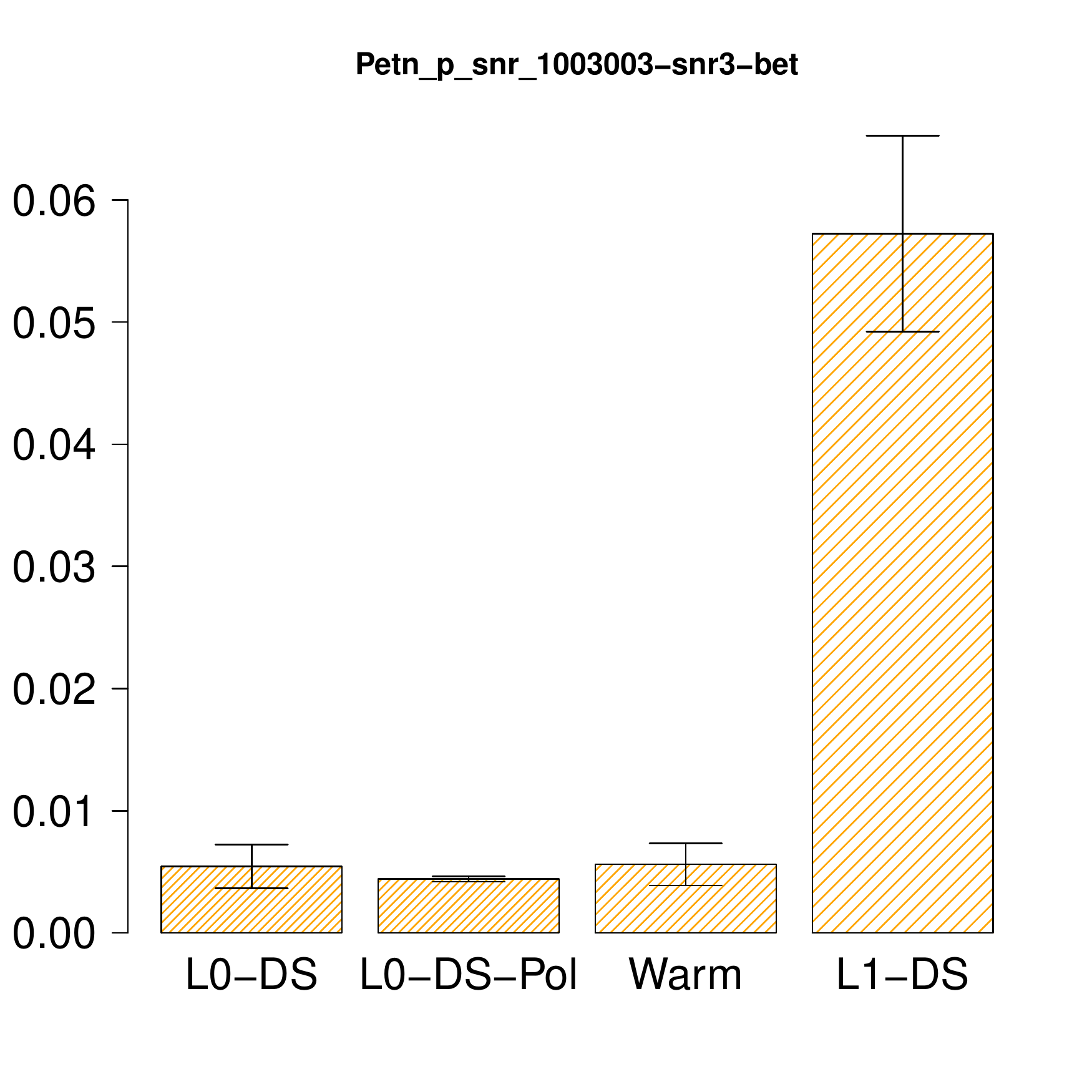}  \vspace{2mm} \\
\rotatebox{90}{\sf {\scriptsize {~~~~~~Error in Variable Selection}}}&
\includegraphics[width=0.23\textwidth,height=0.2\textheight,  trim = 0.2cm 1.5cm .2cm 1.5cm, clip = true ]{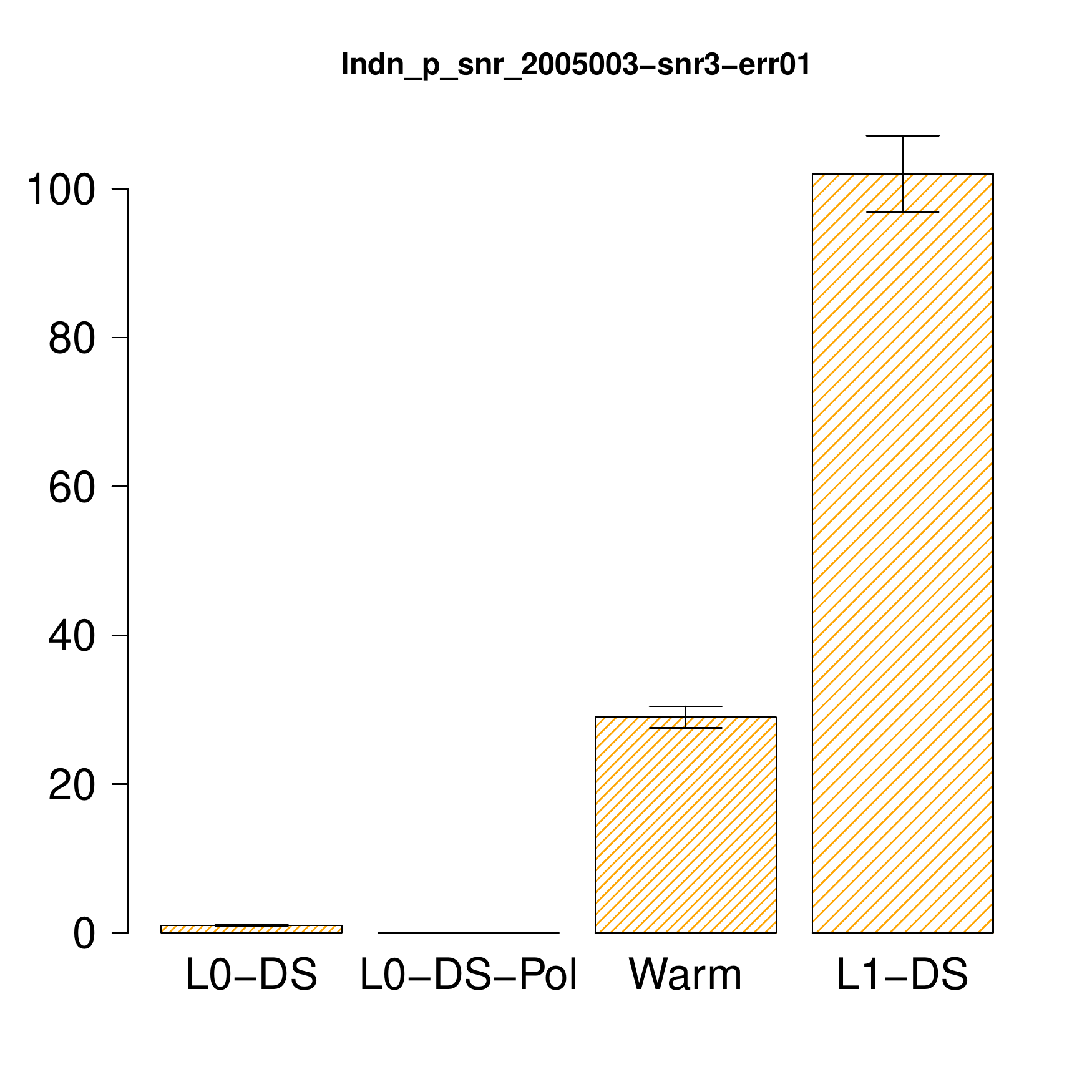}&
\includegraphics[width=0.23\textwidth,height=0.2\textheight,  trim = 0.2cm 1.5cm .2cm 1.5cm, clip = true ]{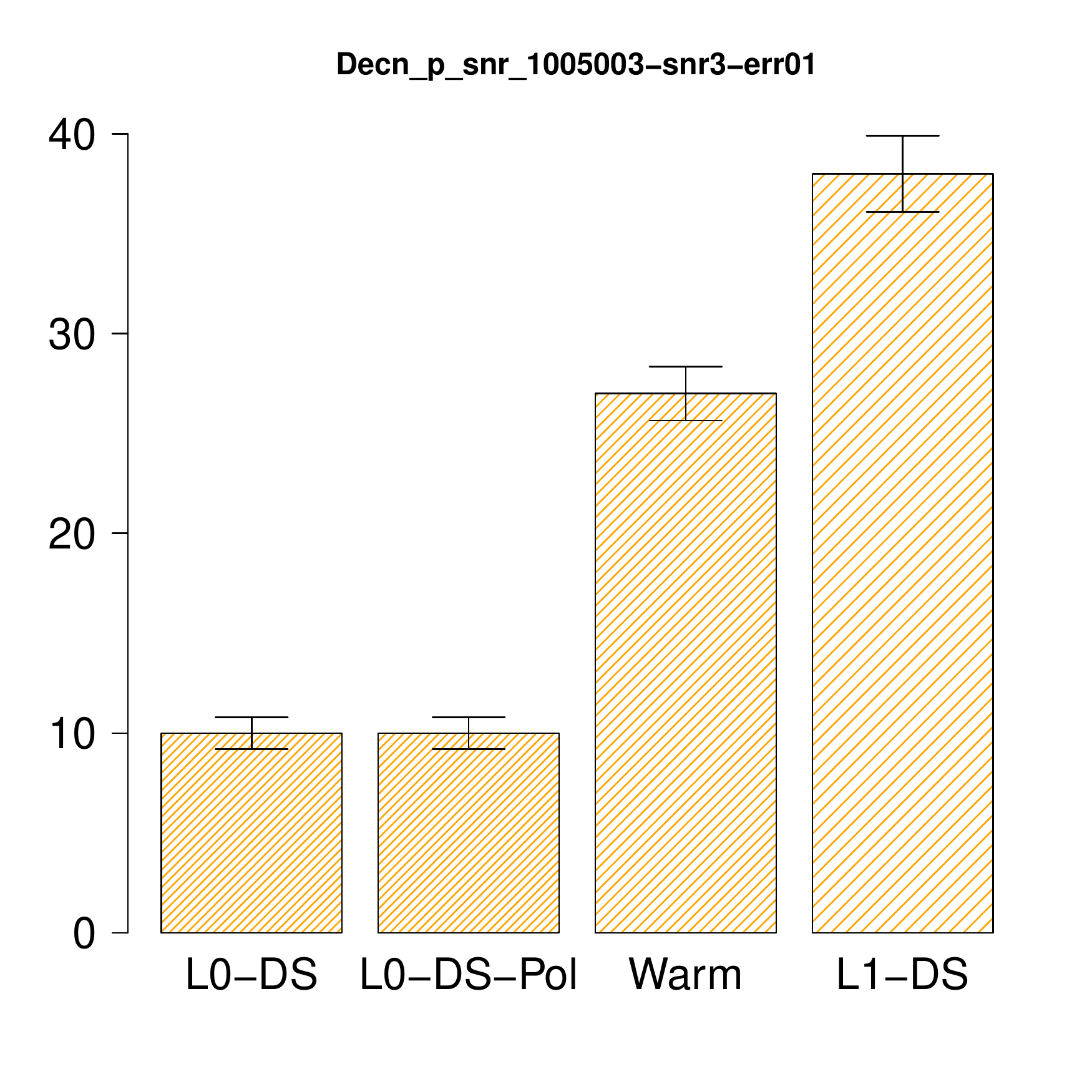}&
\includegraphics[width=0.23\textwidth,height=0.2\textheight,  trim = 0.2cm 1.5cm .2cm 1.5cm, clip = true ]{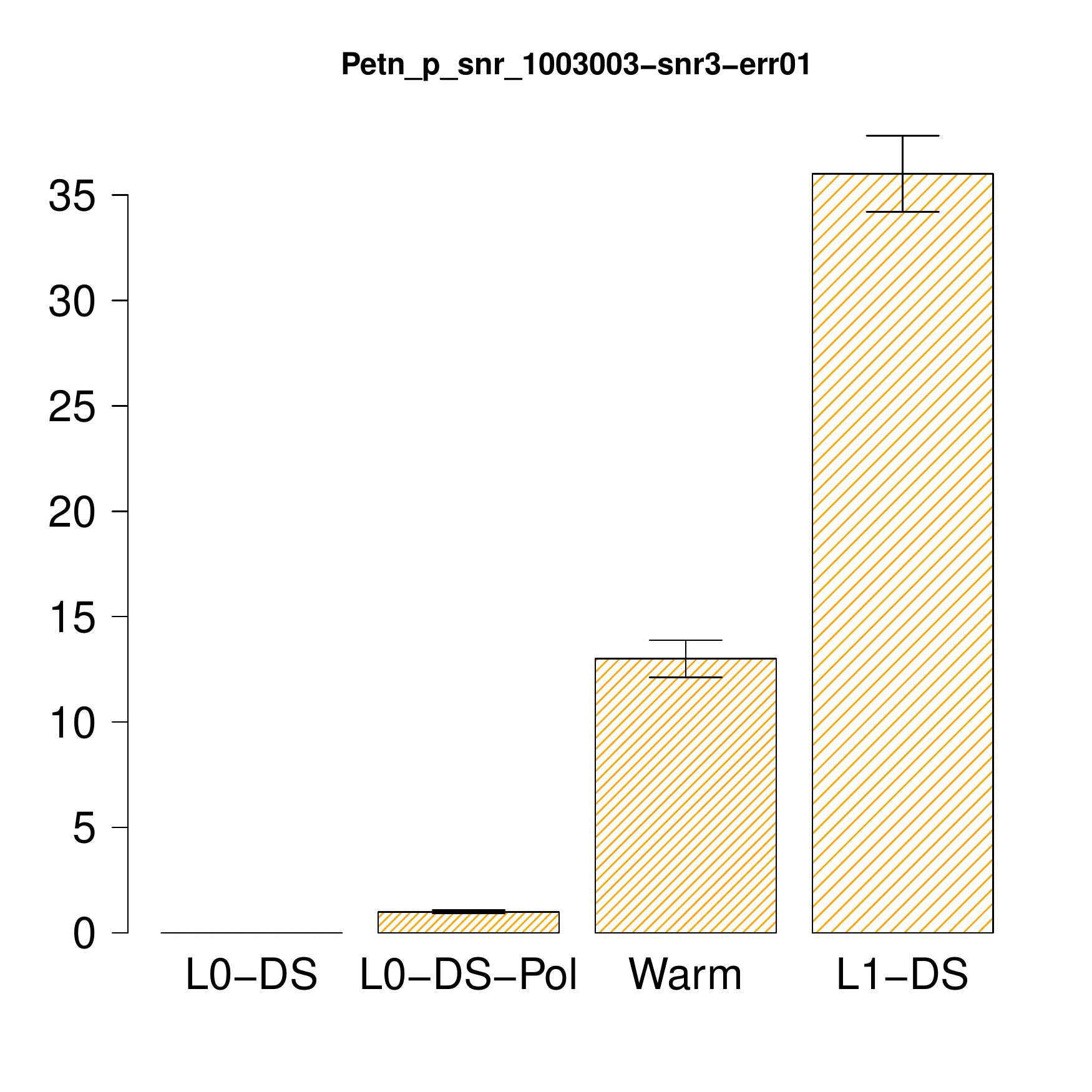}  \vspace{2mm} \\
\rotatebox{90}{\sf {\scriptsize {~~~~~~~~~~Number of nonzeros }}}&
\includegraphics[width=0.23\textwidth,height=0.2\textheight,  trim = 0.2cm 1.5cm .2cm 1.5cm, clip = true ]{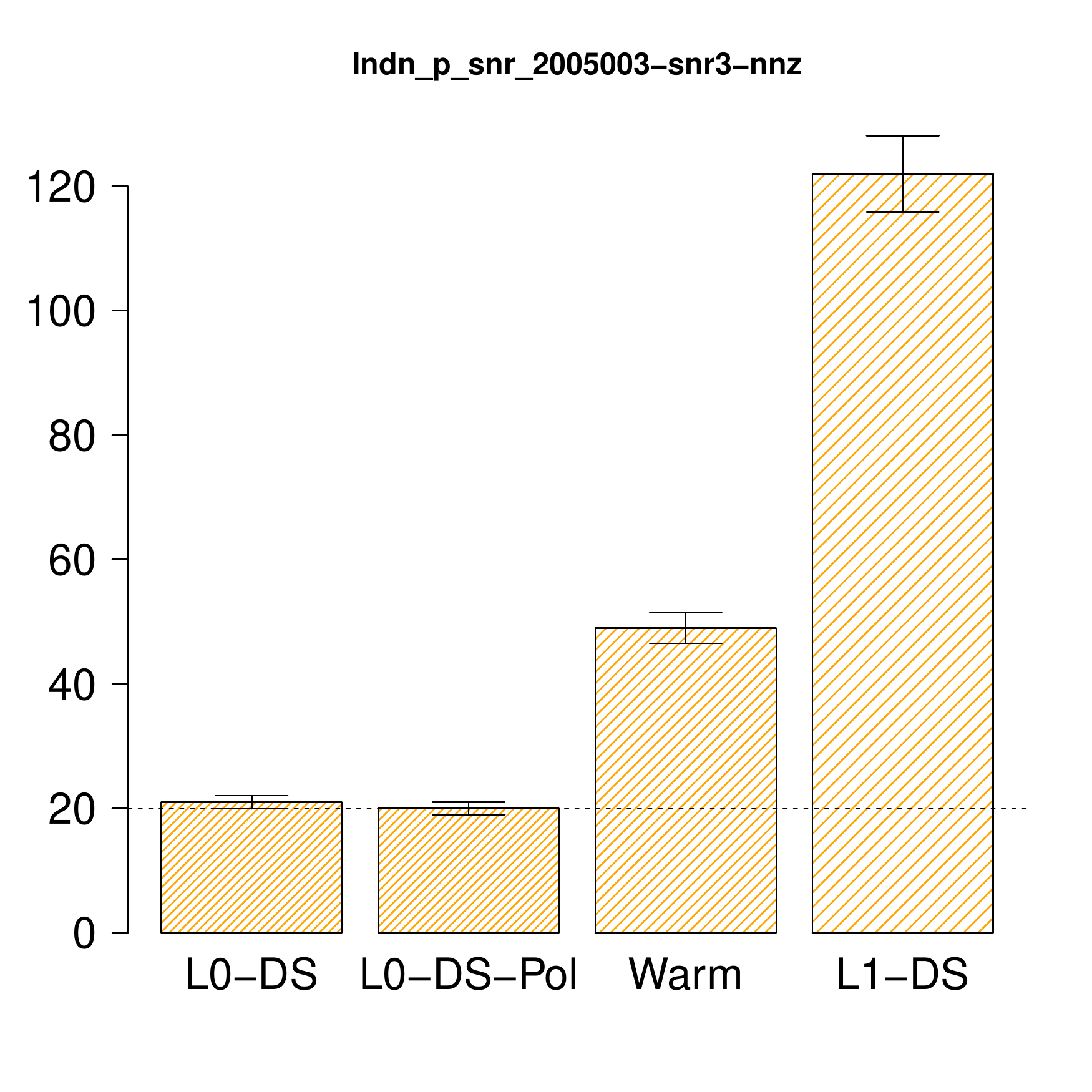}&
\includegraphics[width=0.23\textwidth,height=0.2\textheight,  trim = 0.2cm 1.5cm .2cm 1.5cm, clip = true ]{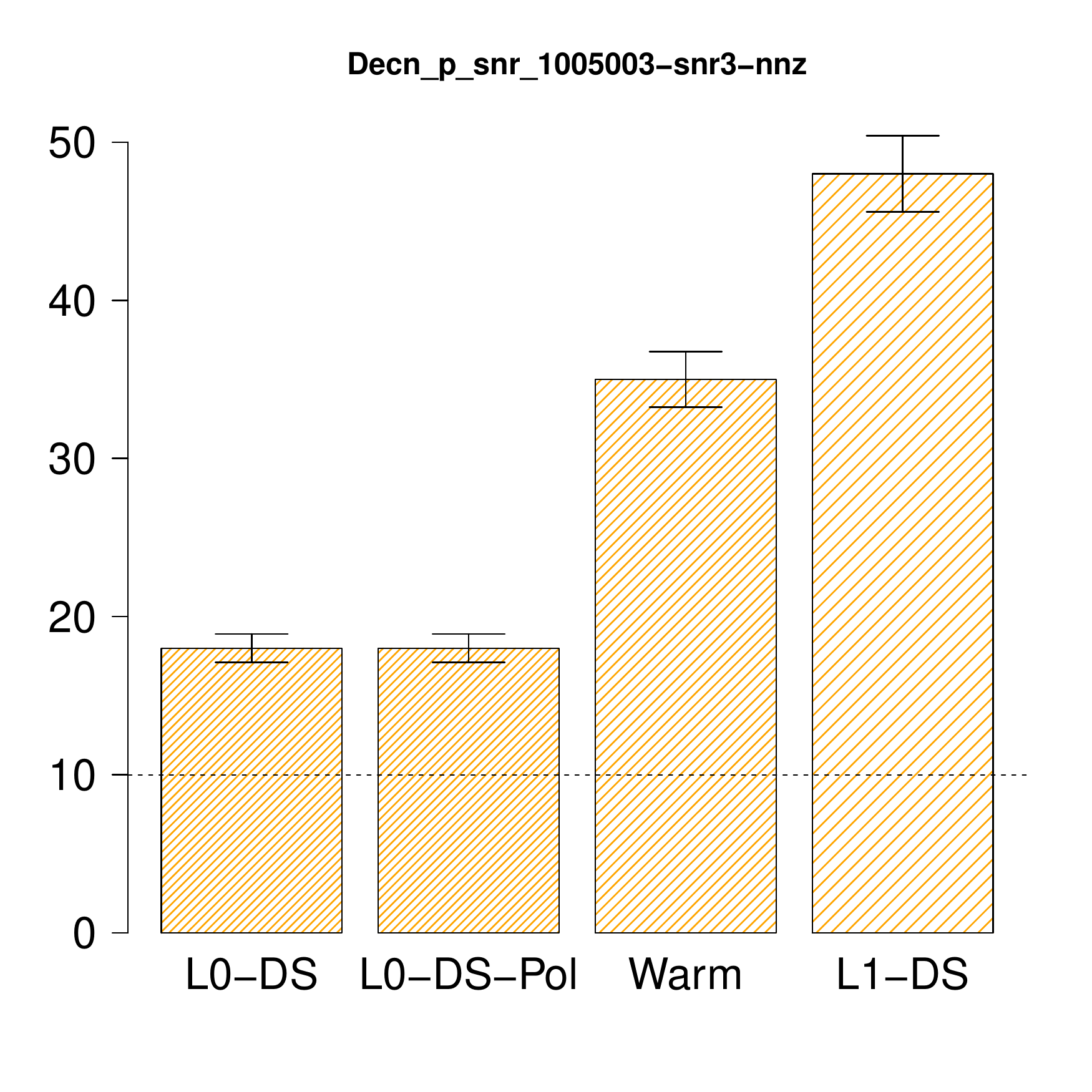}&
\includegraphics[width=0.23\textwidth,height=0.2\textheight,  trim = 0.2cm 1.5cm .2cm 1.5cm, clip = true ]{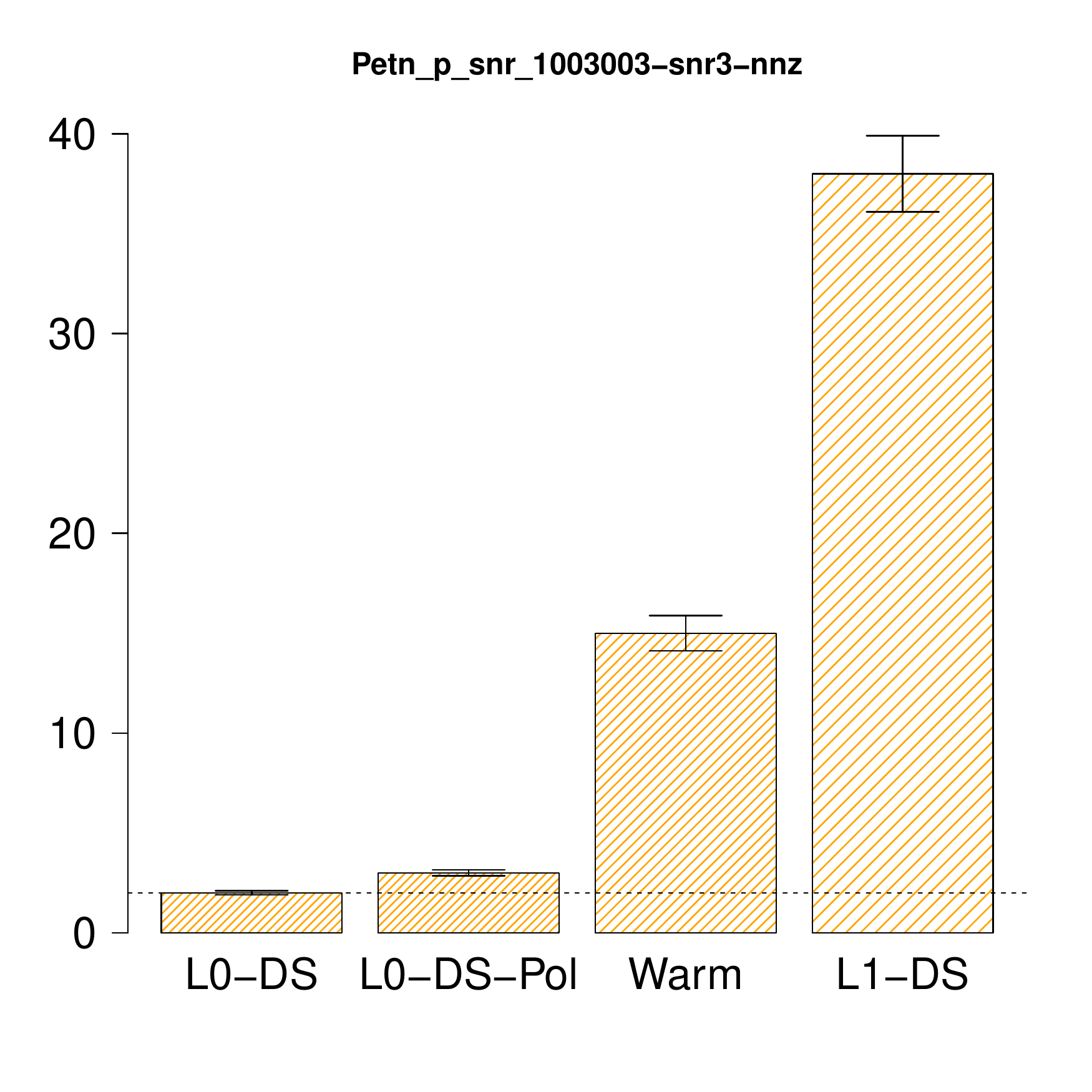}\\
\end{tabular}}
\caption{ {{The statistical performance of the \LDSF~(L0-DS); its polished version, which does a least squares refitting on the support (L0-DS-Pol);
the heuristic estimates delivered by Algorithm~2 (Warm); and the original $\ell_{1}$-Dantzig Selector (L1-DS).  We display three different metrics:
[top panel] squared $\ell_{2}$-error in estimating $\B\beta$,
[middle panel] the 0-1 variable selection error and [bottom panel] the number of nonzeros in the optimally selected model; the horizontal dotted line shows the number of nonzeros in the ``true'' model.  We observe that the \LDSF~based approaches perform very well in terms of obtaining a model with high quality estimation and variable selection properties. Their models are substantially sparser than those for the $\ell_{1}$-based methods.
The heuristic approach, ``Warm'' which approximately optimizes Problem~\eqref{L0-DZ-1}, falls short in obtaining high quality statistical estimates.
A number of additional experiments (with results similar to this figure), are presented in Section~\ref{app-sec:addl-expts} of the Appendix.}}}\label{fig:l1-l0-DS-stat}
\end{figure}

A collection of representative results with SNR=10 is displayed in Figure~\ref{fig:l1-l0-DS-stat}.
The error bars correspond to standard errors, the width being set to $2 \hat{s}/\sqrt{N}$ where,
$\hat{s}$ is the mean absolute deviation around the median and $ {N}$ denotes the number of replicates (here, ten).
A larger display of additional examples with varying SNR values is presented in Table~\ref{tab:stat-results-l0-l1-all} in
   Appendix~\ref{app-sec:addl-expts}, where we also report the ``Prediction Error'', defined as $\| \M{X} \widehat{\B\beta} - \M{X}\B\beta^* \|_{2}^2/\|  \M{X}\B\beta^*\|_{2}^2$.  In Table~\ref{tab:polished-l1-l0}, given in the same section, we provide comparisons with the polished version of the $\ell_{1}$-\DSF.
  Our experiments show that polishing the $\ell_{1}$-\DSF~may lead to marginally better solutions relative to the original \DSF, but the corresponding statistical performance is inferior to that of the estimates based on the \LDSF~estimator.

We note that the performance of the Lasso was found to be quite similar to that of the $\ell_{1}$-\DSF.  The statistical performance
of the subset selection procedure~\eqref{subset-l0-old}, as described in~\cite{bertsimas2015best}, was found to be similar to that of the \LDSF~for $p \leq 1000$. Because the main focus of the paper is to show that \LDSF~is a computationally tractable procedure, which delivers estimates with better statistical properties than its $\ell_{1}$ counterpart, we restrict our numerical studies to the methods listed above.

\pparagraph{Summary of Findings.}
Based on the experimental results, we observe that the \LDSF~and its polished variant perform quite well, when compared to the competing methods in terms of
estimating $\B\beta^*$; they also demonstrate superior variable selection properties (not surprisingly, the $\ell_{0}$ methods obtain the sparsest models across all the examples).
``Warm'' does not perform very well when compared to ``L0-DS", even though both methods attempt to solve Problem~\eqref{L0-DZ-1} --- this
suggests that estimators based on rigorous optimization procedures have better statistical properties.  We observe that ``L0-DS'' and ``L0-DS-Pol''  possess similar variable selection properties, however, the latter may lead to better estimators of
$\B\beta^*$ and $\M{X}\B\beta^*$, due to the least squares post-processing. Polishing of the $\ell_{1}$-\DSF~may not lead to better solutions,
due to the weak variable selection properties of ``L1-DS''.  In some cases, when
the value of $\rho$ is quite large and, consequently, the covariates are highly correlated (see \Examplefour, \Exampleeight),
the basic problem of variable selection becomes difficult: instead of choosing a ``signal'' variable, the ``L0-DS'' chooses its correlated surrogate.
In these cases, as expected, we observe that the ``L0-DS'' incurs relatively large variable selection error--the prediction accuracy of these models however, demonstrate a more optimistic picture than the variable selection properties (See Table~\ref{tab:stat-results-l0-l1-all}).

\begin{table}[]
\begin{center}
\scalebox{0.95}{\begin{tabular}{ c| c cc c } \hline
   Example  & Metric                             & Time (secs) & L0-DS-Pol & MIPGO \\ \hline
   $\rho=0$            & Variable Selection &    60 &   0  &   20 \\
       $n=100$       & Error                    &   200 & 0 &          14\\  \cline{2-5}
        $p=200$            & \multirow{2}{*}{\bet} &    60 &   0.309  &  20.00 \\
         $k^*=20$        &                                         &   200 & 0.309   & 8.79  \\ \hline\hline
   $\rho=0.7$         & Variable Selection &    60 &   0  &   10 \\
     $n=100$            &  Error                                        &   200 & 0 & 4\\   \cline{2-5}
       $p=200$         & \multirow{2}{*}{\bet} &    60 &   0.063  &  0.271  \\
     $k^* = 10$             &                                         &   200 &  0.063 &  0.076   \\  \hline\hline
     $\rho=0.7$         & Variable Selection &    100 &   0  &   9 \\
     $n=300$            &  Error                      &   500 & 0 & 5\\   \cline{2-5}
       $p=600$         & \multirow{2}{*}{\bet} &     100 &   0.094  &  9.142  \\
     $k^* = 20$             &                                         &   500 &  0.094 &  1.548   \\  \hline\hline
           $\rho=0.7$          & Variable Selection &    150 &   0  &   25\\
     $n=300$              &  Error                                        &   1000 & 0 & 16 \\   \cline{2-5}
     $p=1000$         & \multirow{2}{*}{\bet} &    150 &   0.175   & 25.00  \\
     $k^*=25$         &             &  1000 & 0.175   & 14.893   \\  \hline
\end{tabular}}\end{center}
\caption{{\small{In all the above instances, the generated data is of {\texttt{Type-Synth}}, with SNR=10.  Both the \LDSF~and MIPGO (with the MCP penalty) methods were run on twenty different values of the tuning parameter, and the best solutions are reported. L0-DS-Pol refers to the least squares solution obtained on the variables selected by \LDSF. For every value of the tuning parameter, each method was run for a time budget of $t_{1}, t_{2}$  seconds with $t_{1} <t_{2}$ specified in the ``Time'' column.
The \LDSF~method reaches the best solution within $t_{1}$ seconds, much earlier than its competitor. The MIPGO method is seen to take orders
 of magnitude longer to get a solution of the same quality as the \LDSF --- the differences become more pronounced with increasing problem size.}}}\label{tab:mio-mcp-1}
\end{table}

\subsection{Comparisons with Least Squares Subset Selection}\label{sec:com-mipgo}
We discuss some comparisons of our proposal with the recently proposed methods: Problem~\eqref{subset-l0-old} by~\cite{bertsimas2015best}  and MIPGO~\cite{mipgo-li-2016}.

For underdetermined problems (with $n<p$) the authors in~\cite{bertsimas2015best} (see Section~5.3.2 in~\cite{bertsimas2015best}) point out that
{\texttt{MIQO}} solvers take a long time to \emph{certify} optimality, by producing matching upper and lower bounds.
For problems with $n \leq 100$ and $p=1000$, \cite{bertsimas2015best} demonstrated how the {\texttt{MIQO}} methods for
Problem~\eqref{subset-l0-old} could certify \emph{local} optimality\footnote{We note that certifying local optimality i.e., optimality in a neighborhood of a candidate solution is also an NP-hard problem.}
in a (small) bounding box around a candidate solution.
We observed in our computational experiments that Problem~\eqref{L0-DZ-1}
 is orders of magnitude faster than~\eqref{subset-l0-old} for underdetermined problems, in obtaining solutions with \emph{certificates} of global optimality.
On several randomly generated problem instances
generated as per {\texttt{Type-Synth}} with $n=100$,  $p = 200$,  $\rho=0$, $k^*=10$ and SNR=10,
Problem~\eqref{L0-DZ-1} was solved to global optimality, i.e., zero optimality gap with a median time of about 4 minutes. On the same instances,
the {\texttt{MIQO}} formulation for Problem~\eqref{subset-l0-old} took more than 7 hours of computation time to obtain similar optimality certificates.
 In addition, a \MIO formulation for Problem~\eqref{L0-DZ-1} consumes much less memory than a comparable {\texttt{MIQO}} formulation for Problem~\eqref{subset-l0-old}. For example, on problem instances with $n=p \in \{ 1.5, 2, 2.5, 3\} \times 10^3$,
 we observed that Problem~\eqref{subset-l0-old} requires at least twice as much memory as
that for Problem~\eqref{L0-DZ-1}, within the first 800 seconds of computation time. The memory requirement for a {\texttt{MIQO}} for Problem~\eqref{subset-l0-old} with $n=p=3000$ was more than 12GB.


MIPGO~\cite{mipgo-li-2016} is a discrete optimization framework for minimizing a regularized version of the least squares loss, with a
nonconvex quadratic penalty (for example, SCAD or MCP).  This corresponds to a nonconvex \emph{quadratic} optimization problem, which
the authors express as a discrete linear optimization problem via linear complementary constraints~\cite{giannessi1973nonconvex}.  This representation
results in many more binary variables (several multiples of~$p$) than that required for the \LDSF.
For example, in the case of the MCP penalty, the paper~\cite{mipgo-li-2016} presents a \MIO with~$4p$ binary variables and many more continuous variables. In particular, with $n=300$ and $p=1000$,
the MIPGO solver\footnote{We used the code of~\cite{mipgo-li-2016}, obtained from the first author's website.
The numbers are read off from the \textsc{Gurobi} log report.} creates a problem with $38,\!000$ variables and $27,\!000$ equality constraints,
which after presolve reduces to a problem with approximately $7,\!000$ continuous and $4,\!000$ binary variables.  These optimization problems are substantially larger
than~\eqref{L0-DZ-1-mio-M}, which is a \MIO with $p$ continuous and as many binary variables.  As a result, the MIPGO formulation seems to become
computationally expensive as the dimensionality of the problem increases.  This is illustrated in Table~\ref{tab:mio-mcp-1} -- we show on several synthetic instances
that, with a particular time budget, the \LDSF~formulation~\eqref{L0-DZ-1-mio-M} equipped with a warm-start from  Algorithm~2, leads to better solutions than MIPGO. The quality of solutions produced by MIPGO is found to improve with more computation time, but the time taken can be substantially larger than that of the \LDSF.
The synthetic datasets for Table~\ref{tab:mio-mcp-1} were generated the same way as in Section~\ref{sec:numerics-algo}. We set the concavity parameter for the
MCP penalty in the MIPGO code to its default value of $a=2$.


\section*{Acknowledgements}
The authors thank the anonymous referees for their helpful comments that led to improvements in the manuscript.
The authors also thank Emmanuel Candes and Robert Freund for helpful suggestions and encouragement.  R.M. thanks Jonathan Goetz, Juan-Pablo Vielma and Dimitris Bertsimas for helpful discussions.
R.M.'s research was partially supported by ONR N000141512342 and a grant from the Moore Sloan Foundation.
Peter Radchenko's research was partially supported by NSF Grant DMS-1209057.

\begin{appendix}

\section{Proofs for Section~3}\label{sec:proofs-stat-theory}

\noindent {\bf Proof of Theorem~\ref{orthonorm.thm}}.
Let~$\tilde{\epsilon}$ denote the projection of $\M{y}$ onto the orthogonal complement to the space spanned by the predictor vectors $\M{x}_j, j=1,...p$.  Note that
\begin{equation*}
\|\M{y}-\M{X}\B\beta\|^2=\sum_{j=1}^p (c_j-\beta_j)^2+\|\tilde{\epsilon}\|^2.
\end{equation*}
Thus, under the constraint $\| \B\beta \|_{0} \leq k$, the smallest sum of squares is achieved by setting $\beta_j=c_j$ for $j=1,...,k$ and $\beta_j=0$ for $j=k+1,...,p$.  This completes the proof of part 1.

Note that
\begin{equation*}
|\M{x}^\top_j(\M{y}-\M{X} \B\beta)|=|c_j-\beta_j|.
\end{equation*}
Thus, the constraint $\|\M{X}^\top  (\M{y}-\M{X}\B\beta)\|_\infty \leq \delta$ is satisfied if and only if $\beta_j\in[c_j-\delta,c_j+\delta]$ for $j=1,...,p$.  In order to minimize $\|\B\beta\|_0$, coefficients $\beta_j$ for which $0\in[c_j-\delta,c_j+\delta]$ are set to zero.  Thus, $\beta_j=0$ if and only if $|c_j|\le\delta$, which implies $\beta_j=0$ for $j>k$.  This completes the proof of part 2.\\

\noindent {\bf Proof of Theorem~\ref{fixed.p.thm}}.
Throughout this proof we omit the words ``with probability tending to one'' to improve the presentation.
The constraint $\|\M{X}^\top  (\M{y}-\M{X}\B\beta)\|_\infty \leq \delta$ implies $\|\bX^\top\bX(\bbeta^*-\bbeta)+\bX^\top\bepsilon\|_{\infty}\le\delta$.  Recall that $\delta=o(n^{1/2})$ and note that $\|\bX^\top\bepsilon\|_{\infty}=O_p(1)$, due to the scaling of the predictors and the assumptions on the $\bepsilon$.  Consequently,  $\|\bX^\top\bX(\bbeta^*-\bbeta)\|_{2}=o_p(n^{1/2})$.  Because $\bX^\top\bX$ converges to an invertible matrix~$C$, we conclude that there exists a $o_p(1)$ sequence of random variables $b_n$, such that the bound
\begin{equation*}
\|n^{-1/2}\widehat\bbeta-n^{-1/2}\bbeta^*\|_2\le b_n
\end{equation*}
simultaneously holds for all the Discrete Dantzig solutions $\widehat\bbeta$.  Recall that $\bbeta^*=n^{1/2}\tilde{\B\beta}^*$, for some fixed vector $\tilde{\B\beta}^*$.  Consequently, $\beta^*_j\ne0$ implies $\widehat\beta_j\ne0$ for $j=1,...,p$.  In other words the support of each Discrete Dantzig solution contains $J^*$. It is only left to show that the cardinality of each such support cannot be greater than~$|J^*|$. Note that with probability tending to one, $\bbeta^*$ is feasible for the Discrete Dantzig optimization problem.  Indeed,
\begin{equation*}
\|\M{X}^\top  (\M{y}-\M{X}\bbeta^*)\|_\infty=\|\bX^\top\bepsilon\|_{\infty}=O_p(1),
\end{equation*}
which is bounded above by $\delta$, due to the assumption $\delta\rightarrow\infty$.  Thus, inequality $\|\widehat\bbeta\|_0\le |J^*|$ holds for each Dantzig Selector solution, which completes the proof of part 1.

In the paragraph above we deduced that the minimum value of the Discrete Dantzig objective function equals $|J^*|$.  We also showed that~$\bbeta^*$ is feasible for the Discrete Dantzig optimization problem.  A similar argument establishes the feasibility of $\bbeta^O$.    Consequently, $\bbeta^*$ and $\bbeta^O$ are indeed Discrete Dantzig solutions, which completes the proof of part 2.\\

\noindent {\bf Proof of Proposition~\ref{prop.ineqs}}.  Consider an arbitrary nonzero $\btheta\in\RR^p$, such that $\|\btheta\|_0\le 2k$.  Let $J_0$ be the index set of the~$k$ largest, in magnitude, coordinates of~$\btheta$.   Observe that $\|\btheta_{J_0^c}\|_1\le \|\btheta_{J_0}\|_1$ and $\|\btheta\|_2=\|\btheta_{J_{01}}\|_2$, because~$m\ge k$.  Thus
\begin{equation*}
\frac{\|\M{X}\btheta\|_2}{\|\btheta\|_2} =\frac{\|\M{X}\btheta\|_2}{\|\btheta_{J_{01}}\|_2}  \ge \kappa(k,c_0,m),
\end{equation*}
for $c_0\ge1$, which implies $\gamma(2k)\ge\kappa(k,c_0,m)$.  Also note that
\begin{equation*}
2\frac{\|\M{X}\btheta\|^2_2}{\|\btheta\|^2_2} \ge \frac{\|\M{X}\btheta\|^2_2}{\|\btheta_{J_0}\|^2_2} \ge \left[\kappa(k,c_0)\right]^2,
\end{equation*}
for $c_0\ge1$, which gives $\gamma(2k)\ge\kappa(k,c_0)/\sqrt{2}$.  \\

\noindent {\bf Proof of Theorem~\ref{gen.thm} and Corollary~\ref{cor.gen.thm}}.
Note that $\M{X}^\top \bepsilon$ is a mean zero Gaussian vector, such that the variance of each component is~$\sigma^2$.  Consequently, it follows from well-known maximal inequalities for Gaussian variables that the bound $\|\M{X}^\top \bepsilon\|_{\infty}\le \delta$ holds with probability at least $1-(p^a\sqrt{\pi\log p} )^{-1}$.  The rest of the proof is conducted on the set where the above bound is valid.  Note that on this set $\bbeta^*$ is a feasible solution for the optimization problem~\eqref{L0-DZ-1}, which implies $\|\widehat\bbeta\|_0\le \|\bbeta^*\|_0$.  Recall that we denote $\|\bbeta^*\|_0$ by~$\ s^*$ and derive the following inequalities:
\begin{equation*}
\begin{myarray}[1.3]{l l}
 \gamma(2s^*)^2\left\|\widehat\bbeta-\bbeta^* \right\|_2^2
\le  \left\|\M{X}(\widehat\bbeta-\bbeta^*) \right\|_2^2 & \\
\; = \; (\widehat\bbeta-\bbeta^*)^\top  \M{X}^\top  \M{X}(\widehat\bbeta-\bbeta^*)&\\
\; \le \; \left\|\M{X}^\top  \M{X}(\widehat\bbeta-\bbeta^*) \right\|_{\infty}\left\|\widehat\bbeta-\bbeta^*\right\|_1&\\
\; \le \; \left(\left\|\M{X}^\top  (\bY-\M{X}\widehat\bbeta) \right\|_{\infty} +n^{-1}\left\|\M{X}^\top  \bepsilon\right\|_{\infty} \right)\left\|\widehat\bbeta-\bbeta^*\right\|_1.&
\end{myarray}
\end{equation*}
Because both $\|\M{X}^\top  (\bY-\M{X}\widehat\bbeta) \|_{\infty}$ and $\|\M{X}^\top  \bepsilon\|_{\infty}$ are bounded above by~$\delta$, we derive
\begin{equation*}
\left\|\widehat\bbeta-\bbeta^* \right\|_2^2\le \gamma(2s^*)^{-2}2\delta\left\|\widehat\bbeta-\bbeta^* \right\|_1.
\end{equation*}
Applying inequality $\|\widehat\bbeta-\bbeta^* \|_1\le (2s^*)^{1/2} \|\widehat\bbeta-\bbeta^* \|_2$ to either the left or the right hand side of the above display yields the~$\ell_{1}$ and the~$\ell_2$ estimation bounds, respectively, in the statement of Theorem~\ref{gen.thm}.

Finally, to establish the prediction error bound, observe the following inequality:
\begin{equation*}
\left\|\M{X}(\widehat\bbeta-\bbeta^*) \right\|_2^2\le  2\delta\left\|\widehat\bbeta-\bbeta^* \right\|_1,
\end{equation*}
which is is a direct consequence of the two displays given above.  We then complete the proof of Theorem~\ref{gen.thm} by combining the above display with the inequalities
\begin{equation*}
\begin{myarray}[1.2]{r c  l}
\left\|\widehat\bbeta-\bbeta^* \right\|_1 &\le& (2s^*)^{1/2}\left\|\widehat\bbeta-\bbeta^* \right\|_2 \\
&\le& (2s^*)^{1/2}\gamma(2s^*)^{-1}\left\|\M{X}(\widehat\bbeta-\bbeta^*) \right\|_2.
\end{myarray}
\end{equation*}

Corollary~\ref{cor.gen.thm} follows directly from the $\ell_1$ estimation bound in Theorem~\ref{gen.thm}.

\noindent {\bf Proof of Theorem~\ref{lowb.thm}}.   We again focus on the set of high probability, where inequality $\|\M{X}^\top  \bepsilon\|_{\infty}\le \delta$ holds.  Because $\bbeta^*$ is a feasible solution to the optimization problem~\eqref{L0-DZ-1}, we have $\|\bbeta^*\|_0\ge \widehat s_{LB}$.  Thus, $\|\widehat \bbeta\|_0\le (1+\c)\|\bbeta^*\|_0$.  The rest of the proof is identical to the one for Theorem~\ref{gen.thm}, with one exception: the bound $\|\widehat \bbeta\|_0\le \|\bbeta^*\|_0$ contains an additional factor $(1+\c)$.

\section{Additional Algorithm Details and Proofs}

\subsection{Details on Algorithm~1}\label{details-algo-1}
Update~\eqref{line-1-1} in the ADMM algorithm can be performed via a hard thresholding operation~\cite{dono:john:1994}, as it is of the form:
\begin{equation}\label{HT-admm-1}
 \widehat{\B\beta}({\lambda'}) :=\argmin_{\B\beta}  \| \B\beta - \M{c}\|_{2}^2 + \lambda' \| \B\beta \|_{0},
 \end{equation}
for an appropriately chosen $\lambda' =\frac{2}{\lambda}$ and
$\M{c} = \B\alpha + \frac{1}{\lambda} \B\nu$; a solution is given by $\widehat{\beta}_{j}(\lambda') = c_{j}\M{1} ( |c_{j}| > \sqrt{\lambda'}), j = 1, \ldots, p.$
The update step~\eqref{line-1-2} involves the following projection:
\begin{equation}\label{project-DZ-set1}
\begin{aligned}
\min_{\B\alpha}& \;\; f(\B\alpha):= \| \B\alpha - \M{\overline{c}}\|_{2}^2  \\
\sbt &\;\; \|\M{X}^\top (\M{y} - \M{X}\B\alpha)\|_{\infty} \leq \delta_{},
\end{aligned}
\end{equation}
where $\M{\overline{c}} = \B\beta - \B\nu/\lambda$.
While the projection~\eqref{project-DZ-set1} can be computed by using standard quadratic programming methods, in our experience, we found them\footnote{Our reference is \textsc{Gurobi}'s quadratic programming solver.} to be quite time consuming for
larger problems ($p\ge1000$), especially because this projection needs to be computed for every iteration (indexed by $k$) of~\eqref{line-1-1}--\eqref{line-1-33}.
 Thus, we recommend using specialized first-order methods  --- these methods also naturally make use of warm-start information, which is particularly useful to us
due to the iterative nature of the updates~\eqref{line-1-1}--\eqref{line-1-33}.
 Unless $\M{X}$ has uncorrelated columns, it is not straightforward to solve~\eqref{project-DZ-set1} in its primal form --- we thus consider a dual of Problem~\eqref{project-DZ-set1}, for which we apply first-order methods for convex composite minimization~\cite{nesterov2013gradient}. To improve the flow of presentation, we relegate the description of a more general first-order method, which also applies to Problem~\eqref{project-DZ-set1}, to Section~\ref{sec:append-dual-algo-1} in the Appendix.
We repeat steps~\eqref{line-1-1}--\eqref{line-1-33} until an (approximate) convergence criterion is met --- see for example~\cite{lu2013sparse,boyd-admm1nourl} for convergence results for the general method.
We terminate the algorithm as soon as the successive changes in the $\B\beta$ updates become small
 and one has approximate primal feasibility (see Step~(3) in Algorithm~1).

\subsection{Additional Details on Algorithm~2: Solving Problem~\eqref{obj-lla-2}}\label{sec:solvingproblem-22}
Observe that Problem~\eqref{obj-lla-2} is of the composite form~\cite{nesterov2013gradient}:
\begin{equation}\label{gen-comp-form-1}
\min_{\B\theta} \;\;  \; f_{1}(\B\theta) + f_{2}(\B\theta)  \;\;\;
\sbt \;\;\;  \B\theta \in {\mathcal C},
\end{equation}
where the function $f_1(\B\theta)$ is smooth, with its gradient Lipschitz continuous: $\|\nabla f_{1}(\B\theta) - \nabla f_{1}(\B\theta')\| \leq L \| \B\theta - \B\theta'\|$; $f_{2}(\B\theta)$ is nonsmooth and $\mathcal C$ is a convex set.
In our specific case, the smooth component is the zero function, $f_{2}(\B\theta) = \sum_{i=1}^{p} w_{i} | \theta_{i}|$ and ${\mathcal C} = \{ \B\theta: \| \M{X}^\top (\M{y} - \M{X} \B\theta)\|_{\infty} \leq \delta\}.$
 Thus, one may appeal to first-order optimization methods~\cite{fista-09,nesterov2004introductorynew,nesterov2013gradient} for composite function minimization. This requires solving, at every iteration, a problem of the form:
\begin{equation}\label{obj-lla-3}
\begin{myarray}[1.3]{ccc}
\B\theta^{m +1} \in& \argmin\limits_{\B\theta}& \frac{L}{2} \| \B\theta - \overline{\B\theta}^{m} \|_{2}^2  +   \sum\limits_{i=1}^{p} w_{i} |\theta_{i}| \\
&\sbt & \| \M{X}^\top  (\M{y} - \M{X} \B\theta)  \|_{\infty} \leq \delta,
\end{myarray}
\end{equation}
for some choice of $L>0$ and $\overline{\B\theta}^{m}$, and $w_{i}=|\rho'_{\gamma}(|\beta^{k}_{i}|) |$.  If $\overline{\B\theta}^m = \B\theta^m$, then the above update sequence becomes identical to
proximal gradient descent~\cite{fista-09}. One may also use accelerated gradient descent methods, with a momentum term. 
We describe in Section~\ref{sec:append-dual-algo-1} first-order gradient methods that can be used to compute solutions to Problem~\eqref{obj-lla-3}.
The sequence $\B\theta^m$, defined via~\eqref{obj-lla-3}, leads to the solution of Problem~\eqref{obj-lla-2} as $m \rightarrow \infty$, providing a $O(\frac{1}{m})$-suboptimal solution in
$m$ many iterations if one uses standard proximal gradient descent methods; the convergence rate can be improved to $O(\frac{1}{m^2})$ if one uses the accelerated gradient descent version of the algorithm.

Instead of choosing $f_{1}(\B\theta) \equiv 0$ one may also choose $f_{1}(\B\theta)=  \frac{\tau}{2} \| \B\theta \|_{2}^2$, for a small value of $\tau >0$.
Interestingly, for small values of $\tau$ the minimizer to Problem~\eqref{gen-comp-form-1}
is also a minimizer of the problem with the choice $f_{1}(\B\theta) = 0$. This equivalence of solutions which holds true in much more generality is often known as \emph{exact} regularization of convex programs in the mathematical programming literature --- see for example~\cite{friedlander2007exact}.
Even if the two problems are not equivalent, the choice of $f_{1}(\B\theta)=  \frac{\tau}{2} \| \B\theta \|_{2}^2$ always serves as an approximate solution to Problem~\eqref{obj-lla-2}.
With this choice of $f_{1}(\B\theta)$, one needs to solve a problem of the form:
\begin{equation*}
\begin{aligned}
\min_{\B\theta} ~~&~~ \frac{\tau}{2} \| \B\theta \|_{2}^2  +   \sum_{i=1}^{p} w_{i} |\theta_{i}|  \\
  \sbt ~~&~~   \| \M{X}^\top  (\M{y} - \M{X} \B\theta)  \|_{\infty} \leq \delta.
\end{aligned}
\end{equation*}
A solution to the above problem can be computed by considering its dual and applying (accelerated) proximal gradient methods on the dual formulation, as described in Section~\ref{sec:append-dual-algo-1}. In this approach,
 a two-stage iterative algorithm of the form~\eqref{obj-lla-3} described above, is not required.


Algorithm~2 suggests that we solve Problem~\eqref{obj-lla-2} repeatedly for different values of $\gamma$ --- it turns out that the overall cost for solving all these problems is quite small.
This is because (a) the problems do not change much across different values of $\gamma$;  and (b) for a fixed $\gamma$, while moving across different values of $k$, the linear optimization problems are quite similar since
the weights $|\rho'_{\gamma}(|\beta^{k}_{i}|) |$ do not change much across $k$.
Thus the solutions obtained from one linear optimization problem can be used as a warm-start to solve the next linear optimization problem. This is found to reduce the
overall computation time. Both the first-order methods (described above) and simplex methods can gracefully take advantage of warm-starts.

\subsection{Dual Gradient Method} \label{sec:append-dual-algo-1}
Here we describe how to solve a problem of the form:
\begin{equation}\label{min-project-dset-1}
\begin{aligned}
\min_{\B\alpha} & \;\; \half \| \B\alpha - \M{\overline{c}}\|_{2}^2  +  \sum_{i=1}^{p} w_{i} | \alpha_{i}| \\
  \sbt &\;\; \|\M{A}\B\alpha - \M{b} \|_{\infty} \leq \delta_{},
\end{aligned}
\end{equation}
where we assume that $w_{i} \geq 0$ and the set $ \{\B\alpha:  \|\M{A}\B\alpha - \M{b} \|_{\infty} \leq \delta_{} \}$ is nonempty.
Note that the constraint set in~\eqref{min-project-dset-1} makes solving the primal form~\eqref{min-project-dset-1} challenging. However, due to the strong convexity of the objective a dual is smooth (has Lipschitz continuous gradient) and the non differentiability nicely separates across the dual variables -- we thus use dual proximal gradient algorithms to optimize~\eqref{min-project-dset-1}. This trick is often used in optimization and signal processing, see for example~\cite{combettes2010dualization}.


To derive a dual for Problem~\eqref{min-project-dset-1}, we note that it can be written equivalently as:
\begin{equation*}
\begin{aligned}
\min_{\B\alpha,\B\zeta} & \;\;  \half\| \B\alpha - \M{\overline{c}}\|_{2}^2  +  \sum_{i=1}^{p} w_{i} | \alpha_{i}| \\
  \sbt &\;\; \| \B\zeta\|_{\infty} \leq \delta_{}, \\
  & ~~ \B\zeta = \M{A}\B\alpha - \M{b}.
\end{aligned}
\end{equation*}
The minimum of the above problem can be obtained by maximizing a \emph{dual} problem, obtained by dualizing the equality constraints $\B\zeta = \M{A}\B\alpha - \M{b}$;
this consequently leads to the following
problem:
\begin{equation*}
\begin{myarray}[1.3]{r c l }
g(\B\mu) :=& \min\limits_{\B\zeta, \B\alpha: \| \B\zeta\|_{\infty} \leq \delta}& \Big(  \half\| \B\alpha - \M{\overline{c}}\|_{2}^2  +  \left\langle \B\mu, \B\zeta - (\M{A}\B\alpha - \M{b})  \right\rangle  \\
& & + \sum\limits_{i=1}^{p} w_{i} | \alpha_{i}|     \Big).
                \end{myarray}
\end{equation*}
The above can be simplified to:
\begin{equation*}
\begin{aligned}
& g(\B\mu) \\
=& \min_{ \B\alpha} \left(  \half\| \B\alpha - \M{\overline{c}}\|_{2}^2  - \| \B\mu\|_{1} \delta -   \langle \B\mu, (\M{A}\B\alpha - \M{b})  \rangle + \sum_{i=1}^{p} w_{i} | \alpha_{i}|     \right)  \\
=& g_{1} (\B\mu)  - \delta \|\B\mu\|_{1} ,
\end{aligned}
\end{equation*}
where,
\begin{equation*}
 g_{1}(\B\mu) = \min_{\B\alpha} \left (   \half\| \B\alpha - \M{\overline{c}}\|_{2}^2   -   \langle \B\mu, (\M{A}\B\alpha - \M{b})  \rangle + \sum_{i=1}^{p} w_{i} | \alpha_{i}|   \right).
 \end{equation*}
Note that $\widehat{\B\alpha}$,  the unique minimizer of the above problem, is given by:
\begin{equation*}
\begin{aligned}
\widehat{\B\alpha}  =& \argmin_{\B\alpha} \; \half \| \B\alpha - (\overline{\M{c}} + \M{A}^\top \B\mu) \|_{2}^2 + \sum_{i=1}^{p} w_{i} | \alpha_{i}|,~~~~~~~~~~~~~~~~\\
\text{i.e.,}~~~ \widehat{\alpha}_{i}=& \sgn (\overline{c}_{i} + \M{a}_{i}^\top \B\mu) \cdot  \max \left\{ | \overline{c}_{i} + \M{a}_{i}^\top \B\mu | - w_i , 0  \right\},  &
\end{aligned}
\end{equation*}
for $i = 1, \ldots, p,$
where $\M{a}_{i}$ is  the $i$th column of $\M{A}$.
It follows from standard convex analysis~\cite{rock-conv-96} that the function $\B\mu \mapsto g_{1}(\B\mu)$ is differentiable with its gradient given by:
$$\nabla g_{1}(\B\mu) = - (\M{A}\widehat{\B\alpha} - \M{b}),$$
and its gradient is Lipschitz continuous:
$$ \|\nabla g_{1}(\B\mu) - \nabla g_{1}(\B\mu') \| \leq \| \M{A}\|^2_{2} \| \B\mu - \B\mu'\| ,$$
where $\| \M{A} \|_{2}$ denotes the largest singular value of $\M{A}$ and for a vector $\M{u}$, the term $\| \M{u} \|$  denotes the usual $\ell_{2}$-norm of $\M{u}$.

By using standard quadratic programming duality theory~\cite{BV2004}, the minimum of Problem~\eqref{min-project-dset-1} can be obtained by maximizing the unconstrained dual problem $g(\B\mu)$ in the dual variable $\B\mu$, which is equivalent to the following
minimization problem:
\begin{equation}\label{project-DZ-set1-dual-11}
\min_{\B\mu}\; -g(\B\mu) =   \min_{\B\mu} \left (  -g_{1}(\B\mu)  + \delta \|\B\mu\|_{1} \right).
\end{equation}
This problem is of the composite form~\cite{nesterov2013gradient}, and proximal gradient descent methods~\cite{nesterov2004introductorynew,nesterov2013gradient,parikh2013proximal} apply to it directly.

For the special case of Problem~\eqref{project-DZ-set1}, the method described above applies with $w_{i} = 0, i=1, \ldots, p$.
Clearly, \eqref{project-DZ-set1-dual-11} is an $\ell_{1}$-regularized quadratic program, with the primal dual relationship being:
$\B\alpha = \overline{\M{c}} + \M{A}^\top \B\mu$.

Note that Problem~\eqref{min-project-dset-1} needs to be solved several times during the course of Algorithm~1 and Algorithm~2, across the different iterations.
Fortunately, these problems are not completely unrelated, in fact, they are quite  ``similar''. In Algorithm~2  the weights $w_{i}$ change; and in Algorithm~1, the parameter $\overline{\M{c}}$ changes.
Since the problems are similar, it is not unreasonable to expect that the optimal dual variables corresponding to these two problems do not change much.  Thus, it is useful to initialize the dual variable $\B\mu$ for
one instantiation of Problem~\eqref{min-project-dset-1} with the (dual) solution obtained from another instantiation of Problem~\eqref{min-project-dset-1}. This simple strategy leads to substantial performance gains over
solving the problems independent of one another.

\subsection{Proof of Theorem~\ref{thm:conv-rate-1}}\label{sec:thm:conv-rate-1}
\begin{proof}
Note that the sequence $\B\beta^{k}$, defined via~\eqref{obj-lla-2} satisfies the following relationship:
\begin{equation*}
\begin{aligned}
h(\B\beta^{k})  =& \overline{h}(\B\beta^{k}; \B\beta^{k}) \\
 \geq&~ \min\limits_{\B\beta}~  \overline{h}(\B\beta; \B\beta^{k}) ~~\sbt ~~ \| \M{X}^\top  (\M{y} - \M{X} \B\beta)  \|_{\infty} \leq \delta\\
 =&  \overline{h}(\B\beta^{k+1}; \B\beta^{k}).
\end{aligned}
\end{equation*}

Observing that
$$  \overline{h}(\B\beta^{k+1}; \B\beta^{k}) \geq h(\B\beta^{k+1}), $$
 we have:
 \begin{equation}\label{eqn:obj-val-dec-1}
 h(\B\beta^{k}) = \overline{h}(\B\beta^{k}; \B\beta^{k}) \geq   \overline{h}(\B\beta^{k+1}; \B\beta^{k}) \geq h(\B\beta^{k+1}),
 \end{equation}
and, thus, the sequence $h(\B\beta^{k})$ is decreasing. Subtracting $h(\B\beta^{k})$ from all sides of the above inequality, we derive:
$$ 0 \geq  \overline{h}(\B\beta^{k+1}; \B\beta^{k}) - h(\B\beta^{k}) \geq h(\B\beta^{k+1}) - h(\B\beta^{k}). $$
The first part of the above display gives us, using~\eqref{upper-bd-1-h}:
\begin{equation*}
\begin{aligned}
 0 \geq&   \overline{h}(\B\beta^{k+1}; \B\beta^{k}) - h(\B\beta^{k}) \\
 =&  \sum_{i=1}^{p}  \left\langle  \rho'_{\gamma}( |\beta^{k}_{i}|), | \beta^{k+1}_{i}|  - | \beta^{k}_{i}| \right \rangle = \Delta (\B\beta^{k}),
 \end{aligned}
 \end{equation*}
which means $\Delta(\B\beta^{k}) \leq 0$ for all $k$.  If $\Delta(\B\beta^{k}) < 0$, then $\B\beta^{k+1}$ leads to a strictly improved value of the objective function.
If $\Delta(\B\beta^{k}) =0$, then $\B\beta^{k}$ is a fixed point of the above update equation.  Hence, $\Delta(\B\beta^{k})$ is a measure of how far $\B\beta^{k}$ is from a first-order stationary point of Problem~\eqref{obj-lla-1}.

The display in~\eqref{eqn:obj-val-dec-1} shows that the objective values are decreasing, and, because the objective values are all bounded below (by zero), the decreasing sequence converges.

In addition, we have that
$$h(\B\beta^{k}) - h(\B\beta^{k+1}) \geq -\Delta (\B\beta^{k}). $$
Adding the above for $k= 1, \ldots, {\mathcal K},$ we have:
\begin{equation}
\begin{myarray}[1.3]{r c l}
h(\B\beta^1) - h(\B\beta^{\mk+1})  &\geq& \sum\limits_{\mk \geq k \geq 1} \left \{ -\Delta(\B\beta^{k}) \right \}  \\
&\geq& \mk \min\limits_{1\leq k\leq \mk}\left \{ -\Delta(\B\beta^{k}) \right\},
\end{myarray}
\end{equation}
which leads to the following convergence rate:
\begin{align}
 \min_{1\leq k\leq \mk}\left\{ -\Delta(\B\beta^{k}) \right\} \leq& \frac{1}{\mk} \left(h(\B\beta^1) - h(\B\beta^{\mk+1}) \right) \label{lin-last-1-1}\\
 \leq&   \frac{1}{\mk} \left(h(\B\beta^1) - \widehat{h} \right), \label{lin-last-1-2}
  \end{align}
where~\eqref{lin-last-1-2} follows from~\eqref{lin-last-1-1} by using the observation that
$h(\B\beta^k) \downarrow \widehat{h}$.

\end{proof}

\subsection{Additional details on Algorithm~3} \label{sec:details-algo-3}
We seek an upper bound to a simple variant of Problem~\eqref{obj-lla-1}:
\begin{equation}\label{obj-lla-1-mod}
\begin{aligned}
\min\limits_{\B\beta}& \;\; h(\B\beta):= \sum\limits_{i=1}^{p} \rho_{\gamma}( |\beta_{i}|) \;\; \\
\sbt & \;\; \| \M{X}^\top  (\M{y} - \M{X} \B\beta)  \|_{\infty} \leq \delta \\
& \;\; \beta_{i} = 0 , i \in {\mathcal I}^c,
\end{aligned}
\end{equation}
where $\text{Supp}(\widehat{\B\beta}^{(1)}) := \{ i : \widehat{\beta}^{(1)}_{i} \neq 0 , i =1, \ldots, p \} \subset {\mathcal I}$, and ${\mathcal I}^c$ is the complement of ${\mathcal I}$.
We assume, of course, that the feasible set in Problem~\eqref{obj-lla-1-mod} is nonempty.
A simple method for constructing ${\mathcal I}$, which we found to be quite useful in practice, is presented below.
Let $ {\mathcal B} \subset \{1, \ldots, p \}$ and ${\mathcal B}^c$ denote its complement.  We define the following set:
$$ {\mathcal F} ({\mathcal B}) := \left\{  \B\beta : \| \M{X}^\top (\M{y} - \M{X}\B\beta) \|_{\infty} \leq \delta, \beta_{i} = 0 , i \in {\mathcal B}^c \right\}. $$
Let $\widehat{\B\alpha}^{(1)}, \widehat{\B\beta}^{(1)}$ be the solutions produced by Algorithm~1.
Suppose we let $B$ denote the support of $\widehat{\B\beta}^{(1)}$; the size of $B$ is typically much smaller than $p$.
If ${\mathcal F}(B)$ is nonempty, we take ${\mathcal I}  = B$. Note, however, that ${\mathcal F}(B)$ may be empty, because  $\widehat{\B\alpha}^{(1)}, \widehat{\B\beta}^{(1)}$ are only approximately equal:
$\widehat{\B\alpha}^{(1)} \approx \widehat{\B\beta}^{(1)}$.
In this case, we need expand the set $B$, so that the set $\mathcal F(B)$ becomes nonempty.  There may be several ways to do this, but we found the following simple method to be quite useful in our numerical experiments.
\begin{enumerate}
\item If ${\mathcal F}(B)$ is empty, we consider the set $\{ |\widehat{\alpha}^{(1)}_{i}|, i \in  B^c\}$  and find the index of  the largest element in this set, which we denote by:
$\widehat{i} \in \argmax_{ i \in  B^c} \;  |\widehat{\alpha}^{(1)}_{i}|$.
\item Make $B$ larger by including this new feature $\widehat{i}$: we thus have  $ B \leftarrow B \cup \{ \widehat{i} \}$.
\item Check if the resulting set ${\mathcal F}(B)$ is nonempty, if not, we repeat the above steps until ${\mathcal F}(B)$ becomes nonempty.
\item We let $\mathcal I$ be the resulting set $B$ obtained upon termination: ${\mathcal I} = B$.
\end{enumerate}
Problem~\eqref{obj-lla-1-mod}, which is an optimization problem with fewer variables than Problem~\eqref{obj-lla-1} is found to deliver solutions
that are better upper bounds to Problem~\eqref{L0-DZ-1}. This also leads to better and numerically more robust solutions than those available directly from Algorithm~1.
The general algorithmic framework via sequential linear optimization, presented in Section~\ref{sec:weighted-l1}, readily applies to obtain good upper bounds to
Problem~\eqref{obj-lla-1-mod}.

\subsection{Tighter bounds on $\hat{\beta}_{i}$'s}\label{sec:tight-bounds-bet-1}
The bounds described via~\eqref{ubs-data-1} can be sharpened by making use of good upper bounds to the solution of Problem~\eqref{L0-DZ-1}.
Towards this end, we need to reformulate~\eqref{L0-DZ-1-mio-M}. Note that in Problem~\eqref{L0-DZ-1-mio-M}, if we take $\MU$ to be to be sufficiently large
then this will lead to a solution for Problem~\eqref{L0-DZ-1}. We rewrite Problem~\eqref{L0-DZ-1-mio-M} as follows:
\begin{equation}\label{L0-DZ-1-mio-M-mod1}
\begin{myarray}[1.3]{c  c  r}
\min \limits_{\B\beta, \M{z}, \alpha} &  \alpha & \\
\sbt& \sum\limits_{i=1}^{p} z_{i}\leq \alpha \\
&  - \delta  \leq d_{j}  - \langle \M{q}_{j}, \B\beta \rangle  \leq \delta, & j = 1, \ldots, p\\
& -\MU z_{j} \leq  \beta_{j} \leq \MU z_{j},& j = 1, \ldots, p\\
& z_{j} \in \{ 0 , 1 \}, & j = 1, \ldots, p,
\end{myarray}
\end{equation}
where the optimization variables are $\B\beta, \M{z} \in \RR^p$ and $\alpha \in \RR$.


For a fixed $\alpha$, consider the feasible set of Problem~\eqref{L0-DZ-1-mio-M-mod1}:
\begin{equation*}\label{set-s-1}
\begin{aligned}
{\mathcal S}_{\alpha} =
\left \{  \left(\B\beta, \M{z}\right) :
\begin{aligned}
 \sum_{j=1}^{p} z_{j} \leq \alpha,  \;\;  \| \M{X}^\top (\M{y} - \M{X}\B\beta)\|_{\infty} \leq \delta \\
|\beta_{j}| \leq \MU z_{j}, \;\; z_{j} \in \{ 0 , 1 \},  j = 1, \ldots, p
  \end{aligned} \right\}.
\end{aligned}
 \end{equation*}
Observe that
\begin{equation}\label{set-s-1-relax}
{\mathcal S}_{\alpha} \subset {\mathcal {\overline{S} }}_{\alpha},
\end{equation}
where
\begin{equation*}
\begin{aligned}
 {\mathcal {\overline{S} }}_{\alpha} = \left \{  \left(\B\beta, \M{z}\right) :
\begin{aligned}
 \sum_{j=1}^{p} z_{j} \leq \alpha,  \;\;  \| \M{X}^\top (\M{y} - \M{X}\B\beta)\|_{\infty} \leq \delta \\
|\beta_{j}| \leq \MU z_{j}, \;\; z_{j} \in [0 , 1],  j = 1, \ldots, p
  \end{aligned} \right\}
\end{aligned}
\end{equation*}
is obtained by relaxing the binary variables $z_{j} \in \{0, 1\}$ into the continuous variables $z_{j} \in [0, 1],$ for all $j=1, \ldots, p$.
Noting that ${\mathcal S}_{\alpha} \subset {\mathcal S}_{\alpha'}$ for $\alpha \leq \alpha'$; and using this along with~\eqref{set-s-1-relax} we have
$${\mathcal S}_{\alpha^*} \subset {\mathcal S}_{\alpha_{0}} \subset {\mathcal {\overline{S} }}_{\alpha_{0}} , $$
where $\alpha^*$ is the optimum value, and $\alpha_{0}$ is an upper bound to Problem~\eqref{L0-DZ-1-mio-M-mod1}, and hence $\alpha_{0} \geq \alpha^*$.
The above inequality leads to the following chain of inequalities:

\begin{equation*}
\begin{aligned}
\min_{ (\B\beta, \M{z})  \in {\mathcal S}_{\alpha^*} }    \beta_{i} &\geq&  \min_{ (\B\beta, \M{z})  \in {\mathcal S}_{\alpha_{0} }} \beta_{i} &\geq& \min_{ (\B\beta, \M{z})  \in {\mathcal {\overline{S} }}_{\alpha_0} } \beta_{i} :=  \mu^{-}_{i}(\alpha_{0}) \\
\max_{ (\B\beta, \M{z})  \in {\mathcal S}_{\alpha^*} }    \beta_{i} &\leq&  \max_{ (\B\beta, \M{z})  \in {\mathcal S}_{\alpha_{0} }} \beta_{i} &\leq& \max_{ (\B\beta, \M{z})  \in {\mathcal {\overline{S} }}_{\alpha_0} } \beta_{i} :=  \mu^{+}_{i}(\alpha_{0}).
\end{aligned}
\end{equation*}
The quantities  at the right end above, i.e. $\mu^{-}_{i}(\alpha_{0})$ and $\mu^{+}_{i}(\alpha_{0})$, can be computed by solving a pair of linear optimization problems:
%
\begin{equation}\label{bound-s-1-2}
\begin{aligned}
\begin{myarray}[1.3]{c c  c }
 \mu^{+}_{i}(\alpha_{0}) :=&  \max\limits_{\B\beta}  \;\; \beta_{i} & \\
\sbt~~ &  \| \M{X}^\top (\M{y} - \M{X}\B\beta)\|_{\infty} \leq \delta,& \\
& \| \B\beta\|_{\infty} \leq \MU, \\
& \| \B\beta \|_{1} \leq \MU \alpha_0,
\end{myarray} \\
&&\\
\begin{myarray}[1.3]{c c  c }
 \mu^{-}_{i}(\alpha_{0}) :=&  \min \limits_{\B\beta}  \;\; \beta_{i} & \\
\sbt~~ & \|\M{X}^\top (\M{y} - \M{X} \B\beta) \|_{\infty} \leq \delta_{},& \\
& \| \B\beta\|_{\infty} \leq \MU,  \\
& \| \B\beta \|_{1} \leq \MU \alpha_0.
\end{myarray}
\end{aligned}
\end{equation}
The quantities $\mu^{-}_{i}(\alpha_{0})$ and $\mu^{+}_{i}(\alpha_{0})$ are lower and upper bounds, respectively, for $\widehat{\beta}_{i}$ --- the bounds depend upon $\alpha_{0}$ and $\MU$.
Note that $\mu_i(\alpha_0):= \max\left\{\mu^{+}_{i}(\alpha_{0}), -\mu^{-}_{i}(\alpha_{0})\right\}$ provides an upper bound to $|\widehat{\beta}_{i}|$, which consequently leads to an improved estimate
for $\| \widehat{\B\beta} \|_{\infty}$ --- this suggests a way to adaptively refine $\MU$, and, thus, $\mu^{-}_{i}(\alpha_{0})$ and $\mu^{+}_{i}(\alpha_{0})$.

\section{Additional Experiments}\label{app-sec:addl-expts}
This section complements the experimental results shown in the main body of the paper.
Table~\ref{tab:stat-results-l0-l1-all} is an elaborate version of the representative results displayed in Figure~\ref{fig:l1-l0-DS-stat}.
Here, we consider different values of SNR and also display the prediction errors.  The results show that \LDSF~outperforms the $\ell_{1}$-\DSF~based methods in terms of estimating the true underlying regression coefficients, and does so with better variable selection properties.

\begin{table*}[h!]
\begin{center}
\resizebox{\textwidth}{.2\textheight}{\begin{tabular}{ c  c }
{\bf \Exampleone} $(n=200,p=500)$ & {\bf \Examplethree}  $(n=200,p=500)$  \smallskip \\

\begin{tabular}{|cccccc|}
  \hline
\err& SNR & L0-DS & L0-DS-Pol & L1-DS & Warm \\
  \hline
\bet & \snrone & 3.858 (0.466) &  1.123 (0.254) & 5.266 (0.162) & 3.429 (0.19) \\
 \errzone & \snrone & 6 (0.543) & 16 (0.844) & 102 (1.327) & 43 (1.327) \\
 \xbet & \snrone & 0.158 (0.013) & 0.581 (0.038) & 0.18 (0.004) & 0.133 (0.004) \\
  \nnz & \snrone & 24 (0.274) & 35 (0.663) & 122 (1.327) & 63 (1.266) \\  \hline
 \bet & \snrthree & 0.635 (0.053) & 0.172 (0) & 1.58 (0.049) & 0.664 (0.035) \\
 \errzone & \snrthree & 1 (0.137) & 0 (0) & 102 (1.327) & 29 (1.371) \\
 \xbet & \snrthree & 0.031 (0.002) & 0.249 (0) & 0.054 (0.001) & 0.028 (0.001) \\
  \nnz & \snrthree & 21 (0.137) & 20 (0) & 122 (1.327) & 49 (1.371) \\

   \hline
\end{tabular}&
\begin{tabular}{|cccccc|}
  \hline
\err& SNR &L0-DS & L0-DS-Pol & L1-DS & Warm \\
  \hline

\bet & \snrone & 150.386 (8.993) & 64.17 (4.888) & 137.073 (4.511) & 102.583 (5.711) \\
 \errzone & \snrone & 11 (0.247) & 16 (0.573) & 38 (1.266) & 14 (0.658) \\
 \xbet & \snrone & 0.155 (0.007) & 0.109 (0.003) & 0.132 (0.005) & 0.123 (0.005) \\
  \nnz & \snrone & 16 (0.362) & 23 (0.362) & 46 (1.447) & 22 (0.693) \\ \hline
 \bet & \snrthree & 45.599 (2.508) & 13.773 (1.39) & 47.113 (1.016) & 38.256 (2.068) \\
 \errzone & \snrthree & 9 (0.411) & 11 (0.482) & 36 (1.387) & 14 (0.724) \\
 \xbet & \snrthree & 0.045 (0.001) & 0.052 (0.001) & 0.043 (0.001) & 0.035 (0.001) \\
  \nnz & \snrthree & 17 (0.151) & 23 (0.392) & 50 (1.266) & 26 (0.663) \\

   \hline
\end{tabular} \\

&  \\

{\bf \Examplefour}  $(n=100,p=500)$ & {\bf \Exampleeight} $(n=100,p=300)$  \smallskip \\
\begin{tabular}{|cccccc|}
  \hline
\err& SNR &L0-DS & L0-DS-Pol & L1-DS & Warm \\
  \hline

\bet & \snrone & 7.322 (0.65) & 6.823 (0.693) & 6.452 (0.263) & 7.674 (0.347) \\
 \errzone & \snrone & 10 (0.573) & 9 (0.814) & 43 (1.538) & 21 (1.116) \\
 \xbet & \snrone & 0.168 (0.015) & 0.966 (0.051) & 0.172 (0.004) & 0.193 (0.008) \\
  \nnz & \snrone & 13 (0.271) & 13 (0.392) & 47 (1.658) & 22 (1.096) \\ \hline
 \bet & \snrthree & 2.395 (0.379) & 1.988 (0.333) & 3.245 (0.097) & 2.529 (0.357) \\
 \errzone & \snrthree & 10 (0.795) & 10 (0.795) & 38 (1.357) & 27 (1.343) \\
 \xbet & \snrthree & 0.054 (0.004) & 0.515 (0.036) & 0.056 (0.002) & 0.051 (0.004) \\
  \nnz & \snrthree & 18 (0.685) & 18 (0.685) & 48 (1.357) & 35 (1.266) \\

   \hline
\end{tabular}&
\begin{tabular}{|cccccc|}
  \hline
\err& SNR &L0-DS & L0-DS-Pol & L1-DS & Warm \\
  \hline

\bet & \snrone & 0.019 (0.006) & 0.015 (0) & 0.191 (0.027) & 0.018 (0.007) \\
 \errzone & \snrone & 0 (0) & 0 (0) & 36 (0.693) & 12 (0.85) \\
 \xbet & \snrone & 0.033 (0.005) & 0.955 (0) & 0.101 (0.007) & 0.028 (0.003) \\
  \nnz & \snrone & 2 (0) & 2 (0) & 38 (0.693) & 14 (0.85) \\ \hline
 \bet & \snrthree & 0.005 (0.002) & 0.004 (0) & 0.057 (0.008) & 0.006 (0.002) \\
 \errzone & \snrthree & 0 (0) & 1 (0.082) & 36 (0.693) & 13 (0.877) \\
 \xbet & \snrthree & 0.006 (0.002) & 0.527 (0.006) & 0.03 (0.002) & 0.007 (0.001) \\
  \nnz & \snrthree & 2 (0) & 3 (0.082) & 38 (0.693) & 15 (0.877) \\

   \hline
\end{tabular}   \\
\end{tabular}}
\end{center}
\caption{ {{Tables showing the statistical performance of four different methods: ``L0-DS", ``L0-DS-Pol", ``L1-DS'' and ``Warm'', described in Section~\ref{sec:stats-prop-expt}.
 The standard errors are given in parentheses (they are computed in the same fashion as in Figure~\ref{fig:l1-l0-DS-stat}). The \LDSF~based methods deliver models with
good accuracy in estimating the regression coefficients, and the estimated models are sparser than those for the $\ell_{1}$-based method and the method based on Algorithm~2, which is a heuristic strategy to approximate good upper bounds for the \LDSF~problem. 
}}}\label{tab:stat-results-l0-l1-all}
\end{table*}

 An important advantage of the \LDSF~based methods is that
they deliver models that are very sparse.  The polished version of the \LDSF~ is found to exhibit better statistical performance than the original \LDSF~estimator.
Table~\ref{tab:polished-l1-l0} compares the polished versions of the \LDSF~and the $\ell_{1}$-\DSF, and finds that the performance of the former approach is significantly better.

\begin{table}[h!]
\begin{center}
\scalebox{0.95}{\begin{tabular}{c  }
\bf \Exampleone~$(n=200,p=500)$  \medskip \\
\begin{tabular}{|cccc|}
  \hline
\err & SNR & L0-DS-Pol & L1-DS-Pol  \\
  \hline
 \bet & \snrone & 1.123 (0.254) & 2.163 (0.139)   \\
 \errzone & \snrone & 16 (0.844) & 30 (1.116)   \\
  \xbet &\snrone  & 0.581 (0.038) & 0.777 (0.02)   \\
  \nnz & \snrone  & 35 (0.663) & 50 (1.146)   \\ \hline
  \bet &  \snrthree & 0.172 (0) & 0.282 (0.011)  \\
  \errzone  & \snrthree & 0 (0) & 11 (0.392)   \\
  \xbet &  \snrthree& 0.249 (0) & 0.292 (0.003)  \\
  \nnz & \snrthree & 20 (0) & 31 (0.392)   \\    \hline
\end{tabular}
\end{tabular}}\end{center}
\caption{{\upshape{Tables comparing the polished version of \LDSF~with the polished version of $\ell_{1}$-\DSF.  The standard errors are given in parentheses.
The statistical performance of $\ell_{1}$-\DSF~is inferior, most likely due to its weaker variable selection properties.}}} \label{tab:polished-l1-l0}
\end{table}

\end{appendix}

\bibliographystyle{plainnat_my}

\clearpage

\small{\bibliography{rahul_dbm3}}

\begin{thebibliography}{43}
\providecommand{\natexlab}[1]{#1}
\providecommand{\url}[1]{\texttt{#1}}
\expandafter\ifx\csname urlstyle\endcsname\relax
  \providecommand{\doi}[1]{doi: #1}\else
  \providecommand{\doi}{doi: \begingroup \urlstyle{rm}\Url}\fi

\bibitem[sup()]{supercomputer}
{Top500 Supercomputer Sites, Directory page for Top500 lists. Result for each
  list since June 1993}.
\newblock \url {http://www.top500.org/statistics/sublist/}.
\newblock Accessed: 2013-12-04.

\bibitem[Beck and Teboulle(2009)]{fista-09}
A.~Beck and M.~Teboulle.
\newblock A fast iterative shrinkage-thresholding algorithm for linear inverse
  problems.
\newblock \emph{SIAM Journal on Imaging Sciences}, 2\penalty0 (1):\penalty0
  183--202, 2009.

\bibitem[Becker et~al.(2011)Becker, Cand{\`e}s, and Grant]{becker2011templates}
S.~R. Becker, E.~J. Cand{\`e}s, and M.~C. Grant.
\newblock Templates for convex cone problems with applications to sparse signal
  recovery.
\newblock \emph{Mathematical Programming Computation}, 3\penalty0 (3):\penalty0
  165--218, 2011.

\bibitem[Bertsekas(1999)]{bertsekas-99-nourl}
D.~P. Bertsekas.
\newblock \emph{{Nonlinear Programming}}.
\newblock Athena Scientific, Belmont, Massachusetts, 2nd edition, 1999.

\bibitem[Bertsimas and King(2015)]{bertsimas2015or}
D.~Bertsimas and A.~King.
\newblock Or forum -- an algorithmic approach to linear regression.
\newblock \emph{Operations Research}, 2015.

\bibitem[Bertsimas and Mazumder(2014)]{bertsimas2014least}
D.~Bertsimas and R.~Mazumder.
\newblock Least quantile regression via modern optimization.
\newblock \emph{Annals of Statistics}, 42\penalty0 (6):\penalty0 2494--2525,
  2014.

\bibitem[Bertsimas and Weismantel(2005)]{bertsimas2005optimization_new}
D.~Bertsimas and R.~Weismantel.
\newblock \emph{Optimization over integers}.
\newblock Dynamic Ideas Belmont, 2005.

\bibitem[Bertsimas et~al.(2016)Bertsimas, King, and
  Mazumder]{bertsimas2015best}
D.~Bertsimas, A.~King, and R.~Mazumder.
\newblock Best subset selection via a modern optimization lens.
\newblock \emph{Annals of Statistics}, 44(2):\penalty0 813--852, 2016.

\bibitem[Bickel et~al.(2009)Bickel, Ritov, and Tsybakov]{bickel1}
P.~Bickel, Y.~Ritov, and A.~Tsybakov.
\newblock Simultaneous analysis of lasso and dantzig selector.
\newblock \emph{Annals of Statistics}, 37:\penalty0 1705–--1732, 2009.

\bibitem[Bixby(2012)]{bixby}
R.~E. Bixby.
\newblock A brief history of linear and mixed-integer programming computation.
\newblock \emph{Documenta Mathematica, Extra Volume: Optimization Stories},
  pages 107--121, 2012.

\bibitem[Boyd and Vandenberghe(2004)]{BV2004}
S.~Boyd and L.~Vandenberghe.
\newblock \emph{Convex Optimization}.
\newblock Cambridge University Press, Cambridge, 2004.

\bibitem[Boyd et~al.(2011)Boyd, Parikh, Chu, Peleato, and
  Eckstein]{boyd-admm1nourl}
S.~Boyd, N.~Parikh, E.~Chu, B.~Peleato, and J.~Eckstein.
\newblock \emph{Foundations and Trends in Machine Learning}.
\newblock Number 3(1). Now Publishers, 2011.

\bibitem[B{\"u}hlmann and {van-de-Geer}(2011)]{buhlmann2011statistics}
P.~B{\"u}hlmann and S.~{van-de-Geer}.
\newblock \emph{Statistics for high-dimensional data}.
\newblock Springer, 2011.

\bibitem[Burer and Saxena(2012)]{burer2012milp}
S.~Burer and A.~Saxena.
\newblock The {MILP} road to {MIQCP}.
\newblock In \emph{Mixed Integer Nonlinear Programming}, pages 373--405.
  Springer, 2012.

\bibitem[Candes et~al.(2008)Candes, Wakin, and Boyd]{boyd08-new}
E.~Candes, M.~Wakin, and S.~Boyd.
\newblock Enhancing sparsity by reweighted $\ell_1$ minimization.
\newblock \emph{Journal of Fourier Analysis and Applications}, 14\penalty0
  (5):\penalty0 877--905, 2008.

\bibitem[Candes and Tao(2007)]{candes2007dantzig}
E.~Candes and T.~Tao.
\newblock The {D}antzig selector: statistical estimation when p is much larger
  than n.
\newblock \emph{Annals of Statistics}, pages 2313--2351, 2007.

\bibitem[Combettes et~al.(2010)Combettes, D{\~u}ng, and
  V{\~u}]{combettes2010dualization}
P.~L. Combettes, {\ DJ}.~D{\~u}ng, and B.~C. V{\~u}.
\newblock Dualization of signal recovery problems.
\newblock \emph{Set-Valued and Variational Analysis}, 18\penalty0
  (3-4):\penalty0 373--404, 2010.

\bibitem[Donoho and Johnstone(1994)]{dono:john:1994}
D.~Donoho and I.~Johnstone.
\newblock Ideal spatial adaptation by wavelet shrinkage.
\newblock \emph{Biometrika}, 81:\penalty0 425--455, 1994.

\bibitem[Efron et~al.(2004)Efron, Hastie, Johnstone, and Tibshirani]{LARS}
B.~Efron, T.~Hastie, I.~Johnstone, and R.~Tibshirani.
\newblock Least angle regression (with discussion).
\newblock \emph{Annals of Statistics}, 32\penalty0 (2):\penalty0 407--499,
  2004.
\newblock ISSN 0090-5364.

\bibitem[Fan and Li(2001)]{Fan01}
J.~Fan and R.~Li.
\newblock Variable selection via nonconcave penalized likelihood and its oracle
  properties.
\newblock \emph{Journal of the American Statistical Association}, 96\penalty0
  (456):\penalty0 1348--1360(13), 2001.

\bibitem[Freund et~al.(2017)Freund, Grigas, and Mazumder]{freund2015new}
R.~M. Freund, P.~Grigas, and R.~Mazumder.
\newblock A new perspective on boosting in linear regression via subgradient
  optimization and relatives.
\newblock \emph{Annals of Statistics (to appear)}, 2017.

\bibitem[Friedlander and Tseng(2007)]{friedlander2007exact}
M.~P. Friedlander and P.~Tseng.
\newblock Exact regularization of convex programs.
\newblock \emph{SIAM Journal on Optimization}, 18\penalty0 (4):\penalty0
  1326--1350, 2007.

\bibitem[Friedman et~al.(2007)Friedman, Hastie, Hoefling, and
  Tibshirani]{FHT2007}
J.~Friedman, T.~Hastie, H.~Hoefling, and R.~Tibshirani.
\newblock Pathwise coordinate optimization.
\newblock \emph{Annals of Applied Statistics}, 2\penalty0 (1):\penalty0
  302--332, 2007.

\bibitem[Giannessi and Tomasin(1973)]{giannessi1973nonconvex}
F.~Giannessi and E.~Tomasin.
\newblock Nonconvex quadratic programs, linear complementarity problems, and
  integer linear programs.
\newblock In \emph{5th Conference on Optimization Techniques Part I}, pages
  437--449. Springer, 1973.

\bibitem[Gurobi~Optimization(2015)]{gurobi}
I.~Gurobi~Optimization.
\newblock Gurobi optimizer reference manual, 2015.
\newblock URL \url{http://www.gurobi.com}.

\bibitem[Hastie et~al.(2009)Hastie, Tibshirani, and Friedman]{FHT-09-new}
T.~Hastie, R.~Tibshirani, and J.~Friedman.
\newblock \emph{The Elements of Statistical Learning, Second Edition: Data
  Mining, Inference, and Prediction (Springer Series in Statistics)}.
\newblock Springer New York, 2 edition, 2009.
\newblock ISBN 0387848576.

\bibitem[Hastie et~al.(2015)Hastie, Tibshirani, and
  Wainwright]{hastie2015statistical}
T.~Hastie, R.~Tibshirani, and M.~Wainwright.
\newblock \emph{Statistical Learning with Sparsity: The Lasso and
  Generalizations}.
\newblock CRC Press, FL, 2015.

\bibitem[Hemmecke et~al.(2010)Hemmecke, K{\"o}ppe, Lee, and
  Weismantel]{hemmecke2010nonlinear}
R.~Hemmecke, M.~K{\"o}ppe, J.~Lee, and R.~Weismantel.
\newblock Nonlinear integer programming.
\newblock In \emph{50 Years of Integer Programming 1958-2008}, pages 561--618.
  Springer, 2010.

\bibitem[James and Radchenko(2009)]{PVR.gds}
G.~M. James and P.~Radchenko.
\newblock A generalized {D}antzig selector with shrinkage tuning.
\newblock \emph{Biometrika}, 96:\penalty0 323--337, 2009.

\bibitem[James et~al.(2009)James, Radchenko, and Lv]{james2009dasso}
G.~M. James, P.~Radchenko, and J.~Lv.
\newblock Dasso: connections between the {Dantzig} selector and lasso.
\newblock \emph{Journal of the Royal Statistical Society: Series B (Statistical
  Methodology)}, 71\penalty0 (1):\penalty0 127--142, 2009.

\bibitem[J{\"u}nger et~al.(2009)J{\"u}nger, Liebling, Naddef, Nemhauser,
  Pulleyblank, Reinelt, Rinaldi, and Wolsey]{junger200950}
M.~J{\"u}nger, T.~M. Liebling, D.~Naddef, G.~L. Nemhauser, W.~R. Pulleyblank,
  G.~Reinelt, G.~Rinaldi, and L.~A. Wolsey.
\newblock \emph{50 Years of Integer Programming 1958-2008: From the Early Years
  to the State-of-the-art}.
\newblock Springer Science \& Business Media, 2009.

\bibitem[Linderoth and Lodi(2010)]{linderoth2010milp}
J.~T. Linderoth and A.~Lodi.
\newblock {MILP} software.
\newblock \emph{Wiley encyclopedia of operations research and management
  science}, 2010.

\bibitem[Liu et~al.(2016)Liu, Yao, and Li]{mipgo-li-2016}
H.~Liu, T.~Yao, and R.~Li.
\newblock Global solutions to folded concave penalized nonconvex learning.
\newblock \emph{Annals of Statistics}, 44 (2):\penalty0 629--659, 2016.

\bibitem[Lu and Zhang(2013)]{lu2013sparse}
Z.~Lu and Y.~Zhang.
\newblock Sparse approximation via penalty decomposition methods.
\newblock \emph{SIAM Journal on Optimization}, 23\penalty0 (4):\penalty0
  2448--2478, 2013.

\bibitem[Mazumder et~al.(2011)Mazumder, Friedman, and Hastie]{mhf-09-jasa}
R.~Mazumder, J.~Friedman, and T.~Hastie.
\newblock Sparsenet: Coordinate descent with non-convex penalties.
\newblock \emph{Journal of the American Statistical Association},
  117(495):\penalty0 1125--1138, 2011.

\bibitem[Nesterov(2013)]{nesterov2013gradient}
Y.~Nesterov.
\newblock Gradient methods for minimizing composite functions.
\newblock \emph{Mathematical Programming}, 140\penalty0 (1):\penalty0 125--161,
  2013.

\bibitem[Nesterov(2004)]{nesterov2004introductorynew}
Y.~Nesterov.
\newblock \emph{Introductory Lectures on Convex Optimization: A Basic Course}.
\newblock Kluwer, Norwell, 2004.

\bibitem[Parikh and Boyd(2013)]{parikh2013proximal}
N.~Parikh and S.~Boyd.
\newblock Proximal algorithms.
\newblock \emph{Foundations and Trends in optimization}, 1\penalty0
  (3):\penalty0 123--231, 2013.

\bibitem[Rockafellar(1996)]{rock-conv-96}
R.~Rockafellar.
\newblock \emph{Convex Analysis}.
\newblock Princeton University Press, Princeton, 1996.

\bibitem[Tibshirani(1996)]{Ti96}
R.~Tibshirani.
\newblock Regression shrinkage and selection via the lasso.
\newblock \emph{Journal of the Royal Statistical Society, Series B},
  58:\penalty0 267--288, 1996.

\bibitem[Vielma(2015)]{vielma2015mixed}
J.~P. Vielma.
\newblock Mixed integer linear programming formulation techniques.
\newblock \emph{SIAM Review}, 57\penalty0 (1):\penalty0 3--57, 2015.

\bibitem[Williams(2013)]{williams2013model}
H.~P. Williams.
\newblock \emph{Model building in mathematical programming}.
\newblock John Wiley \& Sons, 2013.

\bibitem[Zhang and Huang(2008)]{ZH08}
C.-H. Zhang and J.~Huang.
\newblock The sparsity and bias of the lasso selection in high-dimensional
  linear regression.
\newblock \emph{Annals of Statistics}, 36\penalty0 (4):\penalty0 1567--1594,
  2008.

\end{thebibliography}

\end{document}